\documentclass[a4paper,twoside,prd,showpacs,preprintnumbers,superscriptaddress,nofootinbib]{revtex4-1}
\pdfoutput=1 

\usepackage[left=1.5cm,right=1.5cm,top=2cm,bottom=2cm]{geometry}
\usepackage{amssymb,amsmath,bm,natbib}
\usepackage{color}
\usepackage{slashed}
\usepackage{graphics}
\usepackage{graphicx}
\usepackage[utf8]{inputenc}
\usepackage[caption=false]{subfig}
\usepackage{hyperref}
\usepackage{url}
\usepackage{dsfont}
\usepackage{float} 
\usepackage{cancel}
\usepackage{xcolor,colortbl}
\usepackage{rotating}
\usepackage[export]{adjustbox}
\usepackage{subfig}
\usepackage[capitalize,noabbrev]{cleveref}
\usepackage{booktabs}
\usepackage{multirow}
\usepackage{braket}
\usepackage{float}
\usepackage{soul}
\usepackage{epsfig}
\usepackage[shortlabels]{enumitem}
\usepackage{makecell}
\usepackage[thinlines]{easytable}
\usepackage{breakurl}

\usepackage[formats]{listings}
\usepackage{tcolorbox,enumitem,verbatim,titlesec}
\usepackage{textcomp}
\Crefname{equation}{Eq.}{Eqs.}
\Crefname{table}{Table}{Tables}
\Crefname{figure}{Fig.}{Figs.}

\newcommand{\GeV}{{\rm\ GeV}}

\newcommand{\fb}{{\rm\ fb}}
\newcommand{\Lag}{{\mathcal L}}

\newcolumntype{C}[1]{>{\centering\let\newline\\\vskip 5pt\arraybackslash\hspace{0pt}}m{#1}}

\definecolor{coolblack}{rgb}{0.0, 0.18, 0.39}
\newcommand\cbc[1]{{\color{coolblack}{#1}}}

\allowdisplaybreaks

\begin{document}

\title{Deconstructing squark contributions to di-Higgs production at the LHC}

\author{Stefano Moretti}
\email{s.moretti@soton.ac.uk; stefano.moretti@physics.uu.se}
\affiliation{School of Physics and Astronomy, University of Southampton, Highfield, Southampton SO17 1BJ, UK}
\affiliation{Department of Physics and Astronomy, Uppsala University, Box 516, SE-751 20 Uppsala, Sweden}

\author{Luca Panizzi}
\email{luca.panizzi@physics.uu.se}
\affiliation{Department of Physics and Astronomy, Uppsala University, Box 516, SE-751 20 Uppsala, Sweden}

\author{J\"orgen Sj\"olin}
\email{sjolin@fysik.su.se}
\affiliation{Department of Physics, Stockholm University, 10691, Stockholm, Sweden}

\author{Harri Waltari}
\email{harri.waltari@physics.uu.se}
\affiliation{Department of Physics and Astronomy, Uppsala University, Box 516, SE-751 20 Uppsala, Sweden}

\begin{abstract}

We present a novel approach to the study of di-Higgs production via gluon-gluon fusion at the LHC. The relevant Feynman diagrams involving two Standard Model-like Higgs bosons $hh$ are computed within a simplified model approach that enables one to interpret possible signals of new physics in a model-independent way as well as to map these onto specific theories. This is possible thanks to a decomposition of such a signal process into all its squared amplitudes and their relative interferences, each of which has a well-defined coupling structure. We illustrate the power of this procedure for the case of both a minimal and next-to-minimal representation of 
supersymmetry, for which the new physics effects are due to top squarks entering the loops of $gg\to hh$. The squarks yield both a change of the integrated cross section and peculiar kinematic features in its differential distributions with respect to the Standard Model. These effects can in turn be traced back to the relevant diagrammatic and coupling structures and allow for a detailed analysis of the process. In order to do so, we perform systematic scans of the parameter spaces of such new physics scenarios and identify benchmark points which exhibit potentially observable features during the current and upcoming runs of the LHC.

\end{abstract}

\maketitle

\tableofcontents

\section{Introduction}

The past few years have seen full data taking at Run 1 \& 2 of the Large Hadron Collider (LHC), with Run 3 ongoing, and this has been heralding a new era of precision Higgs physics.
The importance of its detailed understanding cannot be overstated since the Higgs properties, originating from the mass generation mechanism, play a key role in searches for New Physics (NP) which interacts with the Electro-Weak (EW) sector. The measured Higgs couplings so far still allow for a minimal version of the Higgs mechanism, but one should bear in mind that several of the most important properties directly related to the Higgs potential (i.e., the Higgs self-couplings and Higgs-top couplings) are still very weakly constrained. For this reason, Higgs physics remains one of the highest priorities in the global high energy physics programme. 

Of all Higgs channels where Beyond the Standard Model (BSM) effects can be searched for, one which carries particular relevance is di-Higgs production, as it can give direct access to the structure of the scalar potential triggering EW Symmetry Breaking (EWSB). At the LHC, this process primarily takes place via $gg$-fusion \cite{Plehn:1996wb,Dawson:1998py}, i.e., via the loop subprocess $gg\to hh$ involving triangle and box graphs. BSM physics can therefore enter here via loops of new particles at the same perturbative order as those of SM ones, chiefly, the top-quark. In fact, another circumstance which renders these virtual effects from BSM physics potentially accessible specifically in di-Higgs production is that there exists a strong cancellation between the aforementioned triangle and box diagrams involving the top quark \cite{Glover:1987nx} at (or near) the SM limit. 

This is the context of our present study, which therefore dispenses of the case involving `resonant' production of heavy Higgs states decaying into (pairs of) the SM one, as we are concerned here with `non-resonant' di-Higgs production. This has been studied extensively at the LHC to date \cite{CMS:2018sxu,ATLAS:2018ili,CMS:2019noi,ATLAS:2019vwv,CMS:2020tkr,ATLAS:2021ifb,CMS:2022hgz,ATLAS:2022xzm,ATLAS:2023qzf} and the standard approach is to parametrise BSM effects via modifications of the Yukawa coupling of the top quark (entering both the triangle and box diagrams) and/or the Higgs self-coupling (entering solely the triangle ones). In turn, these can be interpreted in terms of an Effective Field Theory (EFT), wherein any new particle entering $gg\to hh$ production via loops is essentially integrated out.

Our approach is different. We allow for a BSM spectrum in which the presence of new states is accounted for exactly, at one-loop level, so that they can have any masses, including those comparable to the dynamical scale of the LHC (of order TeV), for which an EFT approach cannot be adopted. We do so by exploiting a simplified model approach that can easily be translated into any fundamental theory responsible for the EWSB dynamics chosen by Nature. This comes in the form of a numerical toolbox enabling Monte Carlo (MC) studies at a level of sophistication comparable to actual experimental analyses (albeit the version used in the present analysis is one-loop only). Borrowing as reference BSM framework the one of Supersymmetry (SUSY), we will prove that the very same particles (i.e., the top quark companions in SUSY, so-called squarks) responsible for enabling extensions to the SM without hierarchy problems can give sizeable effects in di-Higgs production, when their typical masses are indeed of order TeV. 

The plan of the paper is as follows. In the next section we will introduce the realisations of SUSY that we will be adopting for exemplifying our approach, including both a minimal and non-minimal version \cite{Moretti:2019ulc}. We will then describe our toolbox. Numerical results for $gg\to hh$ will follow,
for both Run 3 and the High-Luminosity LHC (HL-LHC) \cite{Gianotti:2002xx}, in turn preceding our conclusions.

\section{Stops and the Higgs mass}

We shall first consider the Minimal Supersymmetric Standard Model (MSSM), which is based on the superpotential
\begin{equation}
    W_{\mathrm{MSSM}} = y^u QH_{u}U^c + y^d QH_{d}D^c + y^{\ell} LH_{d}E^c + \mu H_{u}H_{d}.
\end{equation}

The Higgs couples strongly to the top-stop sector, so the (s)tops will have a large impact on both the Higgs mass and its production cross sections. The stop mass matrix at tree-level is
\begin{eqnarray}
    M_{\tilde{t}_{L}\tilde{t}_{L}}^{2} & = & m_{\tilde{Q}_{33}}^{2} + m_{t}^2 +m_{Z}^2 \cos 2\beta \left( \frac{1}{2}-\frac{2}{3}\sin^{2} \theta_{W}\right),\\
    M_{\tilde{t}_{L}\tilde{t}_{R}}^{2} & = & m_{t}(\mu \cot \beta -A_{t}),\label{eq:stopmix}\\
    M_{\tilde{t}_{R}\tilde{t}_{R}}^{2} & = & m_{\tilde{U}_{33}}^{2}+m_{t}^2+\frac{2}{3}m_{Z}^{2}\cos 2\beta \sin^{2} \theta_{W}.
\end{eqnarray}
Here $m_{\tilde{Q}}^{2}$ and $m_{\tilde{U}}^{2}$ are the soft SUSY breaking squark masses, $\tan\beta=\langle H_{u}^{0}\rangle/\langle H_{d}^{0}\rangle$ and $A_{t}$ is the trilinear soft SUSY breaking coupling $H_{u}^{0}\tilde{t}_{L}\tilde{t}_{R}$. There are loop corrections to this mass matrix \cite{Donini:1995wh}, which are potentially large.

In the MSSM there is a tree-level bound on the (lightest) SM-like Higgs mass, 
\begin{equation}
    m_{h}^{2}\leq m_{Z}^{2}\cos^{2}2\beta,
\end{equation}
so large loop corrections are needed to produce the observed Higgs mass of $125 \GeV$. Since the top Yukawa coupling is the largest coupling in the MSSM, the most economical way to achieve this is to take $\tan\beta$ to be large (so that the tree-level mass is maximal and the SM-like Higgs boson couples maximally to the top/stop), make the stops heavy (so that the top-stop Supersymmetric cancellation is incomplete) and to introduce large stop mixing (which maximises their loop corrections).

Large mixing in the stop sector creates a large mass splitting between the stops. If one of the soft masses $m_{\tilde{Q}_{33}}^{2}$ or $m_{\tilde{U}_{33}}^{2}$ is somewhat smaller than the other, we might end up with a relatively light stop together with a heavy one. Currently, stop masses down to $600 \GeV$ are allowed if the mass splitting between the stop and the Lightest Supersymmetric Particle (LSP) is small \cite{CMS:2021beq,CMS:2021eha}, since searches based on missing transverse momentum lose their sensitivity in the compressed case.

In the MSSM, a large stop mass splitting is a necessity to achieve a $125 \GeV$ Higgs mass, so searches based on missing transverse momentum are sensitive to the heavier stop. Hence, it must be heavy, the lower bound being around $1250 \GeV$ \cite{CMS:2021eha,ATLAS:2020xzu} 
Since the stop mixing parameter $A_{t}$ needs to be large in the MSSM, the bubble, triangle and box diagrams with trilinear stop couplings (see following section) give a large contribution to Higgs pair production, as known from literature \cite{Cao:2013si,Batell:2015koa,Huang:2017nnw} and which we will elaborate upon.

It is well known that in the MSSM the Higgs trilinear self-coupling is always close to its SM value \cite{Hollik:2001px,Dobado:2002jz}. Numerical scans show that deviations can be at most at the $3\%$ level \cite{Wu:2015nba}. Hence, in the MSSM, the modifications of the Higgs self-coupling will not lead to observable effects in Higgs pair production at the LHC as the predicted precision of the di-Higgs cross section measurement is around $40\%$ \cite{Cepeda:2019klc}.\\

In the Next-to-MSSM (NMSSM), the superpotential is
\begin{equation}
    W = W_{\mathrm{MSSM}}(\mu=0) +\lambda S H_{u} H_{d} +\frac{\kappa}{3}S^{3}.
\end{equation}
In this case there are additional tree-level contributions to the lightest Higgs boson mass so that the bound reads as \cite{Drees:1988fc} 
\begin{equation}\label{eq:NMSSMbound}
    m_{h}^{2} \leq m_{Z}^{2}\left( \cos^{2}2\beta + \frac{2\lambda^{2}}{g^{2}+g'^{2}}\sin^{2}2\beta\right).
\end{equation}

If $1\lesssim \tan \beta \lesssim 3$ and $\lambda$ is large, the Higgs mass can be close to $125 \GeV$ without loop corrections. Hence in the NMSSM one can have two light stops \cite{Beuria:2015mta} but, due to the experimental constraints \cite{CMS:2021eha}, they need to be nearly degenerate with each other and the LSP. In such a case we need to require stop mixing to be minimal, $\mu\cot\beta \simeq A_{t}$ (see \eqref{eq:stopmix}) and hence the contribution from the trilinear bubble and triangle diagrams will be small. 
Furthermore, due to the extended Higgs sector, it is possible to have a large deviation from the SM prediction to the trilinear Higgs self-coupling even if other Higgs couplings are SM-like \cite{Wu:2015nba}. An enhancement can occur when $\tan\beta$ is close to $1$ and $\lambda$ is large while a suppression requires a second, singlet-dominated scalar to be light and $\lambda$ to be small. This second option is incompatible with light squarks. We give a qualitative argument for this behaviour in appendix \ref{sec:trilinear}.\\

In our analysis, we perform a comprehensive scan in the MSSM and then select some benchmark points from the NMSSM to represent cases not present in the 
MSSM, but relevant for the di-Higgs process. The parameter space of the MSSM is chosen as follows.
 We choose the mostly right-handed stop to be the light one and to have a small mass splitting,
\begin{equation}
    m_{\tilde{t}_{1}}-m_{\tilde{\chi}^{0}_{1}} < 10 \GeV,
\end{equation}
with the lightest neutralino, which we take to be higgsino-like. This mass splitting can always be arranged by choosing the value of the $\mu$-parameter. The mostly left-handed squark doublet will then be heavy ($m_{\tilde{t}_{2}}> 1250 \GeV$). We scan over the values of $m_{\tilde{Q}_{33}}^{2}$, $m_{\tilde{U}_{33}}^{2}$, $\tan\beta$ and $A_{t}$, which are the ones that determine the squark contribution to Higgs pair production. The other soft masses are fixed such that the rest of the sparticle spectrum is heavier than the stops and the higgsinos.

We also show a few NMSSM benchmarks representing cases that cannot be realised in the MSSM. One is when we make both stops as light as possible and adjust the Higgs mass to $125 \GeV$ by tuning $\tan\beta$ and $\lambda$. For this benchmark the Higgs trilinear self-coupling is about $50\%$ larger than in the SM, but the trilinear Higgs-stop couplings are much smaller than in the MSSM case. A second benchmark point represents a case where the Higgs trilinear self-coupling is about $60\%$ larger than in the SM and the stop masses and mixing are similar to the MSSM points.

\section{Numerical analysis}

\begin{figure}[h]
\begin{minipage}{.16\textwidth}
\includegraphics[width=\textwidth]{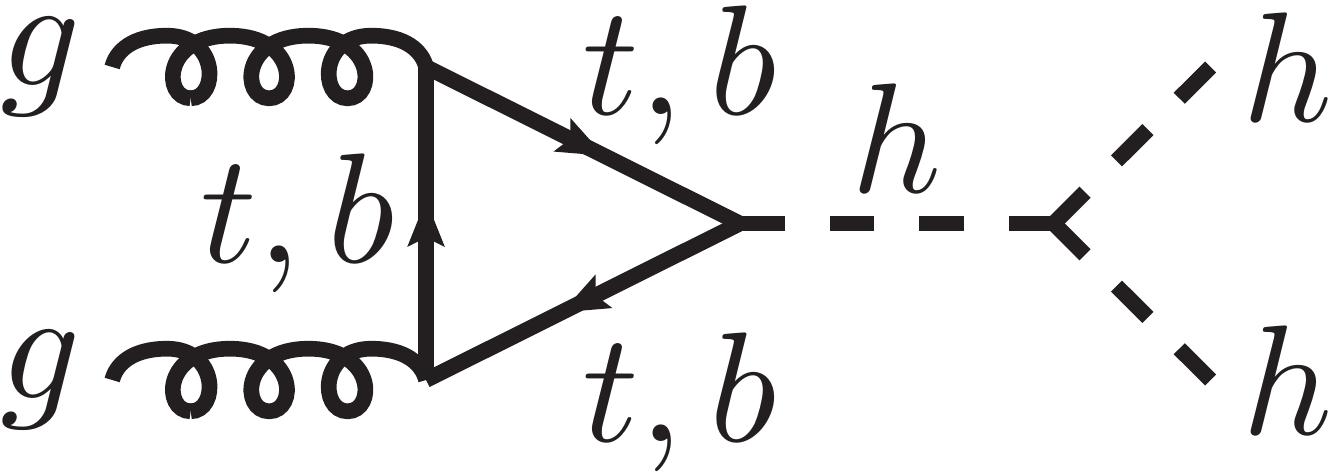}
\end{minipage}\hskip 20pt
\begin{minipage}{.16\textwidth}
\includegraphics[width=\textwidth]{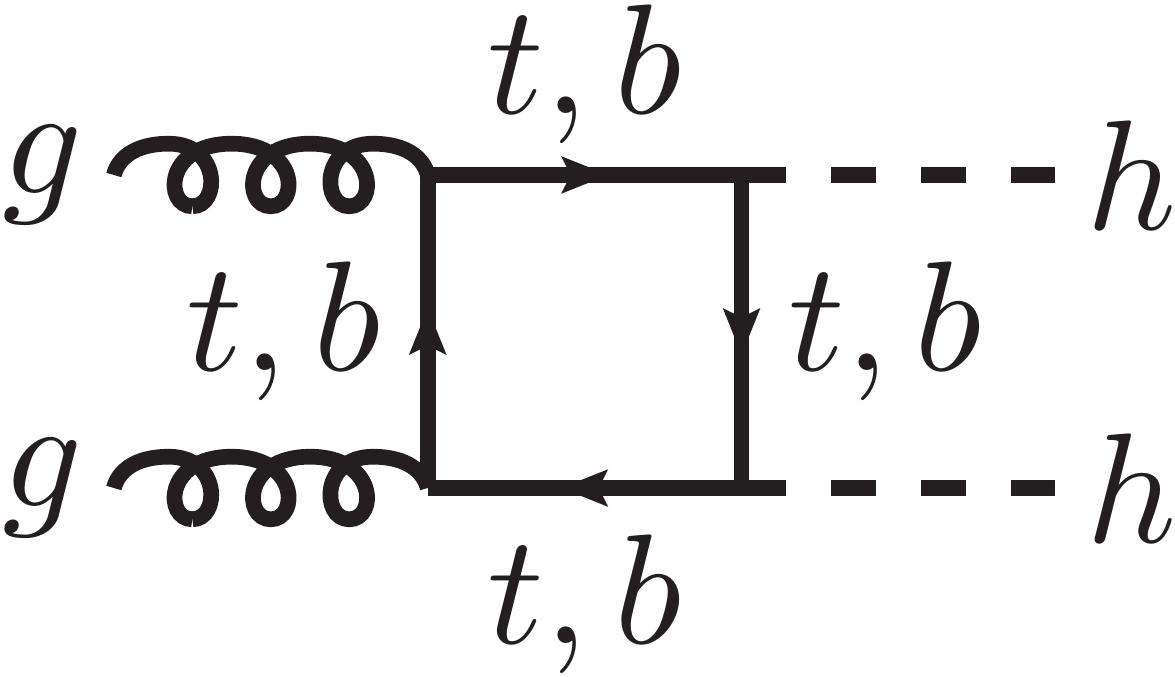}
\end{minipage}
\caption{\label{fig:SMtopologies} SM topologies for di-Higgs production at the LHC.}
\end{figure}

The contributions to the di-Higgs final state from the SM are represented by the well-known, destructively-interfering topologies shown in \Cref{fig:SMtopologies}, where the only coupling parameters entering the amplitudes are proportional to the Higgs trilinear self-coupling $\lambda v$ and the Yukawas $y_{t,b}$. 

The SUSY-inspired parameters that are relevant for the di-Higgs process (squark masses and couplings) will be included in a simplified Lagrangian which assumes that every other SUSY particle does not participate significantly in the process: this means all other scalars are too heavy or their couplings are too small and all other squarks are also decoupled. 

First of all we notice, trivially, that the electric charge of the squarks propagating in the loops does not play any role in the process. What actually matters for a model-independent treatment of the process is that, besides introducing NP modifications of the SM couplings, there are new coloured particles propagating in the loops. Then, the key properties
of such particles are their representation under $SU (3)$ of QCD, their spin, how many they are and the size of their
couplings with the Higgs boson. Only if the new particles are fermions, their electric charge plays a role, due to the fact that heavy quarks with charge $2/3$ or $-1/3$ can mix with the SM quarks.

Limiting our treatment to only two scalars in the fundamental representation of $SU(3)$, the simplified squark Lagrangian we consider in this analysis only contains interactions between those two squarks $\tilde q_i$ (with $i=1,2$ labelling the mass eigenstates), $h$, $t$, and NP contributions to the couplings $h^3$ and $ht\bar t$ (we will neglect modifications to the bottom Yukawa coupling in the following.). The model contains two coloured scalars in the loop so it can be adapted to the SUSY scenarios described in the previous section, and also considering that it is realistically very hard (if not plainly inconsistent) to build scenarios in which one of the two scalar top partners is light enough to produce visible effects at the LHC and the other is too heavy and decoupled from low-energy physics. The simplified interaction Lagrangian therefore generically reads as 
{\setlength{\arraycolsep}{2pt}
\begin{equation}
\Lag_{\text{NP}}^{\tilde q} = -(\lambda^{\rm SM} + \kappa_{hhh}) v h^3 - {1\over\sqrt2}(y_t^{\rm SM} + \kappa_{htt}) h\bar t t + 
v h (\tilde q_1^*~\tilde q_2^*)\left(\begin{array}{cc} 
\kappa_{h\tilde q \tilde q}^{11} & 
\kappa_{h\tilde q \tilde q}^{12} \\[4pt] 
\boldsymbol{\cdot} & 
\kappa_{h\tilde q \tilde q}^{22}\end{array}\right) \left(\begin{array}{c} \tilde q_1 \\[4pt] \tilde q_2 \end{array}\right) +
h h (\tilde q_1^*~\tilde q_2^*)\left(\begin{array}{cc} 
\kappa_{hh\tilde q \tilde q}^{11} & 
\kappa_{hh\tilde q \tilde q}^{12} \\[4pt] 
\boldsymbol{\cdot} & 
\kappa_{hh\tilde q \tilde q}^{22}\end{array}\right) \left(\begin{array}{c} \tilde q_1 \\[4pt] \tilde q_2 \end{array}\right),
\label{eq:LagNPstop}
\end{equation}
}
where the trilinear and quartic couplings are kept independent to account in a model-independent way for further NP effects which may alter the relation between the two. In the following, we identify $\tilde q_{1,2}$ with $\tilde t_{1,2}$, and all the couplings are assumed to be real.

This Lagrangian has been implemented in {\sc Feynrules} \cite{Alloul:2013bka} to obtain a {\sc UFO} \cite{Degrande:2011ua} output suitable for simulations at Next-to-Leading Order (NLO) in QCD with the {\sc MG5\_aMC} \cite{Alwall:2014hca} Monte Carlo (MC) generator\footnote{The model is publicly available on {\sc HEPMDB} \cite{hepmdb} at this link: \href{https://hepmdb.soton.ac.uk/hepmdb:0223.0337}{https://hepmdb.soton.ac.uk/hepmdb:0223.0337} and it actually contains four new coloured scalars in the fundamental of $SU(3)$ (recall that we only consider two of these for this analysis). The additional squarks have been included to account for possible extensions.}. For di-Higgs production this is obviously necessary as the process is at one-loop at leading order (LO). In all our results we used the NNPDF3.0 LO PDF set \cite{NNPDF:2014otw}.

The following analysis is aimed at finding which values of the parameters of NP can alter significantly the signal cross section and at the same time exhibit peculiar kinematic features which could lead to its observation during Run 3 or the HL-LHC.
Crucially, our implementation of the $gg\to hh$ process starting from the aforementioned simplified Lagrangian is such that we are able to identify each component entering the total cross section, i.e., the contributions of the square of
the relevant diagrams as well as the interferences between all of these, thereby affording us with significant diagnostic power of the emerging signal. While this approach, as emphasised, is model independent, we will be applying it to both the MSSM and NMSSM described in the previous section, in order to show its effectiveness. Studies of squark effects of the kind looked for here have been performed in literature before (e.g., see \cite{Huang:2017nnw}), but what this approach adds is the ability to directly reverse engineer the dynamics involved and interpret it already at the experimental analysis stage.

\subsection{Signal cross section and parametric dependence}

We first have performed a parametric scan within the MSSM: the scan parameters and their ranges are given in \Cref{tb:MSSMscan}. The spectra were calculated with \textsc{SPheno v4.0.4} \cite{Porod:2003um,Porod:2011nf}. We required the lighter stop to be heavier than $600\GeV$ and the heavier stop to be heavier than $1250\GeV$ with the Higgs mass being in the interval $125.2\pm0.3\GeV$ ($\sim 2\sigma$ around the central value). We set the $\mu$ parameter so that the higgsino-like neutralino LSP is slightly lighter than the lighter stop, hence, we may use $600\GeV$ as the lower bound for the lighter stop mass. Other parameters are kept fixed such that the rest of the BSM spectrum is decoupled, as previously mentioned.

\begin{table}[h]
    \centering
    \begin{tabular}{l c c}
    \hline
    \hline
\noalign{\vskip 3pt}
      Parameter & minimum & maximum\\
\noalign{\vskip 3pt}
    \hline
\noalign{\vskip 3pt}
      $\tan\beta$  & $7$ & $50$ \\
      $A_{t}$ (GeV)  & $1500$ & $3500$\\
      $m_{\tilde{U}_{33}}^{2}$ (GeV$^{2}$) & $1.35\times 10^{6}$ & $2\times 10^{6}$\\
      $m_{\tilde{Q}_{33}}^{2}$(GeV$^{2}$) & $2.2\times 10^{6}$ & $3.5\times 10^{6}$\\
\noalign{\vskip 3pt}
    \hline
    \hline
    \end{tabular}
    \caption{Parameter ranges for the scan in the MSSM. All other soft scalar masses are larger than those mentioned. The $\mu$-parameter is adjusted so that the higgsino-like LSP is slightly lighter than the lightest stop.}
    \label{tb:MSSMscan}
\end{table}

\begin{figure}[h]
\begin{minipage}{.2825\textwidth}
\includegraphics[width=\textwidth]{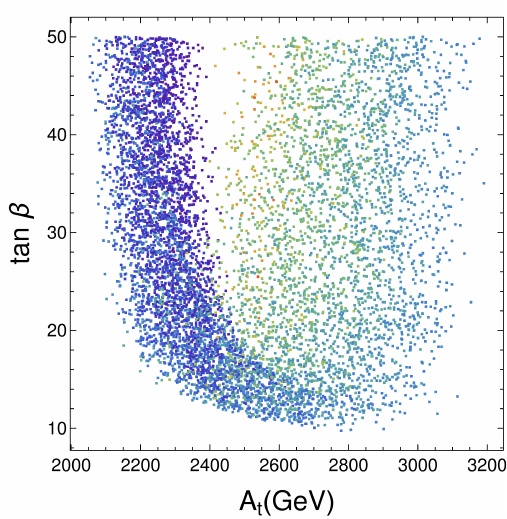}
\end{minipage}\hfill
\begin{minipage}{.2925\textwidth}
\vspace{3pt}
\includegraphics[width=\textwidth]{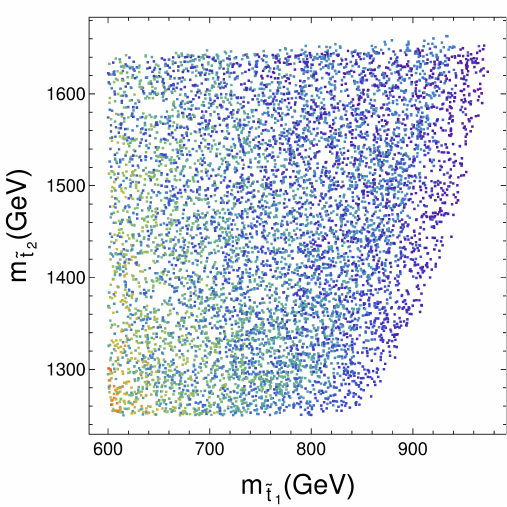}
\end{minipage}\hfill
\begin{minipage}{.3175\textwidth}
\vspace{-5pt}
\hspace{-1pt}\includegraphics[width=\textwidth]{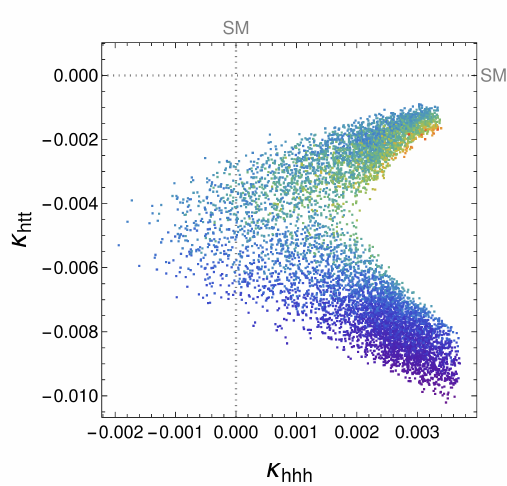}
\end{minipage}\hfill
\begin{minipage}{.1\textwidth}
\vspace{-11pt}\includegraphics[width=\textwidth]{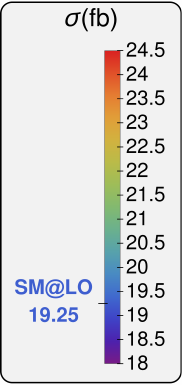}
\end{minipage}\hfill
\caption{\label{fig:scatterplot} Representative examples of LO di-Higgs cross sections (shown as a colour gradient) as function of pairs of MSSM input parameters (\ul{Left}), stop masses (\ul{Centre}), or the parameters of the simplified Lagrangian~\Cref{eq:LagNPstop} (\ul{Right}), highlighting their correlations.}
\end{figure}

The resulting MSSM couplings and masses have been translated into parameters of the simplified Lagrangian \Cref{eq:LagNPstop}. With the obtained set of benchmark points we have performed MC simulations and determined correlations between parameters which maximise the cross section, as shown in Fig. \ref{fig:scatterplot}.

We found preferred directions in the parameter space for which the cross section increases, around specific combinations of the stop masses and for specific values of the trilinear parameters, especially the modifiers of SM couplings $\kappa_{hhh}$ and $\kappa_{htt}$.

In \Cref{fig:scatterplot} we show some of the parametric dependencies. Obviously light squark masses lead to larger deviations from the SM, so the maximal cross sections are seen close to $m(\tilde{t}_1)=600\GeV$, $m(\tilde{t}_2)=1250\GeV$, the lighter stop mass being the more important one in increasing the cross section. Large $\tan\beta\gtrsim 20$ is preferred as the top-stop correction to the Higgs mass is maximal and the $125\GeV$ Higgs mass can then be reached with lighter stops than is the case for low $\tan\beta$.

In the MSSM with fixed squark masses and $\tan\beta$ you either have an interval of possible $A_{t}$ values leading to an acceptable Higgs mass (at low $\tan\beta$) or two intervals of $A_{t}$, where between these intervals the Higgs is too heavy. In the lower of these intervals the squark bubbles and triangles give a smaller contribution and the cross section is SM-like. In the higher interval significant enhancements are possible, if the squarks are light. The region of allowed values for $A_{t}$ depends on the average stop mass\footnote{This is essentially to avoid a color-breaking vacuum. In general a large ratio of $X_{t}/M_{S}$, where $X_{t}=A_{t}-\mu\cot\beta$ and $M_{S}$ is the average tree-level stop mass, can lead to a color-breaking vacuum, though the relation is not so simple if loop corrections are taken into account \cite{Camargo-Molina:2013sta,Camargo-Molina:2014pwa}. Our scan ranges have $X_{t}/M_{s}<2.5$, which is considered to lead to a stable enough color-conserving vacuum.}: the heavier the stops, the larger the maximum value for $A_{t}$. Therefore, the maximal cross sections are obtained with values of $A_{t}$ which are intermediate within the scan range, but which are maximal for the given stop masses and $\tan\beta$.

The point which produces the highest cross section, defined in \Cref{tab:MSSMBP}, will be considered in the following as benchmark point for the kinematic analysis. 

\begin{table}[h]
\centering
\begin{tabular}{lc}
\hline
\hline
\noalign{\vskip 3pt}
\multicolumn{2}{c}{\textbf{MSSM Benchmark Point}}\\
\noalign{\vskip 3pt}
\hline
\noalign{\vskip 3pt}
Input parameter & Value\\
\noalign{\vskip 3pt}
\hline
\noalign{\vskip 3pt}
$\tan\beta$ & $45.4$\\
\noalign{\vskip 3pt}
$A_{t}$ (GeV) & $2595$\\
\noalign{\vskip 3pt}
$m_{\tilde{U}_{33}}^{2}$ (GeV$^{2}$) & $1.547\times 10^{6}$ \\
\noalign{\vskip 3pt}
$m_{\tilde{Q}_{33}}^{2}$(GeV$^{2}$) & $2.447\times 10^{6}$\\
\noalign{\vskip 3pt}
\hline
\hline
\end{tabular} \hskip 20pt
\begin{tabular}{lc}
\hline
\hline
\noalign{\vskip 3pt}
\multicolumn{2}{c}{\textbf{MSSM Benchmark Point}}\\
\noalign{\vskip 3pt}
\hline
\noalign{\vskip 3pt}
Masses and couplings & Value\\
\noalign{\vskip 3pt}
\hline
\noalign{\vskip 3pt}
$m_{\tilde t_1}$ (GeV) & 600.6 \\
\noalign{\vskip 3pt}
$m_{\tilde t_2}$ (GeV) & 1301.0 \\
\noalign{\vskip 3pt}
$\kappa_{hhh}$ & $3.34\times 10^{-3}$\\
\noalign{\vskip 3pt}
$\kappa_{htt}$ & $-1.68\times 10^{-3}$\\
\noalign{\vskip 3pt}
$\left(\begin{array}{cc} 
\kappa_{h\tilde t\tilde t}^{11} & \kappa_{h\tilde t\tilde t}^{12}\\
\boldsymbol{\cdot} & \kappa_{h\tilde t\tilde t}^{22} \end{array}\right)$ & $\left(\begin{array}{cc} 
-6.690 & 7.228 \\
\boldsymbol{\cdot} & 8.519 \end{array}\right)$\\
\noalign{\vskip 3pt}
$\left(\begin{array}{cc} 
\kappa_{hh\tilde t\tilde t}^{11} & \kappa_{hh\tilde t\tilde t}^{12}\\
\boldsymbol{\cdot} & \kappa_{hh\tilde t\tilde t}^{22} \end{array}\right)$ & 
$\left(\begin{array}{cc} 
-0.6702 & -0.0174\\
\boldsymbol{\cdot} & -0.6374 \end{array}\right)$ \\
\noalign{\vskip 3pt}
\hline
\hline
\end{tabular}
\caption{\label{tab:MSSMBP} Benchmark point for the MSSM. \underline{Left}: Treel-level input parameters for SPheno. The loop corrections to the stop mass matrix are large, so the soft masses plus the loop corrections are approximately $m_{\tilde{U}_{33}}^{2}=5.3\times 10^{5}$~GeV$^{2}$ and $m_{\tilde{Q}_{33}}^{2}=1.43\times 10^{6}$~GeV$^{2}$ and the effective value for $A_{t}=2750$~GeV after loop corrections. \underline{Right}: Mass spectrum and couplings, as defined in the simplified model Lagrangian of \Cref{eq:LagNPstop}. For di-Higgs production the $\kappa_{hh\tilde t\tilde t}^{12}$ coupling is not relevant.}
\end{table}

\subsection{Kinematic distributions}

\subsubsection{Deconstruction of the signal}

To perform a differential analysis of the signal associated with the propagation of stops in the loops, we deconstruct the signal into basic independent components. The total signal is then obtained as a weighted sum of such components. This procedure allows us to analyse separately their peculiar kinematic features, assess their relevance for the final result and understand semi-analytically which range of parameters maximises the potential observability of the signal.

Limiting our treatment to only two stops we can build a limited number of elements. Labelling with a black dot ({\Large $\bullet$}) the NP contributions to SM couplings, with a red dot ({\Large \color{red}{$\bullet$}}) the purely NP couplings and with red Feynman propagators the new particles ($\tilde t_{1,2}$), the signal amplitudes can be deconstructed in a small number of elements, listed in  \Cref{tab:deconstructedtopologies}.

\begin{table}[]
\centering
{\setlength\arraycolsep{5pt}
\begin{tabular}{cccc}
\hline\hline
\noalign{\vskip 3pt}
& \textbf{Topology type} & \textbf{Feynman diagrams} & \textbf{Amplitude} \\
\noalign{\vskip 3pt}
\hline
\noalign{\vskip 5pt}
\textbf{1} & \textbf{Modified Higgs trilinear coupling} &
\raisebox{-.5\height}{\includegraphics[width=.16\textwidth]{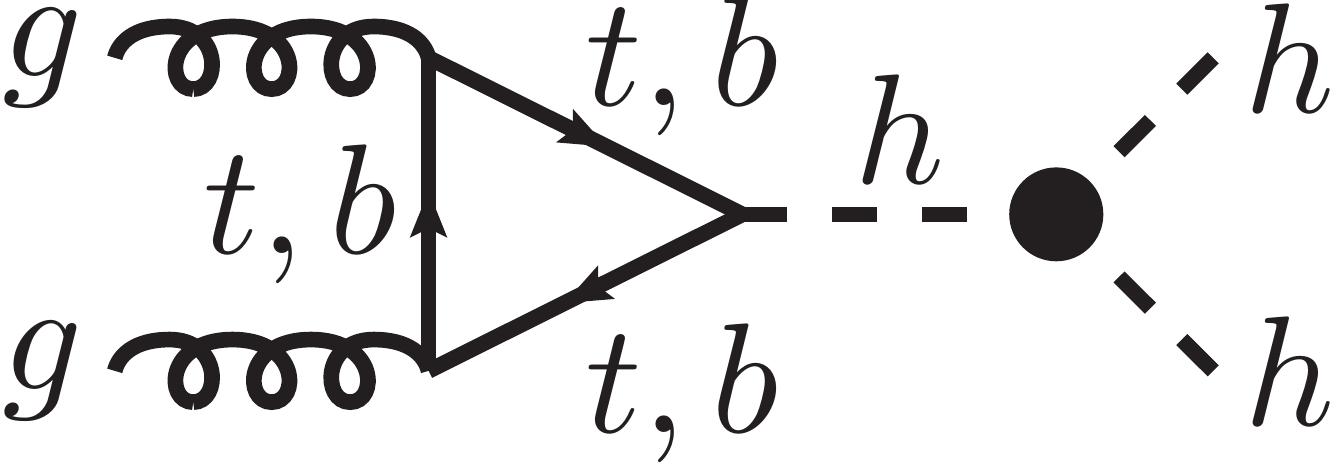}}
&
$\mathcal A_i \propto \kappa_{hhh}$\\
\noalign{\vskip 3pt}
\hline
\noalign{\vskip 3pt}
\textbf{2} & \textbf{One modified Yukawa coupling} &
\raisebox{-.35\height}{\includegraphics[width=.16\textwidth]{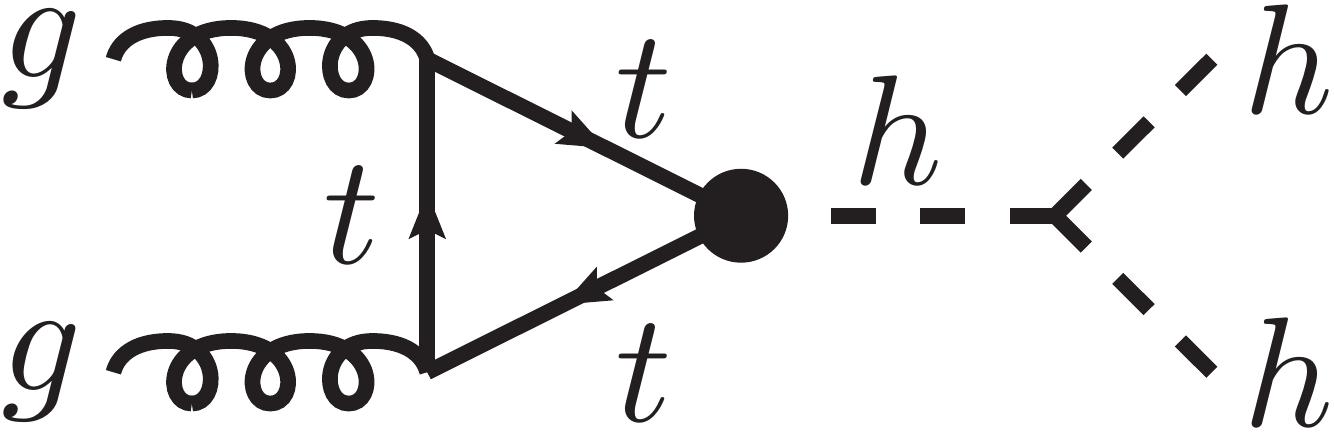}}
\raisebox{-.45\height}{\includegraphics[width=.16\textwidth]{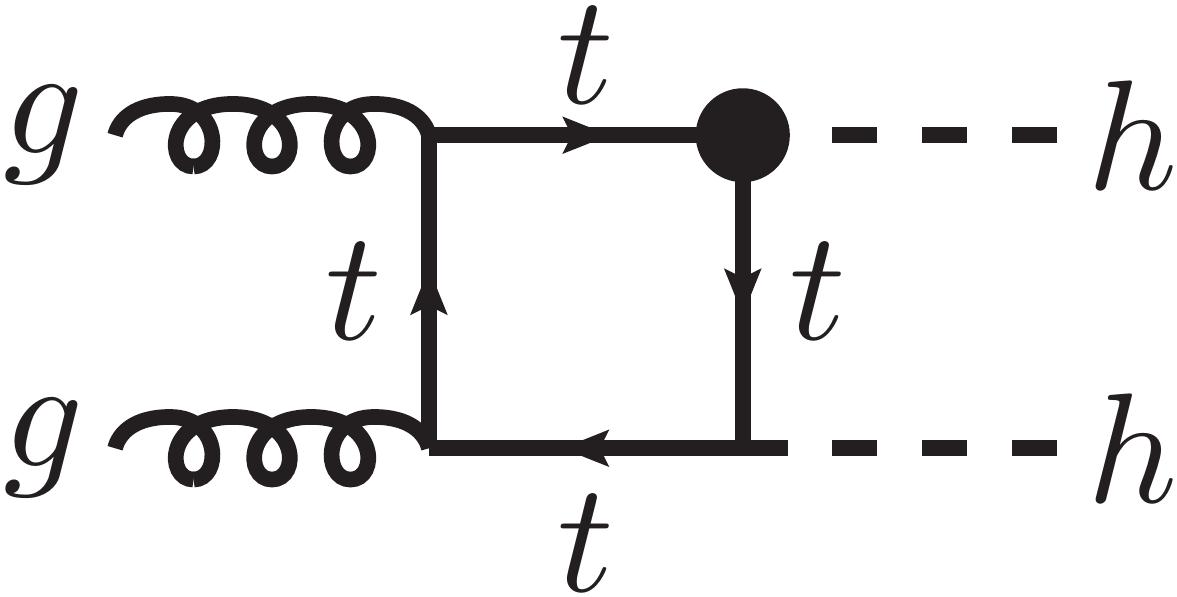}}
&
$\mathcal A_i \propto \kappa_{htt}$ \\
\noalign{\vskip 3pt}
\hline
\noalign{\vskip 5pt}
\textbf{3} & 
$\begin{array}{c}
\text{\bf Modified Higgs trilinear coupling}\\
\text{\bf and modified Yukawa coupling}
\end{array}$ 
&
\raisebox{-.4\height}{\includegraphics[width=.16\textwidth]{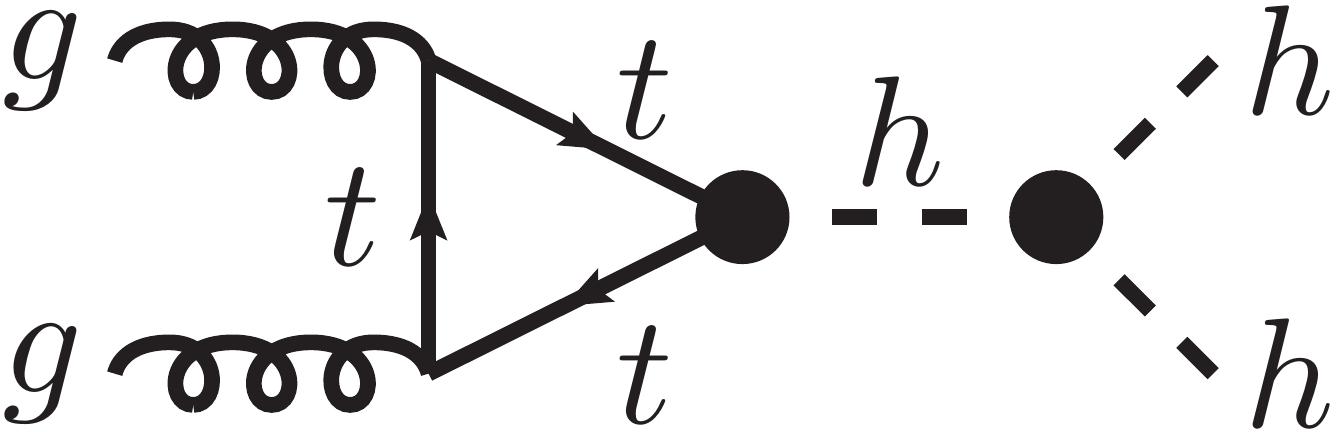}} &
$\mathcal A_i \propto \kappa_{hhh} \kappa_{htt}$\\
\noalign{\vskip 5pt}
\hline
\noalign{\vskip 3pt}
\textbf{4} & \textbf{Two modified Yukawa couplings}&
\begin{minipage}{.16\textwidth}
\includegraphics[width=\textwidth]{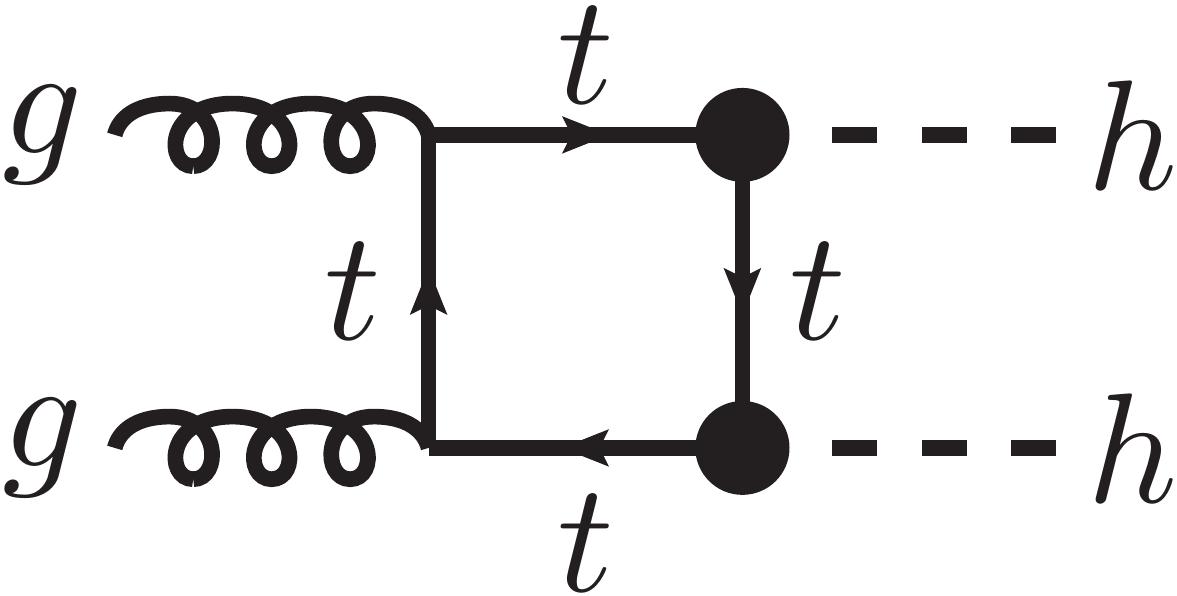} 
\end{minipage} &
$\mathcal A_i \propto \kappa_{htt}^2$ \\
\noalign{\vskip 2pt}
\hline
\noalign{\vskip 3pt}
\textbf{5} & 
$\begin{array}{c}
\text{\bf Bubble and triangle}\\
\text{\bf with $h\tilde t \tilde t$ couplings}
\end{array}$ 
&
\raisebox{-.5\height}{\includegraphics[width=.16\textwidth]{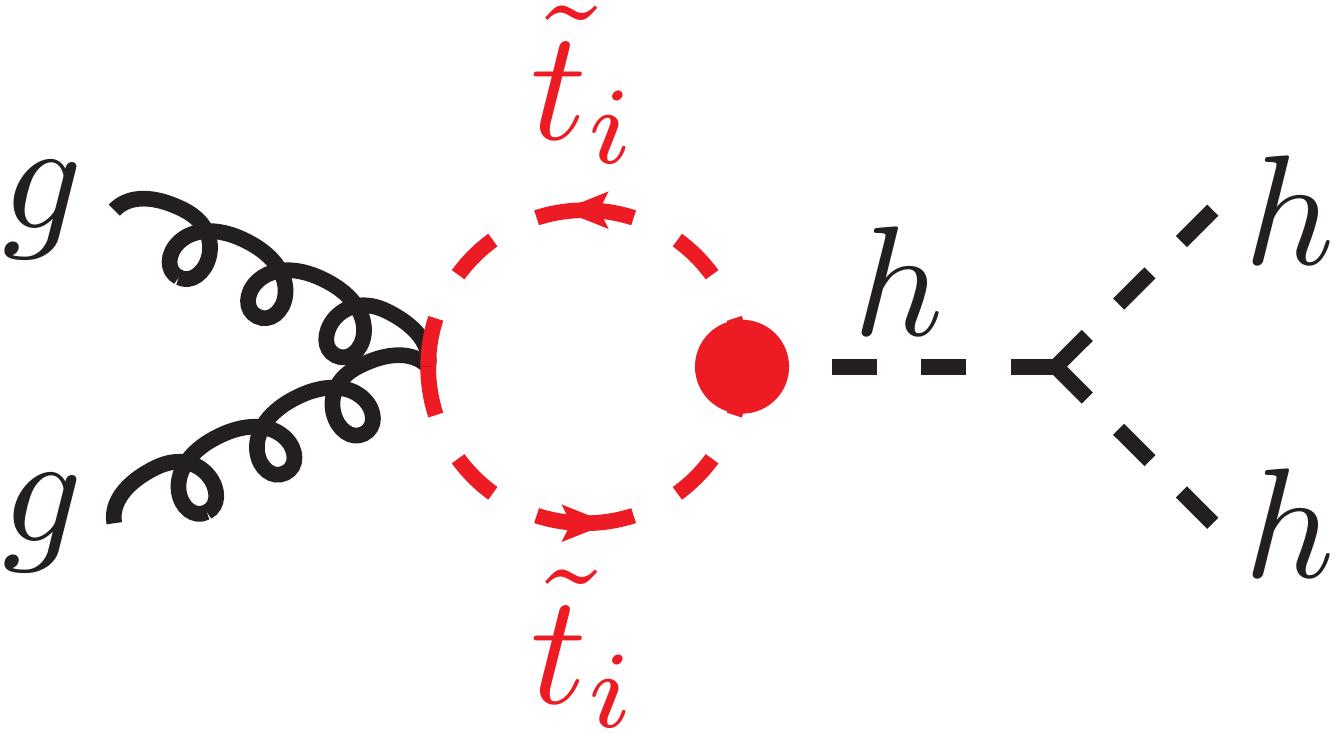}}
\raisebox{-.52\height}{\includegraphics[width=.16\textwidth]{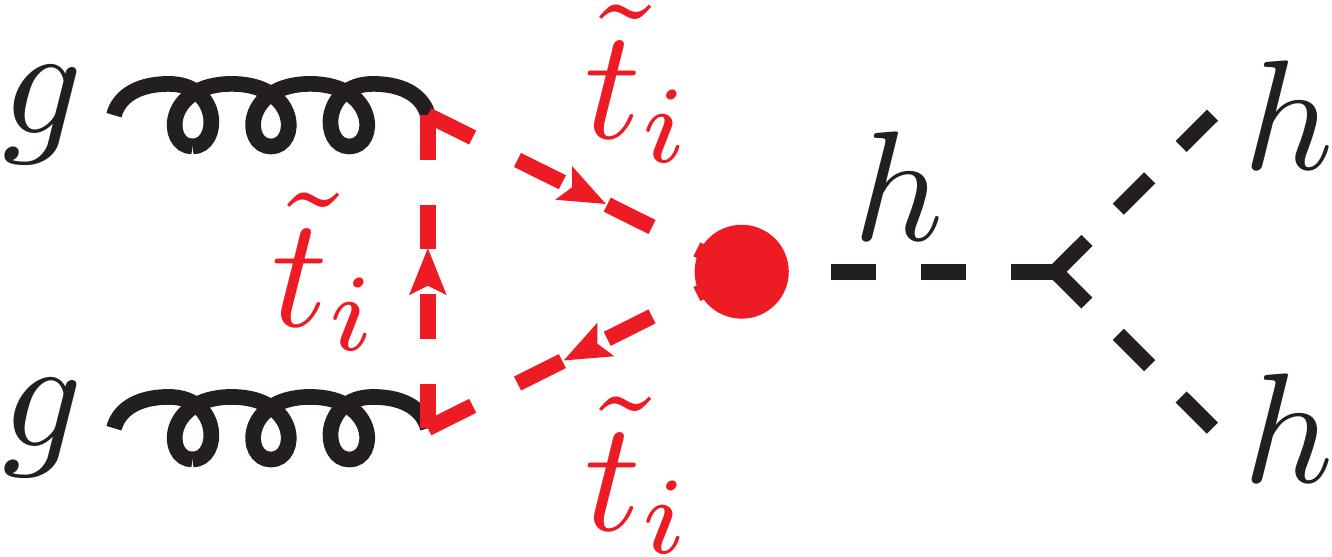}} &
$\mathcal A_i \propto \kappa_{h\tilde t\tilde t}^{ii}$ \\
\multicolumn{4}{c}{\makecell{This class of topologies involves only diagonal couplings between the Higgs and the squarks,\\ due to the absence of FCNCs in strong interactions and the presence of one $h\tilde t\tilde t$ coupling.}} \\
\noalign{\vskip 3pt}
\hline
\noalign{\vskip 3pt}
\textbf{6} & 
$\begin{array}{c}
\text{\bf Modified Higgs trilinear coupling}\\
\text{\bf +}\\
\text{\bf Bubble and triangle}\\
\text{\bf with $h\tilde t \tilde t$ coupling}
\end{array}$ & 
\raisebox{-.45\height}{\includegraphics[width=.16\textwidth]{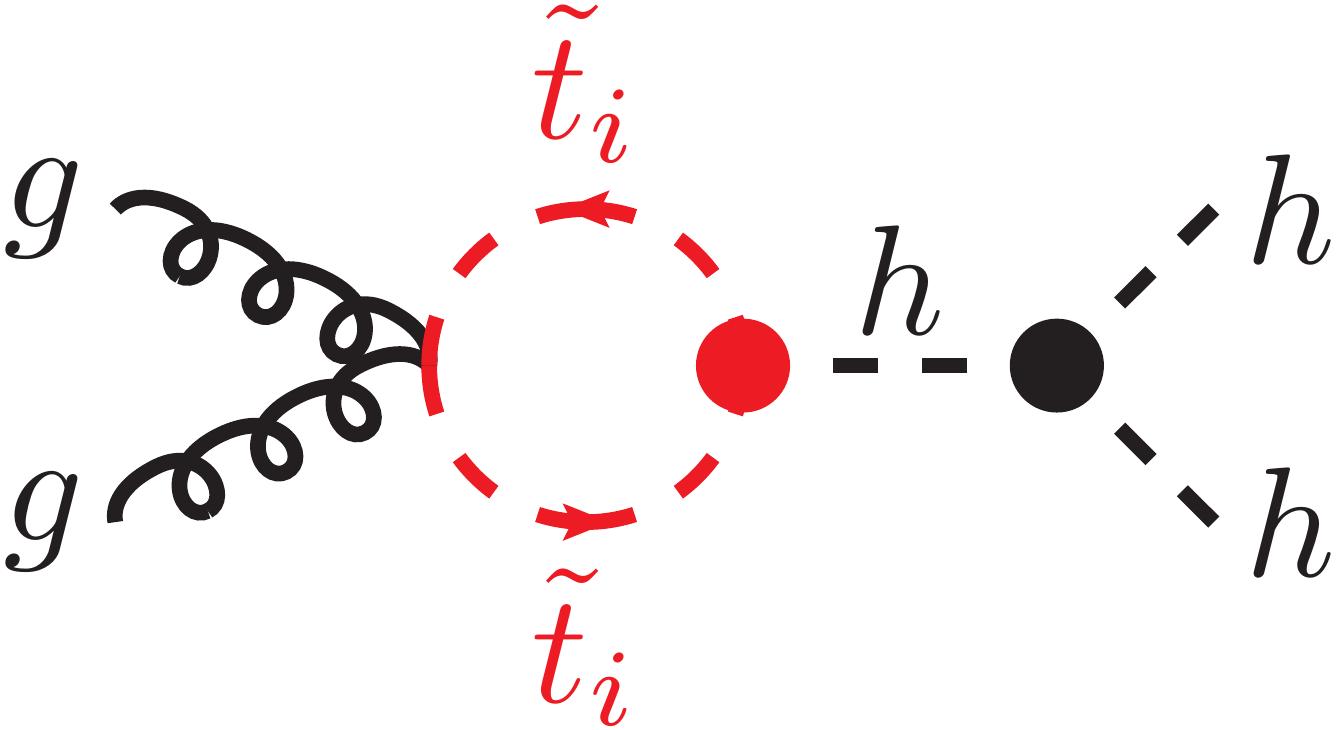}}
\raisebox{-.47\height}{\includegraphics[width=.16\textwidth]{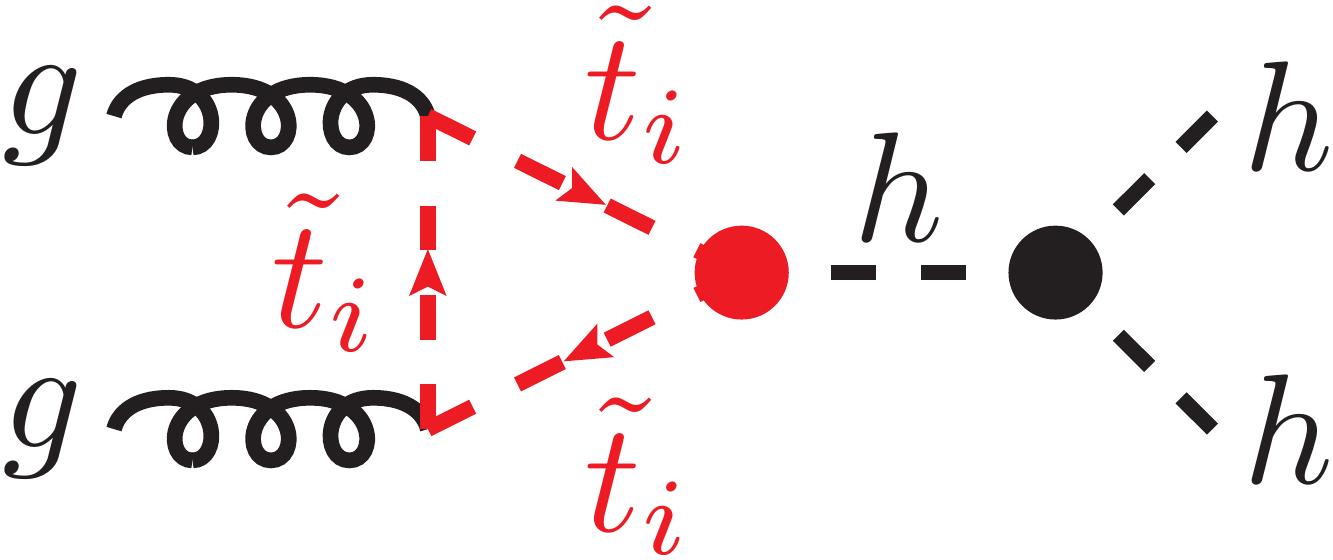}} &
$\mathcal A_i \propto \kappa_{hhh}\kappa_{h\tilde t\tilde t}^{ii}$ \\
\multicolumn{4}{c}{Only diagonal couplings between the Higgs and the squarks due to the strong interaction.} \\
\noalign{\vskip 3pt}
\hline
\noalign{\vskip 3pt}
\textbf{7} & 
$\begin{array}{c}
\text{\bf Triangle and box}\\
\text{\bf with two $h\tilde t \tilde t$ couplings}
\end{array}$ &
\begin{tabular}{c}
\raisebox{-.45\height}{\includegraphics[width=.16\textwidth]{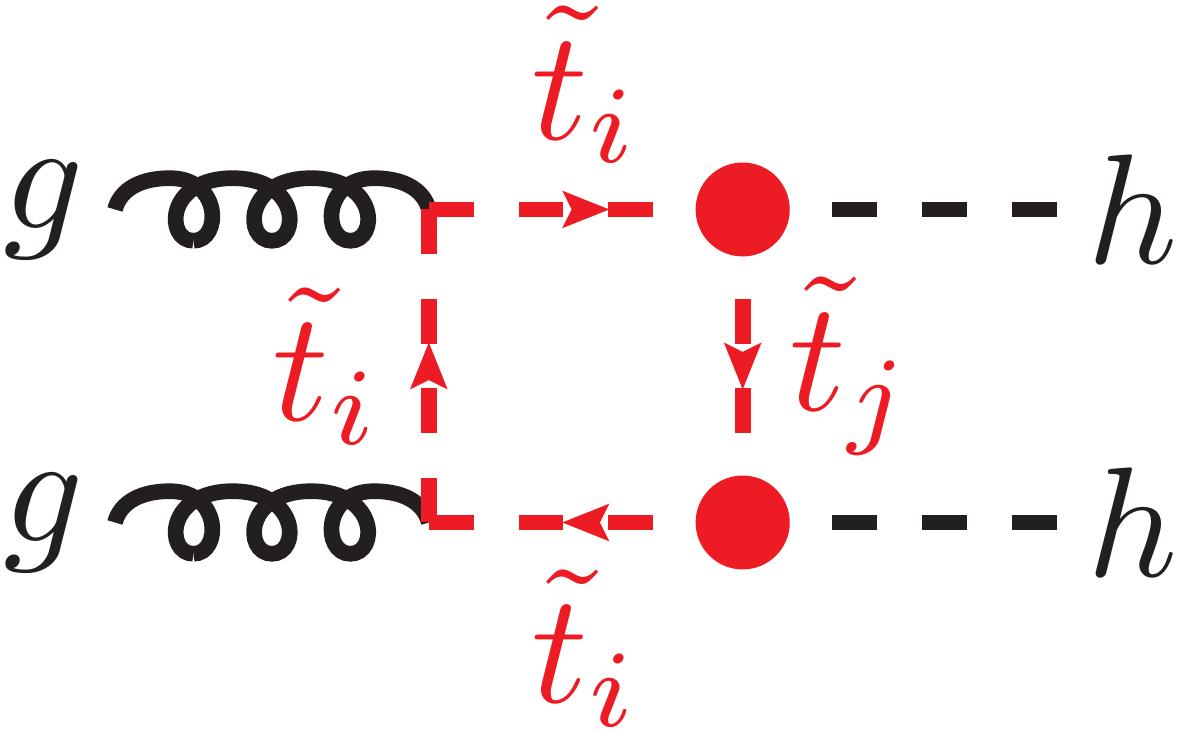}}
\raisebox{-.48\height}{\includegraphics[width=.16\textwidth]{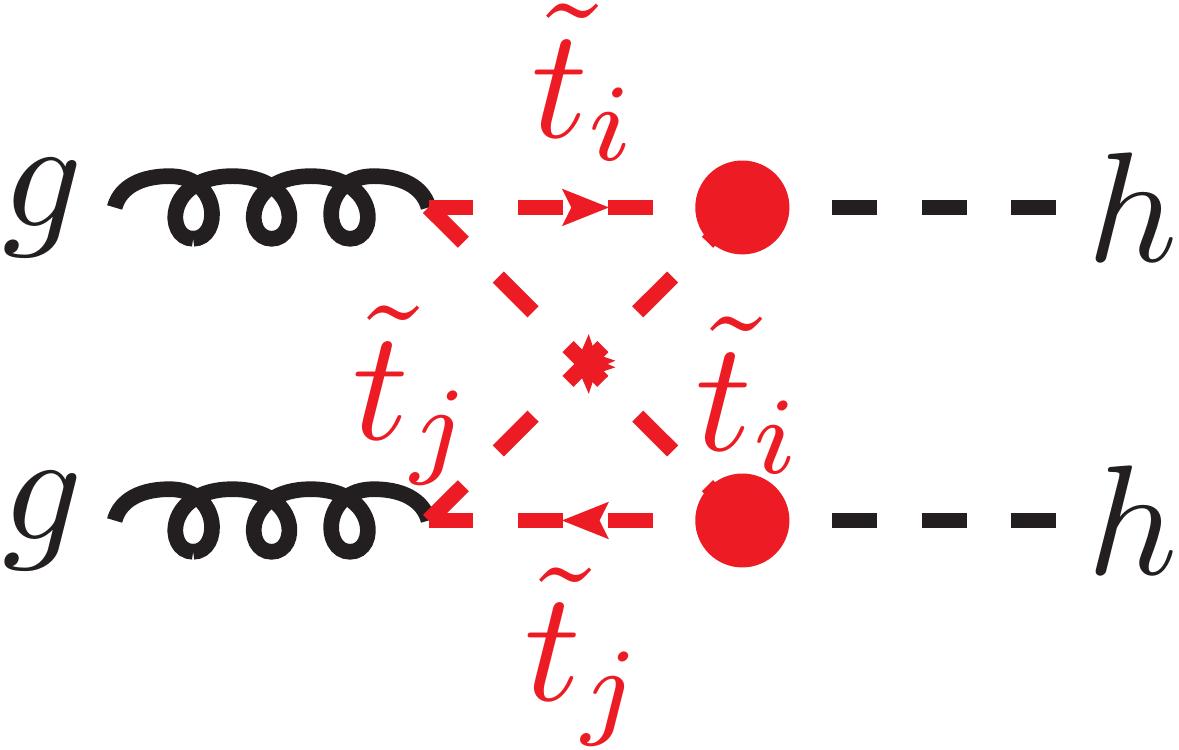}} \\
\raisebox{-.45\height}{\includegraphics[width=.16\textwidth]{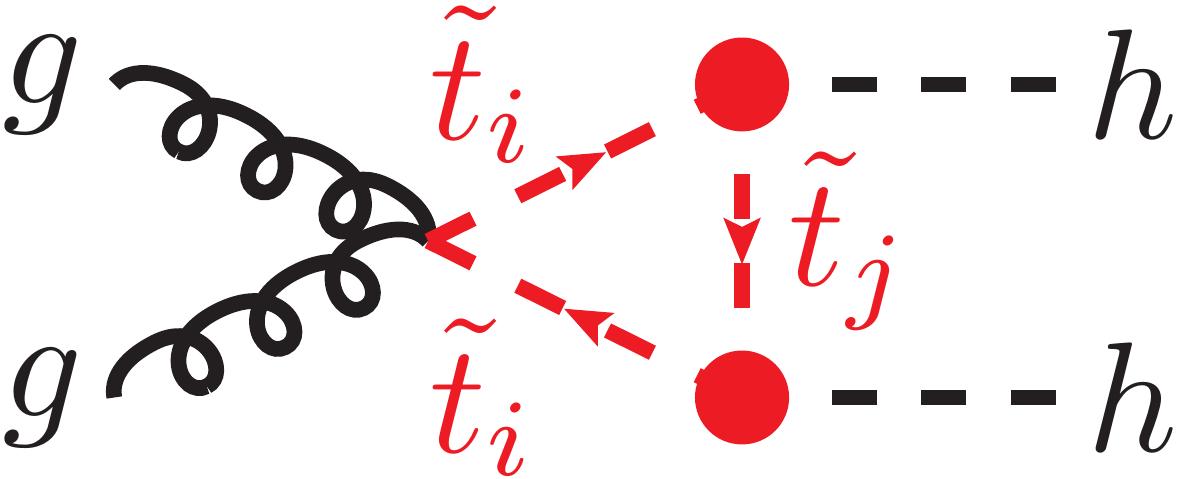}} 
\end{tabular} &
$\mathcal A_i \propto |\kappa_{h\tilde t\tilde t}^{ij}|^2$ \\
\noalign{\vskip 3pt}
\hline
\noalign{\vskip 3pt}
\textbf{8} & 
$\begin{array}{c}
\text{\bf Bubble and triangle}\\
\text{\bf with $hh\tilde t \tilde t$ coupling}
\end{array}$ &
\raisebox{-.46\height}{\includegraphics[width=.16\textwidth]{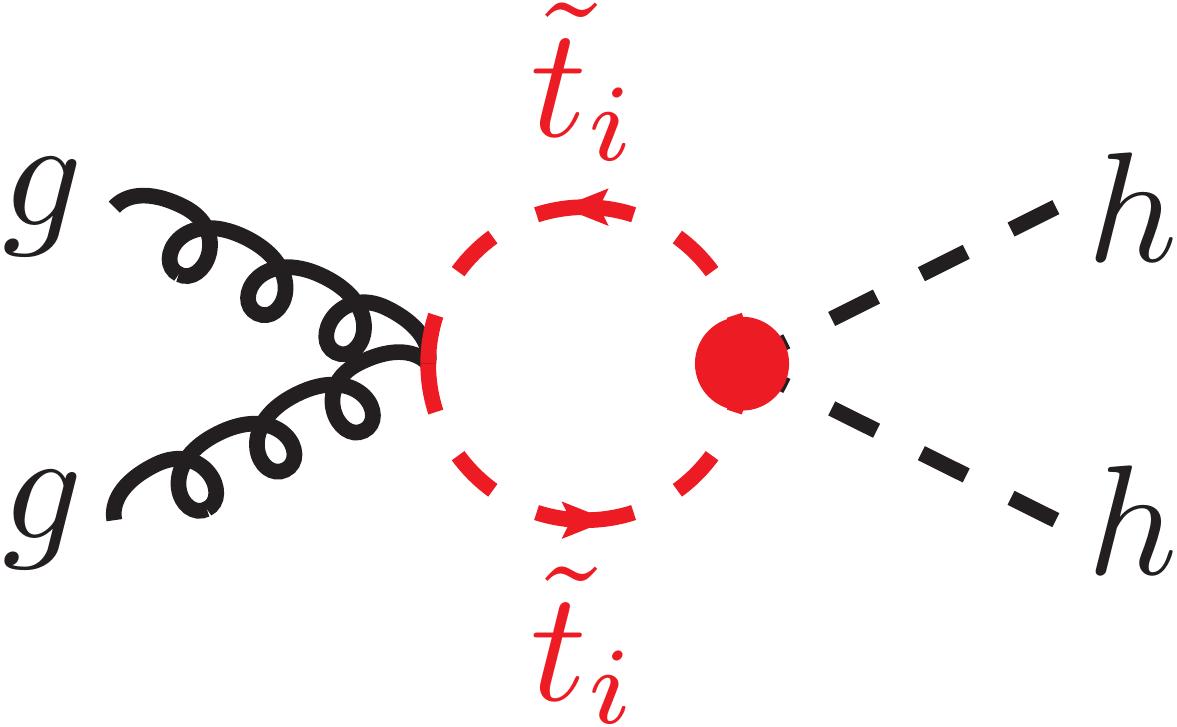}}
\raisebox{-.5\height}{\includegraphics[width=.16\textwidth]{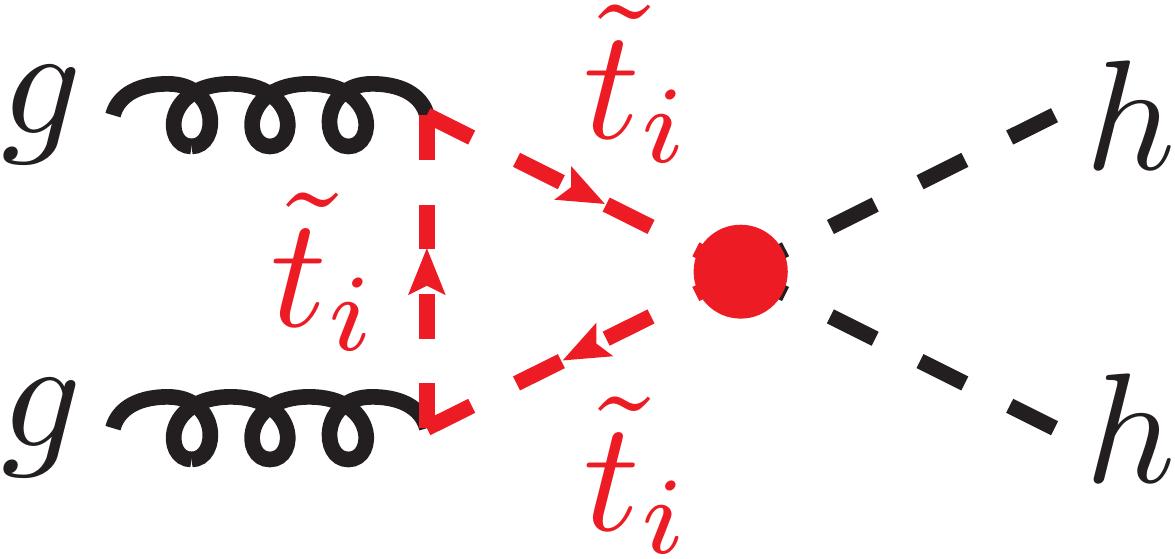}} &
$\mathcal A_i \propto \kappa_{hh\tilde t \tilde t}^{ii}$
\\
\multicolumn{4}{c}{Only diagonal couplings between the Higgs and the squarks due to the strong interaction.} \\
\noalign{\vskip 3pt}
\hline\hline
\end{tabular}
}
\caption{\label{tab:deconstructedtopologies} Complete list of topologies describing di-Higgs production with modified SM couplings $hhh$ and $htt$, and loop propagation of any number of stop squarks $\tilde t_i$. The topologies are classified according to the different products of new couplings (defined in \Cref{eq:LagNPstop}) to which the amplitudes are proportional.}
\end{table}

The signal contributions and their interferences can be parameterised in terms of a sum of different terms, proportional to unique functions of the couplings and to {\it reduced} cross sections $\hat\sigma$ depending exclusively on the stop masses. Labelling as "$\rm B$" the SM irreducible background, as "$\rm M$" the contribution of the topologies containing only modified SM couplings (\textbf{1} to \textbf{4} in \Cref{tab:deconstructedtopologies}) and as "$\rm S$" the contribution of topologies associated with the propagation of squarks (\textbf{5} to \textbf{8}) the complete set of contributions is:
{\setlength{\arraycolsep}{1pt}
\begin{subequations}
\begin{align}
\sigma_{\rm M} &= 
\kappa_{hhh}^2 \hat\sigma_1 + 
(\kappa_{hhh}\kappa_{htt})^2 \hat\sigma_3 + 
\kappa_{htt}^4 \hat\sigma_4 \;, \\
~
\sigma_{\rm S} &= \sum_{i=1,2} \Bigg[
\kappa_{h\tilde t \tilde t}^{ii} \sum_{j>i}\kappa_{h\tilde t \tilde t}^{jj}\hat\sigma_{5i}^{\rm int}(m_{\tilde t_{i,j}}) + 
\kappa_{hhh}^2(\kappa_{h\tilde t \tilde t}^{ii})^2\hat\sigma_{6d}(m_{\tilde t_{i}}) + 
\kappa_{hhh}^2\kappa_{h\tilde t \tilde t}^{ii}\sum_{j>i}\kappa_{h\tilde t \tilde t}^{jj}\hat\sigma_{6i}^{\rm int}(m_{\tilde t_{i,j}}) \nonumber\\
&+
(\kappa_{h\tilde t \tilde t}^{ii})^4\hat\sigma_{7d}(m_{\tilde t_{i}}) + 
\sum_{j>i}(\kappa_{h\tilde t \tilde t}^{ij})^4\hat\sigma_{7o}(m_{\tilde t_{i,j}}) + 
(\kappa_{h\tilde t \tilde t}^{ii})^2\sum_{j>i}(\kappa_{h\tilde t \tilde t}^{jj})^2\hat\sigma_{7idd}^{\rm int}(m_{\tilde t_{i,j}}) + 
(\kappa_{h\tilde t \tilde t}^{ii})^2\sum_{j\neq i}(\kappa_{h\tilde t \tilde t}^{ij})^2\hat\sigma_{7ido}^{\rm int}(m_{\tilde t_{i,j}}) \nonumber\\
&+ 
(\kappa_{hh\tilde t \tilde t}^{ii})^2\hat\sigma_{8d}(m_{\tilde t_{i}}) +
\kappa_{hh\tilde t \tilde t}^{ii}\sum_{j>i}\kappa_{hh\tilde t \tilde t}^{jj}\hat\sigma_{8i}^{\rm int}(m_{\tilde t_{i,j}}) \Bigg] \;, \label{eq:sigmaS}\\
~
\sigma_{\rm MB}^{\rm int} &=
\kappa_{hhh} \hat\sigma_{1B} + 
\kappa_{htt} \hat\sigma_{2B} \;, \\
~
\sigma_{\rm SB}^{\rm int} &= \sum_{i=1,2} \left[
\kappa_{h\tilde t \tilde t}^{ii}\hat\sigma_{5B}^{\rm int}(m_{\tilde t_i}) + 
\sum_{j>i}(\kappa_{h\tilde t \tilde t}^{ij})^2\hat\sigma_{7oB}^{\rm int}(m_{\tilde t_{i,j}}) + 
\kappa_{hh\tilde t \tilde t}^{ii}\hat\sigma_{8B}^{\rm int}(m_{\tilde t_i}) \right]\;, \\
~
\sigma_{\rm MM}^{\rm int} &= 
\kappa_{hhh}^2\kappa_{htt}\hat\sigma_{1,3}^{\rm int} + 
\kappa_{hhh}\kappa_{htt}^2\hat\sigma_{1,4-2,3}^{\rm int} + 
\kappa_{htt}^3\hat\sigma_{2,4}^{\rm int} + 
\kappa_{hhh}\kappa_{htt}^3\hat\sigma_{3,4}^{\rm int}\;, \\
~
\sigma_{\rm SS}^{\rm int} &= \sum_{i=1,2} \Bigg[
\kappa_{hhh}\kappa_{h\tilde t \tilde t}^{ii}\sum_{j> i}\kappa_{h\tilde t \tilde t}^{jj}\hat\sigma_{5,6i}^{\rm int}(m_{\tilde t_{i,j}}) +
(\kappa_{h\tilde t \tilde t}^{ii})^3\hat\sigma_{5,7d}^{\rm int}(m_{\tilde t_{i}}) + 
\kappa_{h\tilde t \tilde t}^{ii}\sum_{j\neq i}(\kappa_{h\tilde t \tilde t}^{jj})^2\hat\sigma_{5,7i}^{\rm int}(m_{\tilde t_{i,j}}) +
\kappa_{h\tilde t \tilde t}^{ii}\sum_{j\neq i}(\kappa_{h\tilde t \tilde t}^{ij})^2\hat\sigma_{5,7o}^{\rm int}(m_{\tilde t_{i,j}}) 
\nonumber\\
&+
(\kappa_{h\tilde t \tilde t}^{ii})^2\hat\sigma_{5,8d}^{\rm int}(m_{\tilde t_{i}}) + 
\kappa_{h\tilde t \tilde t}^{ii}\sum_{j\neq i}\kappa_{hh\tilde t \tilde t}^{jj}\hat\sigma_{5,8i}^{\rm int}(m_{\tilde t_{i,j}}) +
\kappa_{hhh}(\kappa_{h\tilde t \tilde t}^{ii})^3\hat\sigma_{6,7d}^{\rm int}(m_{\tilde t_{i}}) +
\kappa_{hhh}\kappa_{h\tilde t \tilde t}^{ii}\sum_{j\neq i}(\kappa_{h\tilde t \tilde t}^{jj})^2\hat\sigma_{6,7i}^{\rm int}(m_{\tilde t_{i,j}}) \nonumber\\
&+ 
\kappa_{hhh}\kappa_{h\tilde t \tilde t}^{ii}\sum_{j\neq i}(\kappa_{h\tilde t \tilde t}^{ij})^2\hat\sigma_{6,7o}^{\rm int}(m_{\tilde t_{i,j}}) +
\kappa_{hhh}(\kappa_{h\tilde t \tilde t}^{ii})^2\hat\sigma_{6,8d}^{\rm int}(m_{\tilde t_{i}}) +
\kappa_{hhh}\kappa_{h\tilde t \tilde t}^{ii}\sum_{j\neq i}\kappa_{hh\tilde t \tilde t}^{jj}\hat\sigma_{6,8i}^{\rm int}(m_{\tilde t_{i,j}}) 
\nonumber\\
&+ 
(\kappa_{h\tilde t \tilde t}^{ii})^2 \kappa_{hh\tilde t \tilde t}^{ii}\hat\sigma_{7d,8}^{\rm int}(m_{\tilde t_{i}}) + 
(\kappa_{h\tilde t \tilde t}^{ii})^2 \sum_{j\neq i}\kappa_{hh\tilde t \tilde t}^{jj}\hat\sigma_{7i,8}^{\rm int}(m_{\tilde t_{i,j}}) + 
\sum_{j\neq i}(\kappa_{h\tilde t \tilde t}^{ij})^2 \kappa_{hh\tilde t \tilde t}^{ii}\hat\sigma_{7o,8}^{\rm int}(m_{\tilde t_{i,j}}) \Bigg] \;,\label{eq:sigmaSS}\\
~
\sigma_{\rm MS}^{\rm int} &= \sum_{i=1,2} \Bigg[
\kappa_{hhh}^2\kappa_{h\tilde t \tilde t}^{ii} \hat\sigma_{1,6}^{\rm int}(m_{\tilde t_i}) + 
\kappa_{hhh}\sum_{j>i}(\kappa_{h\tilde t \tilde t}^{ij})^2 \hat\sigma_{1,7o}^{\rm int}(m_{\tilde t_{i,j}}) + 
\kappa_{hhh}\kappa_{hh\tilde t \tilde t}^{ii}\hat\sigma_{1,8}^{\rm int}(m_{\tilde t_i}) + 
\kappa_{htt}\kappa_{h\tilde t \tilde t}^{ii} \hat\sigma_{2,5}^{\rm int}(m_{\tilde t_i}) \nonumber\\
&+
\kappa_{hhh}\kappa_{htt}\kappa_{h\tilde t \tilde t}^{ii} \hat\sigma_{2,6-3,5}^{\rm int}(m_{\tilde t_i}) +
\kappa_{htt}(\kappa_{h\tilde t \tilde t}^{ii})^2 \hat\sigma_{2,7d}^{\rm int}(m_{\tilde t_{i}}) + 
\kappa_{htt}\sum_{j>i}(\kappa_{h\tilde t \tilde t}^{ij})^2 \hat\sigma_{2,7o}^{\rm int}(m_{\tilde t_{i,j}}) + 
\kappa_{htt}\kappa_{hh\tilde t \tilde t}^{ii}\hat\sigma_{2,8}^{\rm int}(m_{\tilde t_i}) \nonumber\\
&+
\kappa_{hhh}^2\kappa_{htt}\kappa_{h\tilde t \tilde t}^{ii} \hat\sigma_{3,6}^{\rm int}(m_{\tilde t_i}) +
\kappa_{hhh}\kappa_{htt}(\kappa_{h\tilde t \tilde t}^{ii})^2 \hat\sigma_{3,7d}^{\rm int}(m_{\tilde t_i}) + 
\kappa_{hhh}\kappa_{htt}\sum_{j>i}(\kappa_{h\tilde t \tilde t}^{ij})^2 \hat\sigma_{3,7o}^{\rm int}(m_{\tilde t_{i,j}}) \nonumber\\
&+
\kappa_{hhh}\kappa_{htt}\kappa_{hh\tilde t \tilde t}^{ii} \hat\sigma_{3,8}^{\rm int}(m_{\tilde t_i}) +
\kappa_{htt}^2\kappa_{h\tilde t \tilde t}^{ii} \hat\sigma_{4,5}^{\rm int}(m_{\tilde t_i}) + 
\kappa_{hhh}\kappa_{htt}^2\kappa_{h\tilde t \tilde t}^{ii} \hat\sigma_{4,6}^{\rm int}(m_{\tilde t_i}) +
\kappa_{htt}^2(\kappa_{h\tilde t \tilde t}^{ii})^2 \hat\sigma_{4,7d}^{\rm int}(m_{\tilde t_i}) \nonumber\\
&+ 
\kappa_{htt}^2\sum_{j>i}(\kappa_{h\tilde t \tilde t}^{ij})^2 \hat\sigma_{4,7o}^{\rm int}(m_{\tilde t_{i,j}}) + 
\kappa_{htt}^2\kappa_{hh\tilde t \tilde t}^{ii} \hat\sigma_{4,8}^{\rm int}(m_{\tilde t_i})
\Bigg] \;, \\
~
\sigma_{\rm MSB}^{\rm int} &=
\kappa_{hhh}\kappa_{htt} \hat\sigma_{1,2-3B}^{\rm int} + 
\kappa_{htt}^2 \hat\sigma_{2-4B}^{\rm int} \nonumber\\
&+
\sum_{i=1,2}\Bigg[\kappa_{hhh} \kappa_{h\tilde t \tilde t}^{ii}\hat\sigma_{1,5-6B}^{\rm int}(m_{\tilde t_i}) + 
\kappa_{hhh}(\kappa_{h\tilde t \tilde t}^{ii})^2 \hat\sigma_{1,7d-5,6}^{\rm int}(m_{\tilde t_{i}}) +
(\kappa_{h\tilde t \tilde t}^{ii})^2\hat\sigma_{5d-7dB}^{\rm int}(m_{\tilde t_i})\Bigg]
\;. \label{eq:sigmaMSB}
\end{align}
\label{eq:sigmahats}
\end{subequations}
}

The reduced cross sections have been labelled according to the numbering scheme of \Cref{tab:deconstructedtopologies}: $\hat\sigma_{x,y}^{\rm int}$ and $\hat\sigma_{xB}^{\rm int}$ represent, respectively, the interference terms between the $x$ and $y$ classes of topologies and between $x$ and the SM background, while $\hat\sigma_{x,y-j,k}^{\rm int}$ or $\hat\sigma_{x,y-jB}^{\rm int}$ signal combinations proportional to the same function of couplings. Further, $\hat\sigma_{xd}$ means that the topology class $x$ contains couplings between same quark flavour, $\hat\sigma_{xo}$ between different ones and finally, $\hat\sigma_{xi}$ or $\hat\sigma_{x,yi}^{\rm int}$ mean interference between topologies where two different squarks circulate in the loops. Factors of 2 from interference terms have been included into the $\hat\sigma$'s.

The total cross section induced by NP is given by the sum of all the terms of \Cref{eq:sigmahats}. This parametrisation allows for a model-independent analysis of the process: the kinematics of each individual term of \Cref{eq:sigmahats} is determined exclusively by the masses of the particles circulating in the loops, while the couplings only affect the relative proportion between the various terms. Total and differential quantities can be thus determined for multiple benchmarks characterised by the same stop masses and different couplings by performing a unique set of numerical simulations (one for each term of \Cref{eq:sigmahats}) for those masses. The simulation syntax is explained in \Cref{app:MCsyntax}\footnote{Notice that, if more coloured scalars are present, other terms depending on more than two masses would have to be considered, such as $\sum_{i,k}(\kappa_{h\tilde t \tilde t}^{ii})^2\sum_{j,k\neq i}(\kappa_{h\tilde t \tilde t}^{kj})^2\hat\sigma_{7id_io_{jk}}(m_{\tilde t_{i,j,k}})$, and further simulated samples would have to be included in the database. Considering that the topologies of \Cref{tab:deconstructedtopologies} contain at most {\it two} different squark propagators, a complete set of $\hat\sigma$ elements can be obtained considering four coloured scalars. The particle content of our {\sc UFO} model indeed allows one to explore such possibilities.}.

In the following sections, the analysis of a specific distribution -- the invariant mass of the di-Higgs system -- is discussed in detail following the aforementioned procedure, both before and after the decays of the Higgs bosons.

\subsubsection{Invariant mass of the di-Higgs system before Higgs decay}

One of the key observables to explore the contributions of NP in di-Higgs production at the LHC is the invariant mass of the di-Higgs system. Its shape is affected by the modification of SM couplings and by the presence of new particles in the loop. In this section we will consider the MSSM benchmark point in \Cref{tab:MSSMBP} to describe the analysis strategy.

The invariant mass distribution is built from the individual components of \Cref{eq:sigmahats}, evaluated for $m({\tilde t_{1,2}})=\{600,1300\}$ GeV, where the masses are approximated to the values on the simulation grid. 

\begin{figure}[h]
\epsfig{file=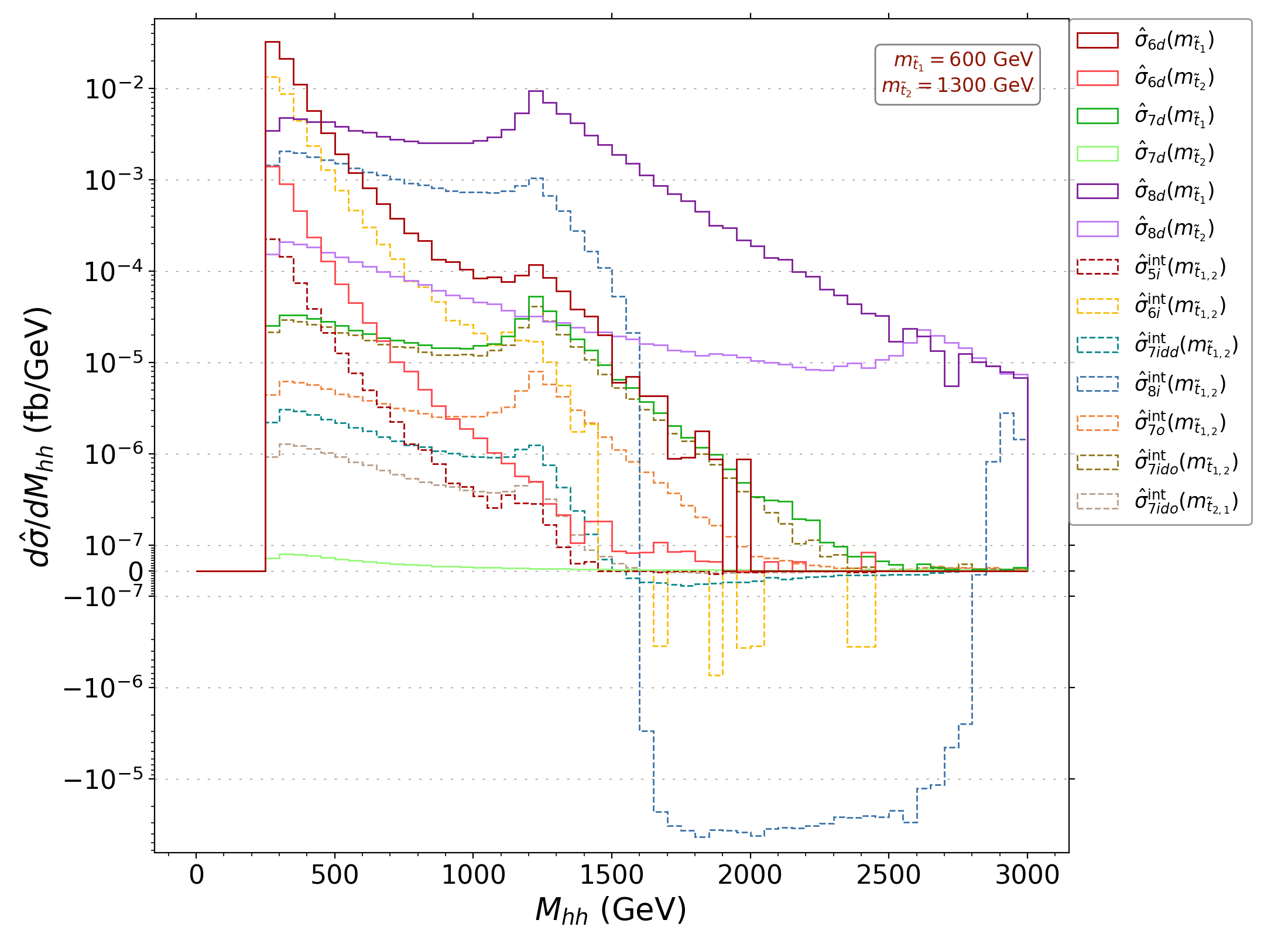,width=.47\textwidth}
\epsfig{file=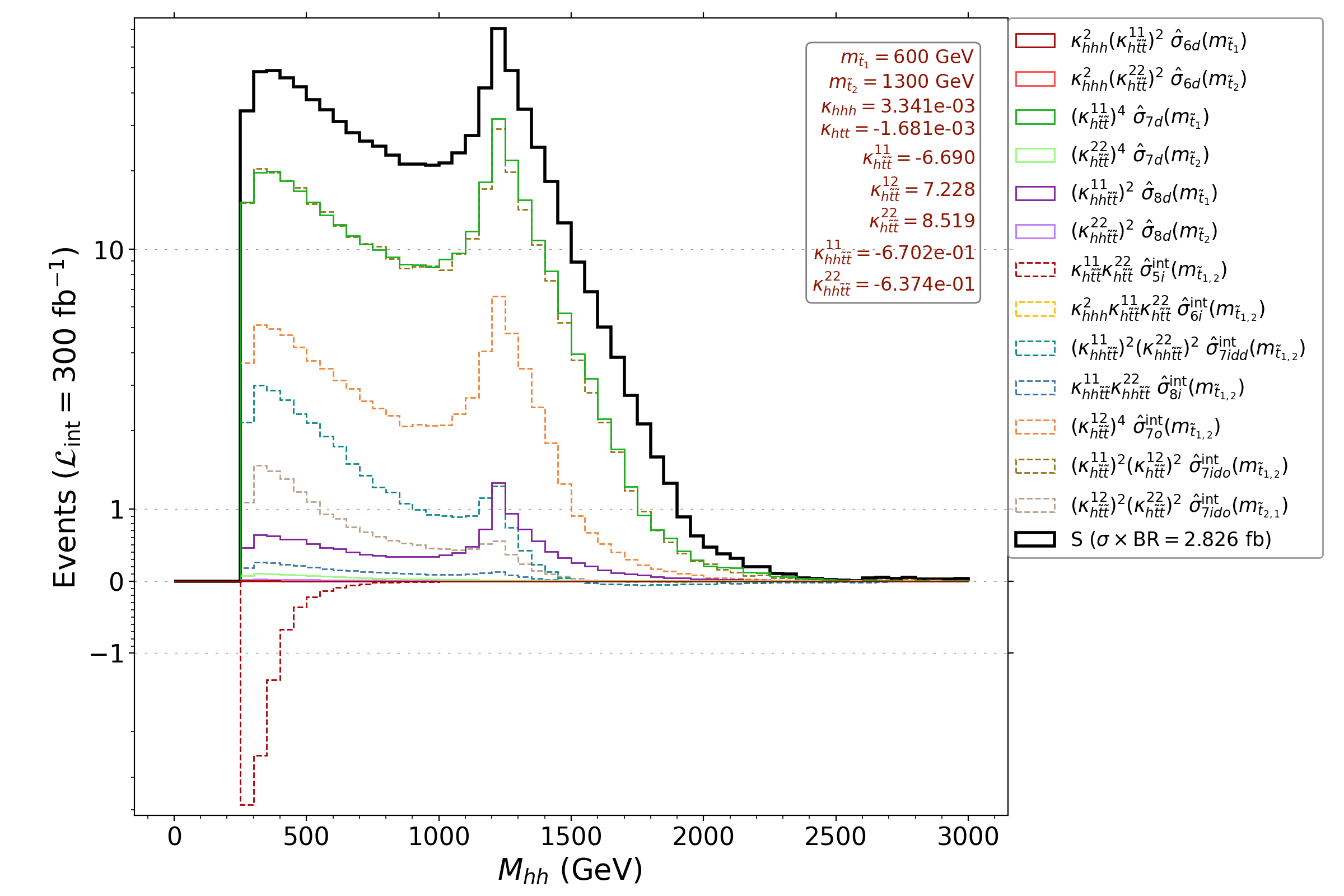,width=.51\textwidth}
\caption{\label{fig:mhhonlyS}{\bf Left panel}: $d\hat\sigma/d M_{hh}$ distributions of all components of $\sigma_S$. {\bf Right panel}: Each component of the left panel is weighted with the associated coupling products corresponding to the MSSM benchmark point of \Cref{tab:MSSMBP} and converted into the number of expected events with at the end of Run 3 with nominal luminosity of 300 fb$^{-1}$. The black curve corresponds to the sum of all components and represents the $M_{hh}$ distribution of \Cref{eq:sigmaS}.}
\end{figure}

Considering \Cref{eq:sigmaS} as an example, one can evaluate each differential cross section corresponding to the various $\hat\sigma$ terms using the corresponding MC samples, as shown in the left panel of \Cref{fig:mhhonlyS}. From this, the differential distribution of $\sigma_S$ can be evaluated by weighting each component with the numerical values of the coupling product factor and sum all terms, as shown in the right panel of \Cref{fig:mhhonlyS}, where the distributions are also multiplied by the nominal luminosity at the end of Run 3 of the LHC to evaluate the number of expected physical events.
From the distributions evaluated before their weighting, it is possible to notice the threshold effects around $m_{hh}=1200\GeV$ corresponding to the propagation of the $600 \GeV$ stop, and the presence of regions with negative interference. The importance of each term is however reshuffled by the couplings: for example, the negative value of the $\hat\sigma_{5i}^{\rm int}(m_{\tilde t_{1,2}})$ contribution is entirely due to the negative coupling product ($\kappa_{h\tilde t\tilde t}^{11}\kappa_{h\tilde t\tilde t}^{22}$ with $\kappa_{h\tilde t\tilde t}^{11}<0$) in front of it. However, when summed with the other contributions, the overall net effect is a positive cross section in the whole $m_{hh}$ range, with a relevant enhancement of the $1200 \GeV$ threshold peak.

\begin{figure}[h]
\epsfig{file=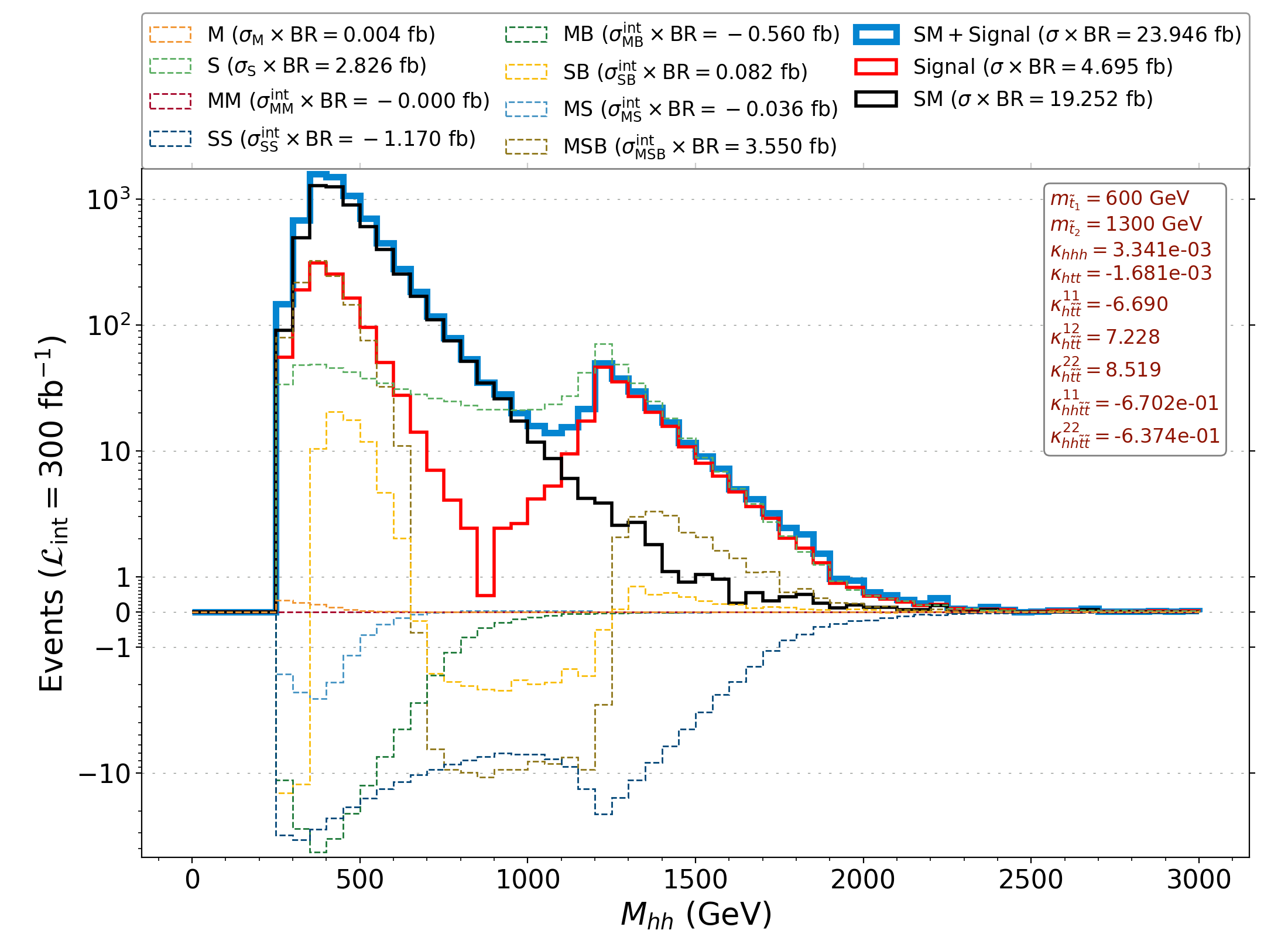,width=.75\textwidth}
\caption{\label{fig:mhh_MSSM} Invariant mass distribution of the di-Higgs system at parton level and before Higgs decay for the MSSM benchmark point of \Cref{tab:MSSMBP}, displaying the number of expected events at the end of Run 3 with nominal luminosity of 300 fb$^{-1}$. Curves for the SM intrinsic background (black), pure signal (red) and their sum (blue) are shown as solid lines. For the signal, the individual contributions of its components are also shown as dashed lines, to assess their relative role in the construction of its shape.}
\end{figure}

The very same procedure is applied to each term in \Cref{eq:sigmahats} and the final result for the MSSM benchmark point is shown in \Cref{fig:mhh_MSSM}. In general there are three types of contributions. Those that involve only modifications to the SM couplings ($\sigma_{\rm M}$, $\sigma_{\rm MM}$ and $\sigma_{\rm MB}$) have a peak at low $M_{hh}$ and then decay exponentially. Such contributions are small in the MSSM, but can be more pronounced in, {e.g.}, NMSSM, where the trilinear Higgs coupling can deviate from its SM value. Second, we have squark contributions squared ($\sigma_{\rm S}$ and $\sigma_{\rm SS}$), which are relatively flat over a large range and peak at $M_{hh}=2m_{\tilde{t}}$. Finally we have interference contributions between squarks and the SM-type contributions ($\sigma_{\rm SB}$, $\sigma_{\rm MS}$ and $\sigma_{\rm MSB}$), which are positive at low $M_{hh}$, negative between $m_{\tilde{t}}<M_{hh}<2m_{\tilde{t}}$ and turn positive again at high $M_{hh}$. This leads to an almost complete cancellation of BSM effects in the range $m_{\tilde{t}}<M_{hh}<2m_{\tilde{t}}$.

Looking at the MSSM benchmark of \Cref{fig:mhh_MSSM} multiple effects are in place: the threshold peak generated by the terms in \Cref{eq:sigmaS} is clearly standing out with respect to the SM irreducible background, even if negative interference contributions from \Cref{eq:sigmaSS} slightly reduce its impact, but the increase in the total cross section (SM+Signal $\simeq24.5\fb$) is mostly due to the excess of events on the global peak of the distribution, generated by the interference terms of \Cref{eq:sigmaMSB}. Inspecting the elements of \Cref{eq:sigmaMSB}, analogously to what was done in \Cref{fig:mhhonlyS}, the largely dominant contribution (in the whole $M_{hh}$ range) is given by the term $(\kappa_{h\tilde t \tilde t}^{11})^2\hat\sigma_{5d-7dB}^{\rm int}(m_{\tilde t_1})$, which sums the pure signal contribution from topology 5 of \Cref{tab:deconstructedtopologies} and the interference between topology 7 and the SM background, both proportional to the same product of new couplings. Given the small modifications to the SM couplings, the topologies proportional to $\kappa_{hhh}$ and $\kappa_{htt}$ have an almost negligible impact on the determination of the final shape of $d\hat\sigma_S / d M_{hh}$.

\begin{figure}[h]
\epsfig{file=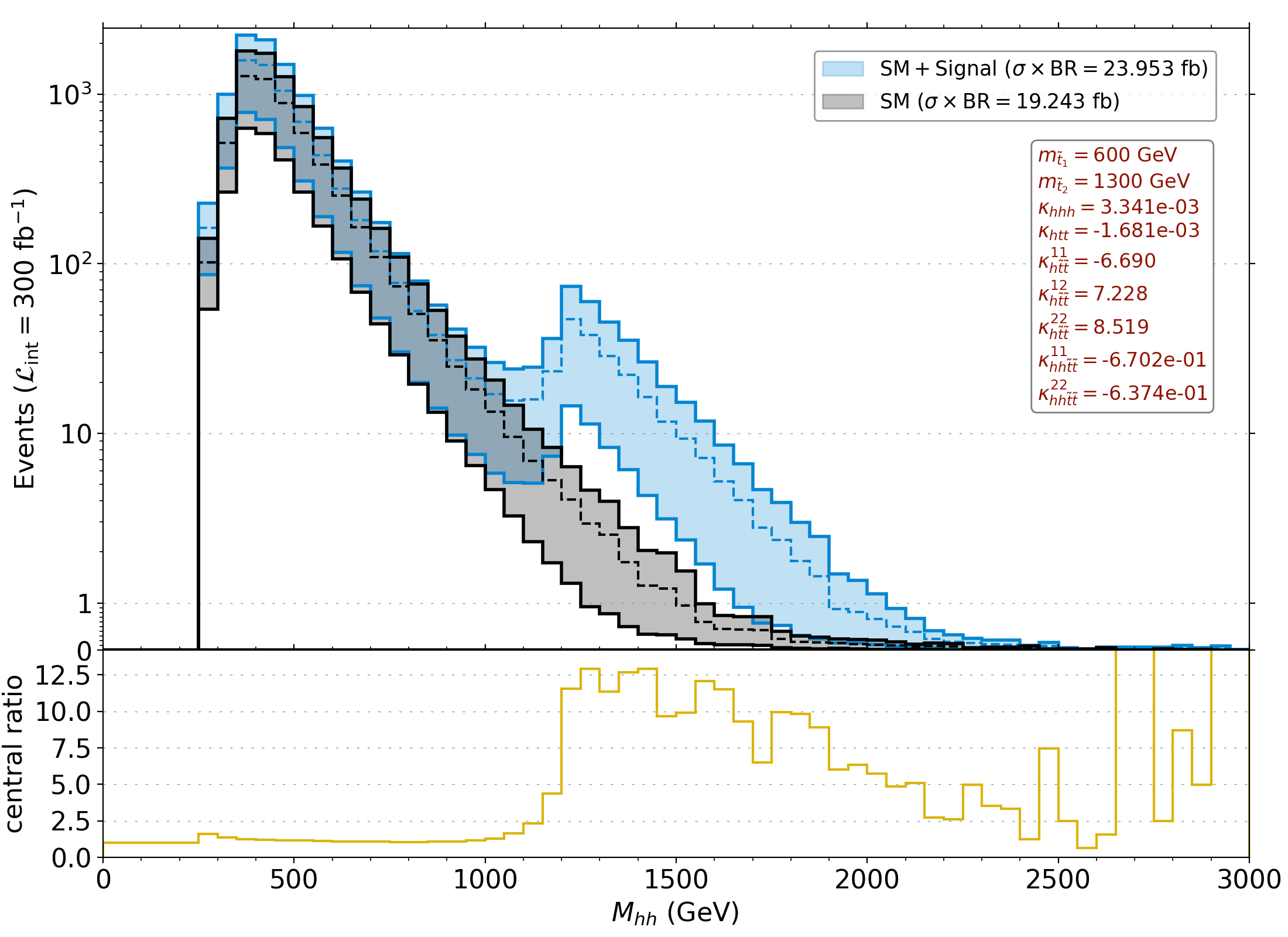,width=.6\textwidth}
\caption{\label{fig:mhh_MSSM_errorbands} Same as \Cref{fig:mhh_MSSM}, but showing only the SM and Signal+SM curves with 1$\sigma$ uncertainty bands representing the combined scale and PDF systematics. In the bottom panel, the ratio between central values, (SM+Signal)/SM, is displayed.}
\end{figure}

Despite the excess in the global peak and the presence of a relatively sizeable threshold peak with respect to the background distribution, an estimation of the systematic uncertainties is in order to assess if the different shapes can be potentially discernible using real data. A dedicated MC simulation has been performed for this estimation, validating the results obtained with the deconstruction method as a by-product. The combination of scale and Parton Distribution Function (PDF) systematics is done using the same method of \cite{Deandrea:2021vje}: bin-by-bin, the asymmetric scale uncertainties are obtained by considering the largest deviations from the central value, while PDF uncertainties are obtained using the standard deviation on the whole PDF set, as described in \cite{Butterworth:2015oua}. The scale and PDF systematics are then linearly summed.
In \Cref{fig:mhh_MSSM_errorbands} the distributions for the SM background and the sum of signal and background are shown: it is possible to see that while the excess on the global peak is completely hidden within the uncertainties, the threshold peak can indeed be potentially discriminated. 

So far, only distributions before Higgs decays have been considered. If no shape discrimination was possible already at this stage, any further investigation would have been pointless. Given the optimistic results, however, the next step of the analysis involves letting the Higgs decay and identify promising final states for experimental detection.

\subsubsection{Invariant mass of the di-Higgs system after Higgs decay}
\label{sec:resultsreco}

In this part of the analysis we consider three final states, motivated either by large branching ratios or by cleanliness of detection. In all cases we select final states where at least one of the Higgs boson decays to 2 bottom quarks, and the other to either 2 photons, 2 $\tau$ leptons or other 2 bottom quarks. 
The reconstruction of final state objects is done using {\sc fastjet}\cite{Cacciari:2011ma} through {\sc MadAnalysis 5}\cite{Conte:2012fm,Conte:2014zja,Araz:2020lnp}, using the anti-$k_T$ jet clustering algorithm \cite{Cacciari:2008gp} with jet radius parameter $R=0.4$.

\begin{figure}[h]
\epsfig{file=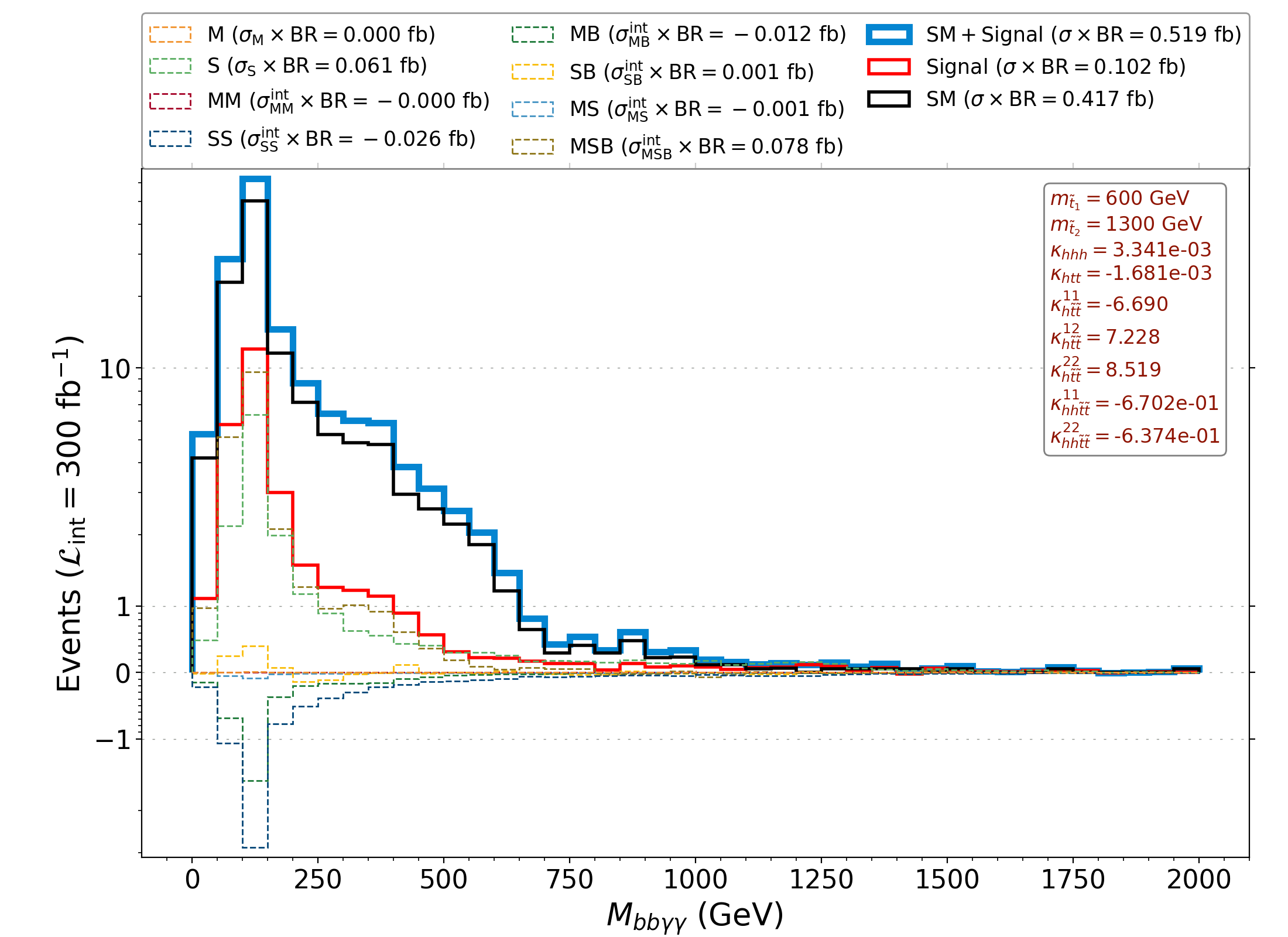,width=.48\textwidth}
\epsfig{file=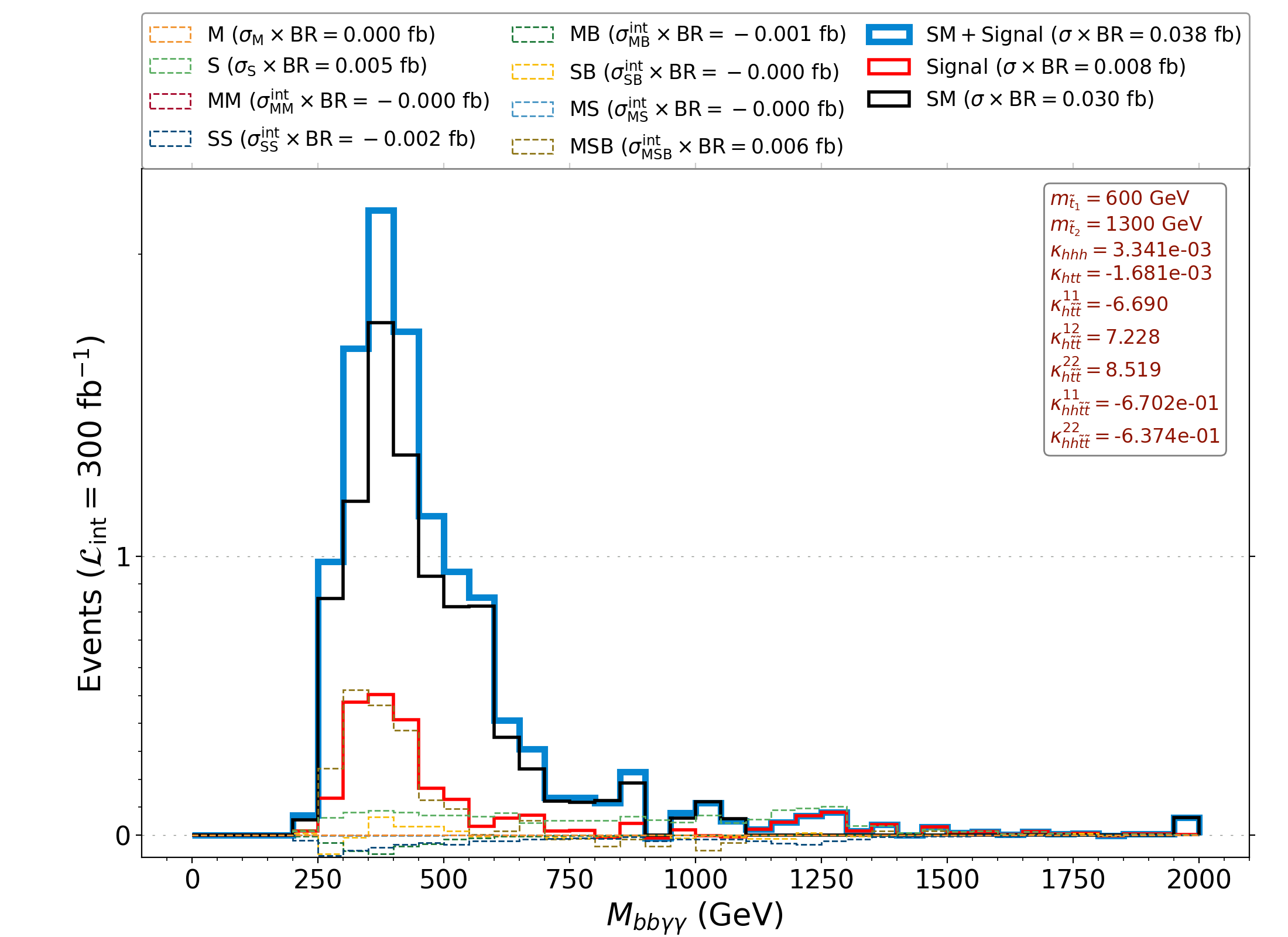,width=.48\textwidth}
\caption{\label{fig:distributionsrecobbaa}Invariant mass distribution of the $bb\gamma\gamma$ system at reconstruction level (\ul{left}) and after the cuts of \Cref{tab:cuts} (\ul{right}).}
\end{figure}

\begin{figure}[h]
\epsfig{file=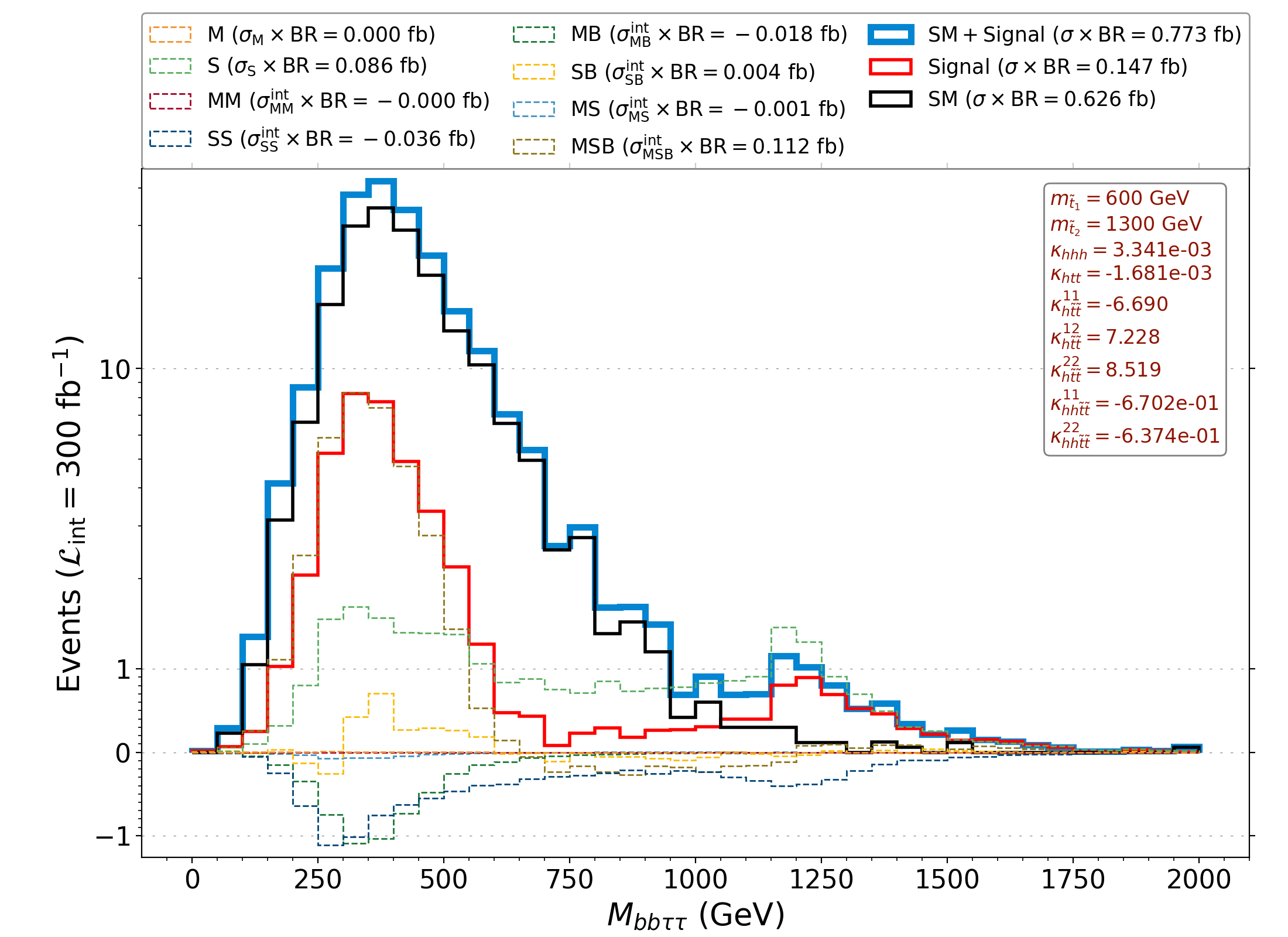,width=.48\textwidth}
\epsfig{file=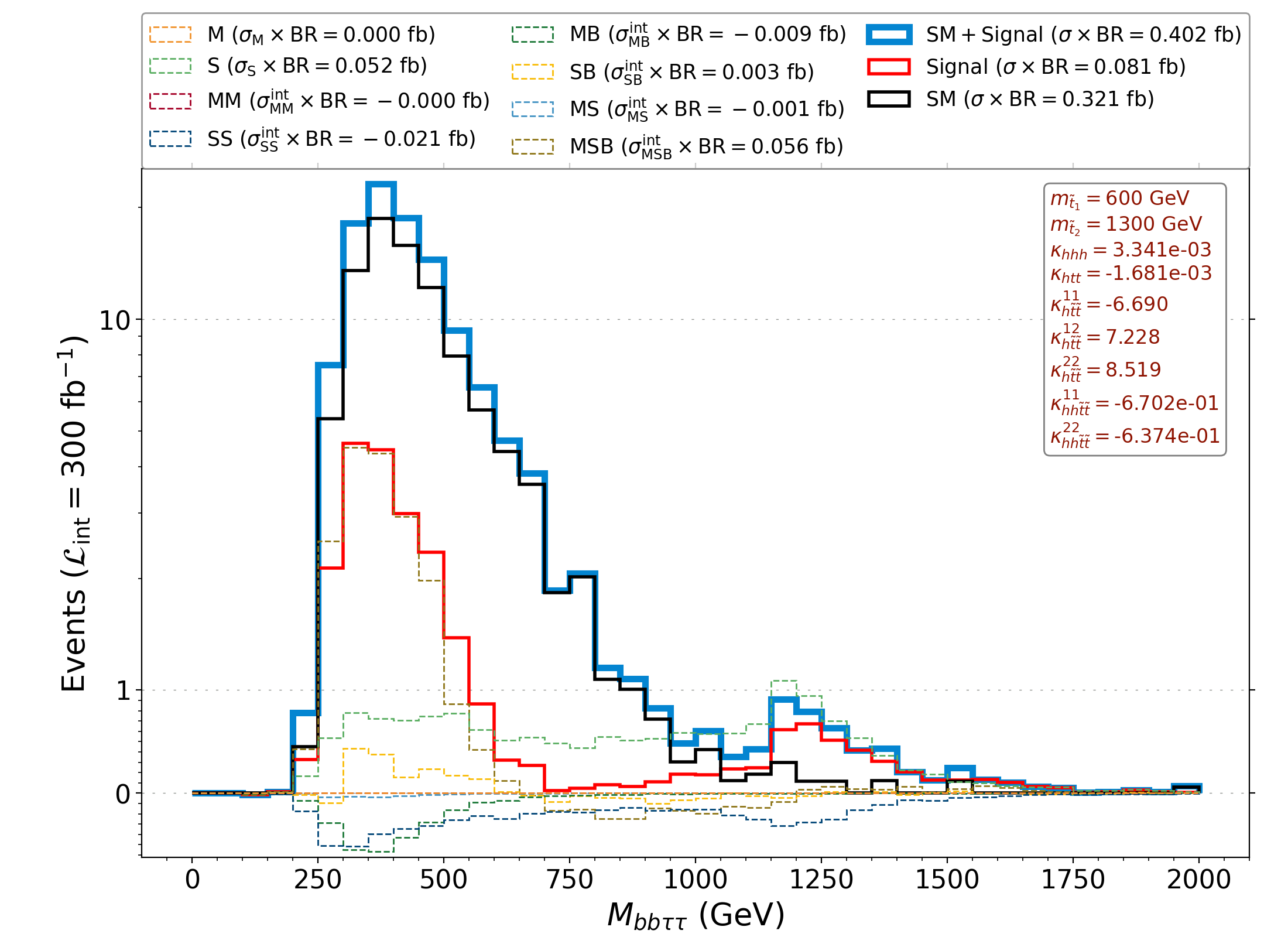,width=.48\textwidth}
\caption{\label{fig:distributionsrecobbtata}Same as \Cref{fig:distributionsrecobbaa} but for the $bb\tau\tau$ final state.}
\end{figure}

\begin{figure}[h]
\epsfig{file=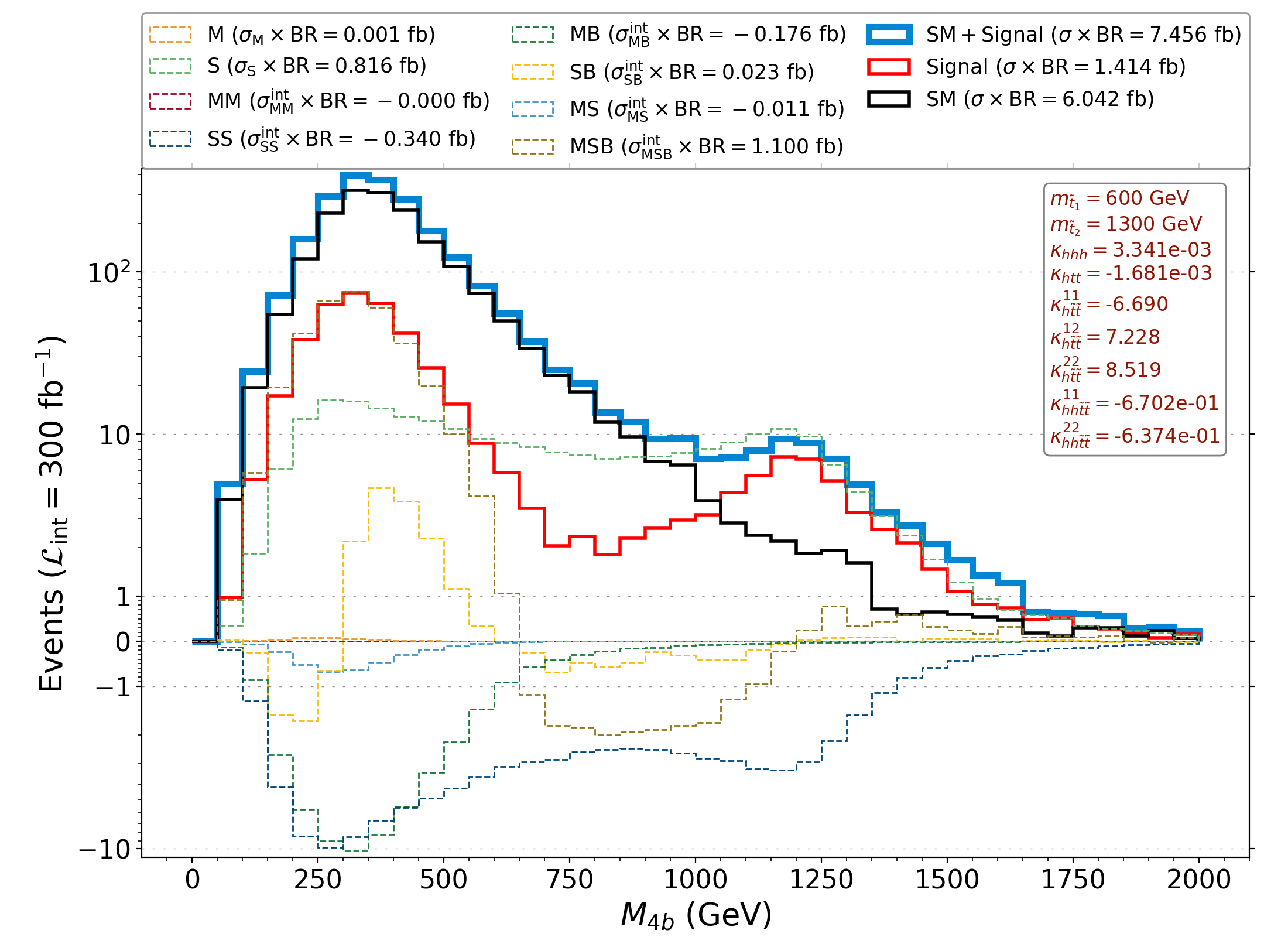,width=.48\textwidth}
\epsfig{file=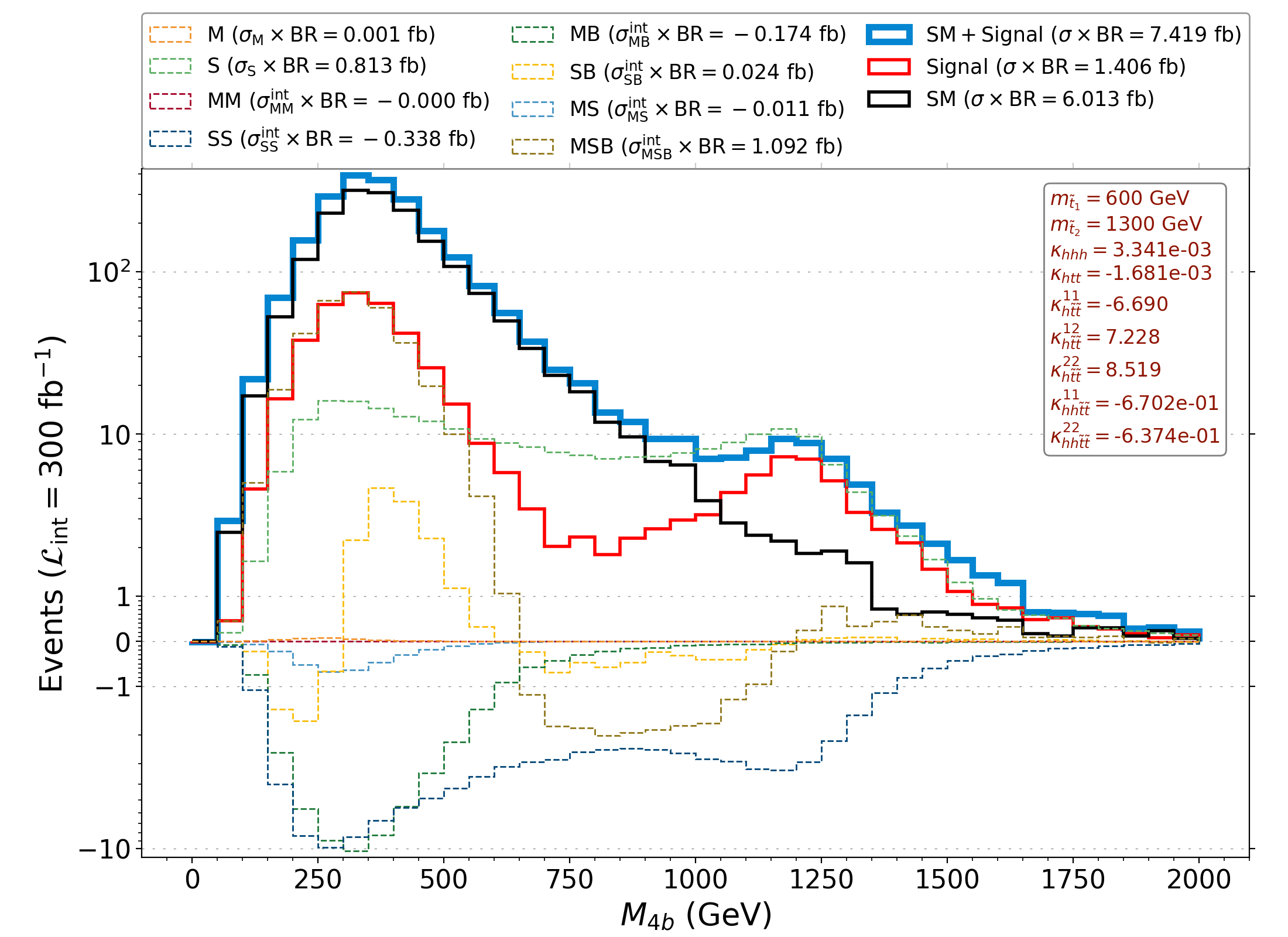,width=.48\textwidth}
\caption{\label{fig:distributionsrecobbbb}Same as \Cref{fig:distributionsrecobbaa} but for the $4b$ final state.}
\end{figure}

The corresponding distributions of physical events as function of the invariant mass of the di-Higgs decay products are shown in \Cref{fig:distributionsrecobbaa,fig:distributionsrecobbtata,fig:distributionsrecobbbb}: in all figures, the left panels show results at reconstruction level without any cut, while the right panels correspond to the basic selection cuts reported in \Cref{tab:cuts}. The preselection cuts are inspired by current experimental prospects \cite{ATL-PHYS-PUB-2022-053}. For the $bb\gamma\gamma$ final state a di-photon invariant mass cut around the Higgs boson mass has been applied to reduce the contribution from radiated photons and enhance the contribution of photons from Higgs decay.

\begin{table}[h]
{\setlength\tabcolsep{10pt}
\begin{tabular}{ccc}
\hline\hline
\noalign{\vskip 3pt}
$bb\gamma\gamma$ & $bb\tau\tau$ & $bbbb$ \\
\hline
\noalign{\vskip 3pt}
$N(b)>1$ & $N(b)>1$ & $N(b)>3$ \\ 
$N(\gamma)>1$ & $N(\tau)>1$ & -- \\ 
$p_T(b)>45~(20) \GeV$ & $p_T(b)>45~(20) \GeV$ & $p_T(b)>40 \GeV$ \\
$|\eta(b)|<2.5$ & $|\eta(b)|<2.5$ & $|\eta(b)|<2.5$ \\
$|\eta(\gamma)|<2.5$ & $|\eta(\tau)|<2.5$ & -- \\
$120\GeV < M(\gamma\gamma) < 130\GeV$ & -- & -- \\[2pt]
\hline\hline
\end{tabular}
}
\caption{\label{tab:cuts} Selection and kinematic cuts for the three final states considered in the analysis.}
\end{table}


The depletion in number of events in the $bb\gamma\gamma$ and $bb\tau\tau$ case makes it impossible to discriminate any NP contribution induced by the scenarios we are considering during the Run 3 of the LHC. While the $bbbb$ final state looks more promising, a careful evaluation of all sources of backgrounds (especially QCD-induced ones) must be performed in order to establish potential discrimination possibilities.

\section{Deviations beyond the MSSM}

After the first two runs of the LHC, the parameter space of the MSSM has become constrained, especially when it comes to parameters related to Higgs pair production. We know that both the top Yukawa coupling and the Higgs trilinear coupling must be close to the SM values, so deviations can mainly arise from a light stop around the threshold of $2m_{\tilde{t}}$ and the interference effects between squark and SM diagrams. This is not the case in non-minimal Supersymmetric models, as an extended Higgs sector can lead to deviations from the SM in the Higgs trilinear coupling. The top Yukawa is more constrained experimentally as the $t\bar th$ production rate depends on it \cite{CMS:2020mpn}.

The squark contribution arises mainly from the diagrams involving trilinear Higgs-stop-stop couplings. The trilinear coupling at tree-level is proportional to $X_{t}=A_{t}-\mu\cot\beta$. Large stop mixing is needed to get a $125\GeV$ mass for the SM-like Higgs in the MSSM, so the stop contribution is always relatively large in the MSSM. Large stop mixing also enhances the mass splitting between the stops, so one of the stops needs to be heavy so it will not contribute to the threshold excess.

In the NMSSM one can achieve the $125\GeV$ Higgs mass already at tree-level, so one may try a setup with $X_{t}\simeq 0$, which would allow for two light stops, both being insensitive to searches targeting missing transverse momentum. The requirement of $m_h=125\GeV$ requires low $\tan\beta$ and large $\lambda$ so the term $|\lambda|^{2}|H_{u}H_{d}|^{2}$ leads to an enhancement of the triple Higgs coupling. A larger Higgs trilinear coupling can lead to a first order EW phase transition. In the EFT limit, where all BSM particles are decoupled, a $50\%$ enhancement is needed for a first order phase transition \cite{Reichert:2017puo}. Additional light degrees of freedom, such as light neutralinos \cite{Akula:2017yfr}, could change the needed enhancement. Nevertheless, an enhancement up to $100\%$ is possible in the NMSSM \cite{Wu:2015nba} so it is natural to ask how well could we distinguish such a case.

\begin{figure}[h]
\epsfig{file=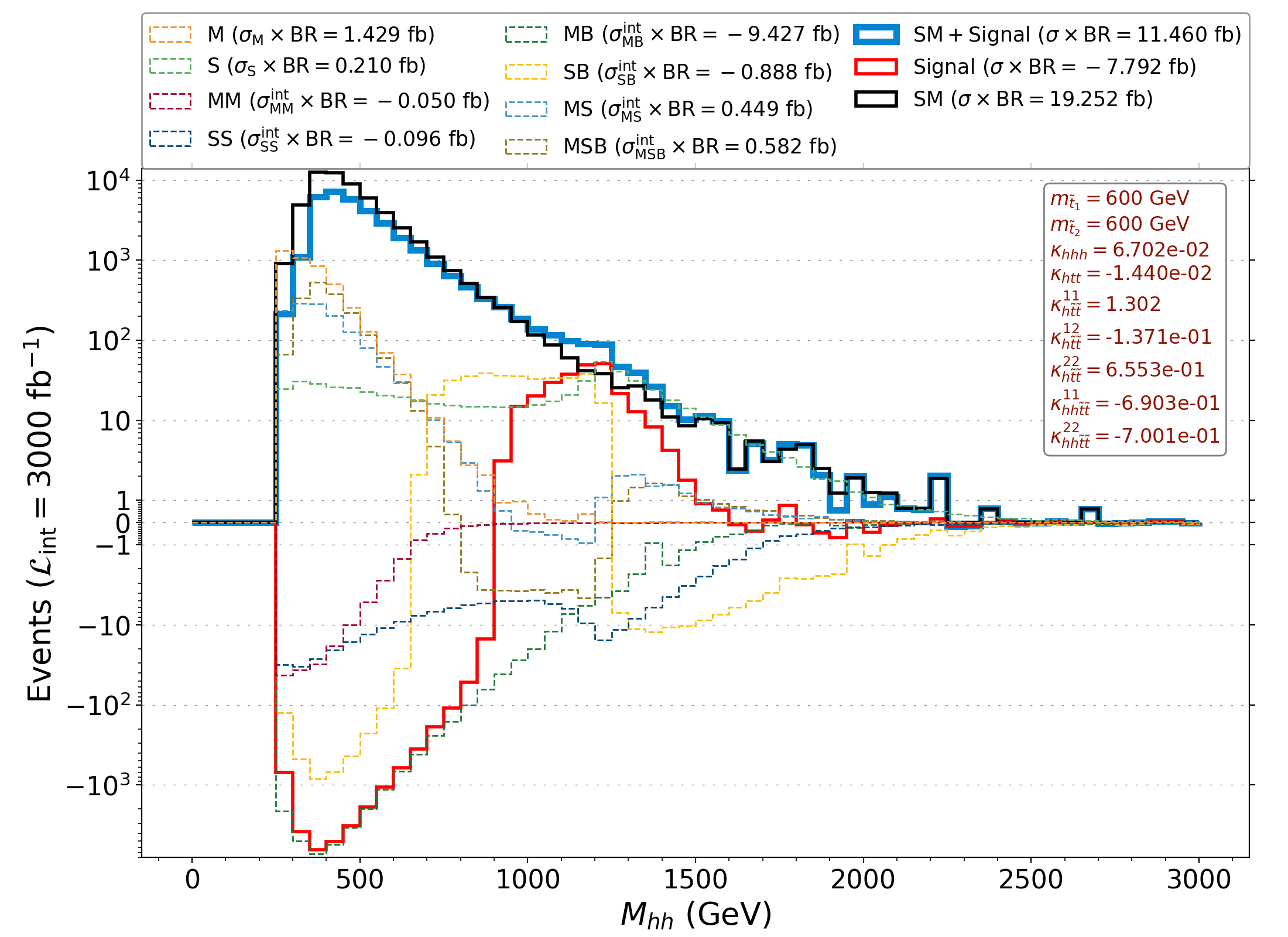,width=.49\textwidth}\hfill
\epsfig{file=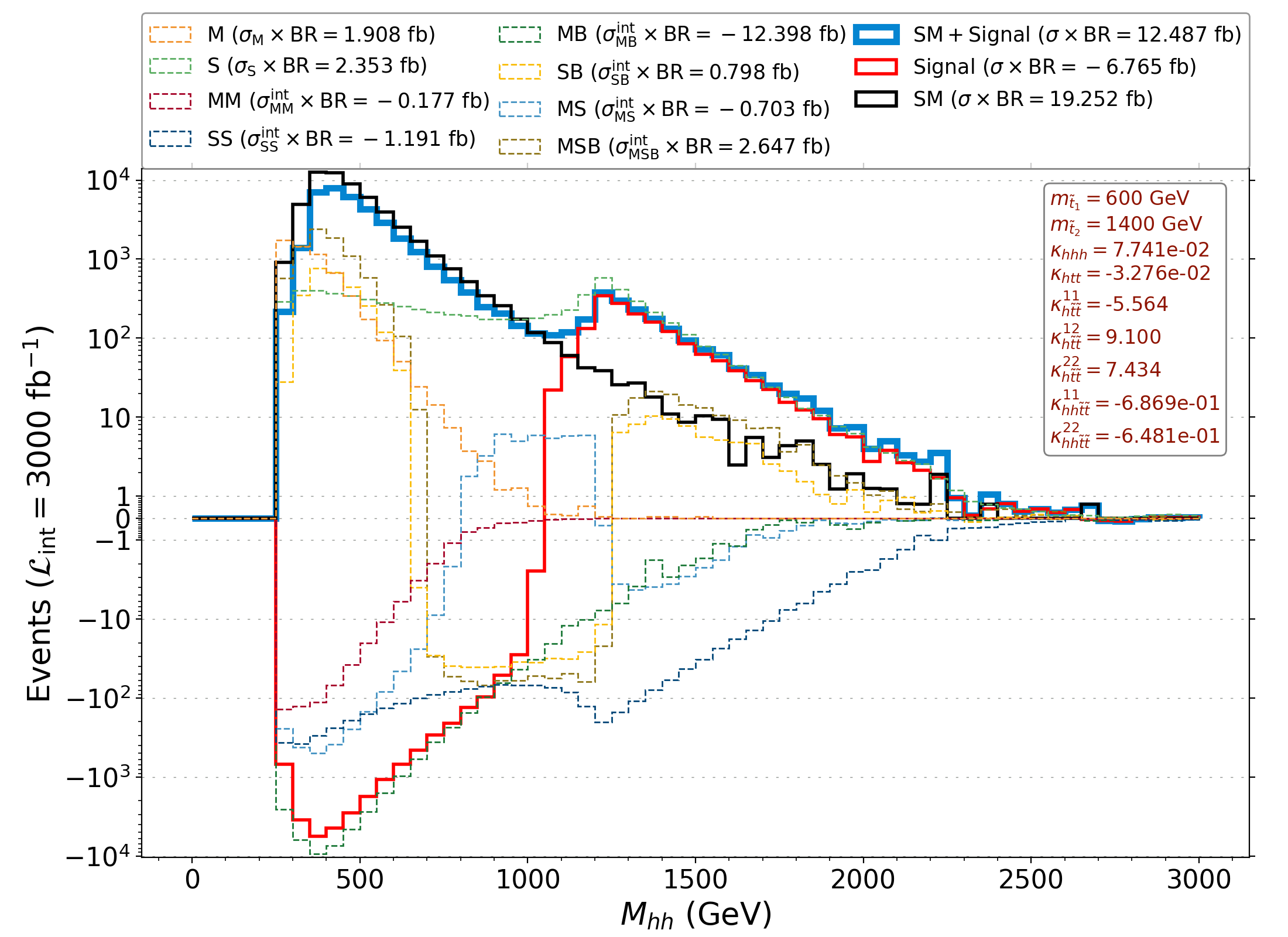,width=.49\textwidth}\\
\hfill
\epsfig{file=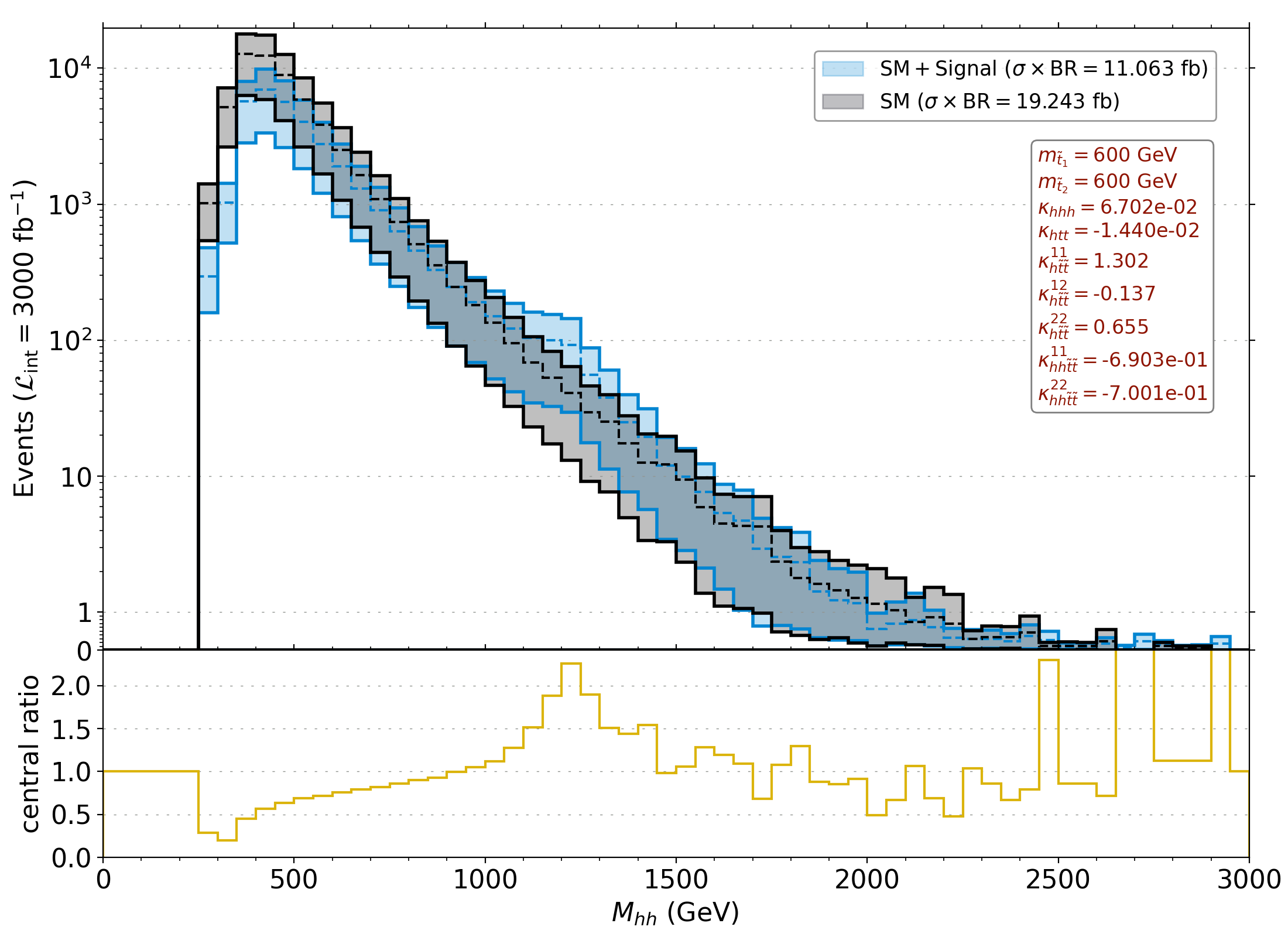,width=.475\textwidth}\hskip 18pt
\epsfig{file=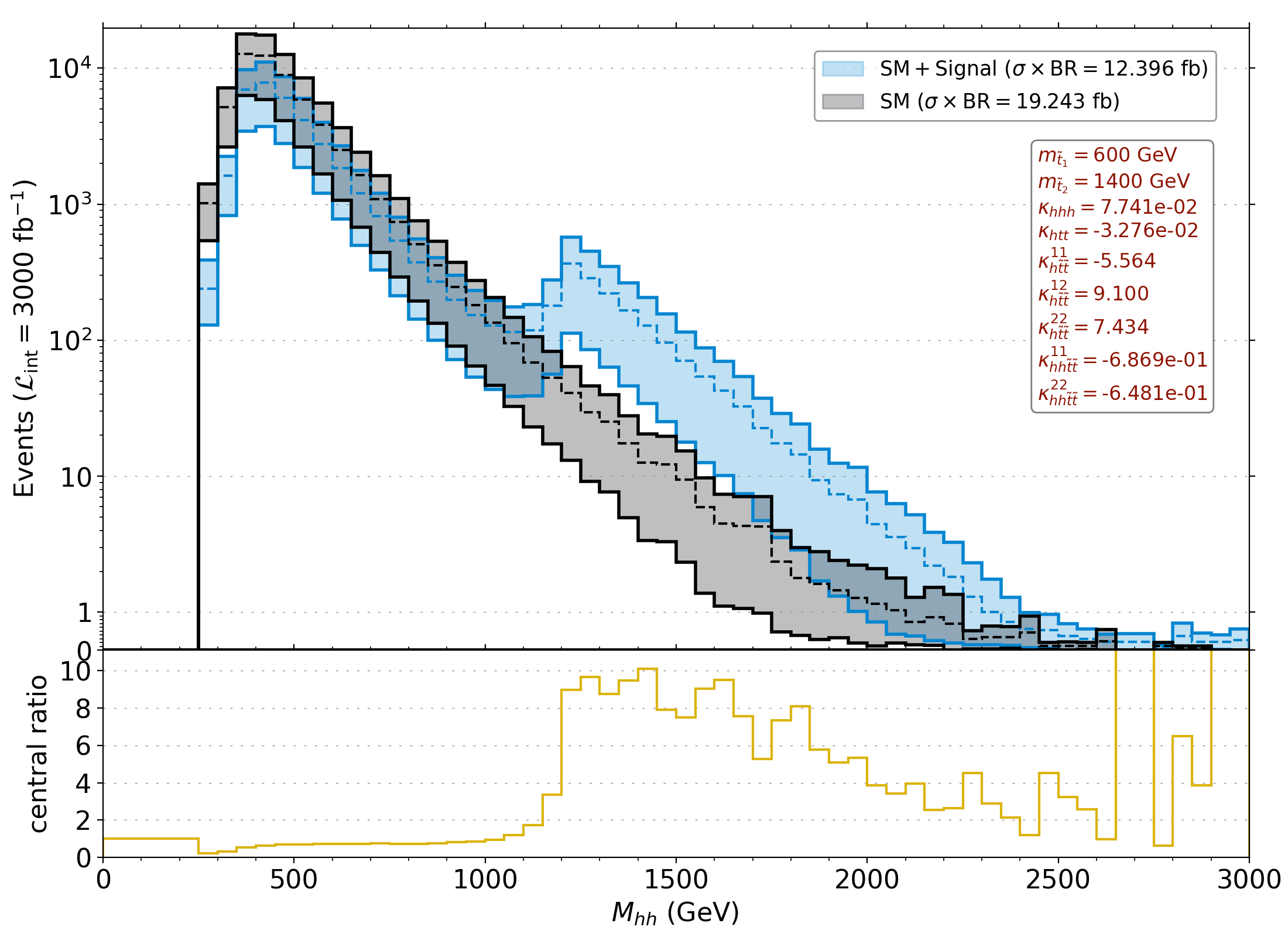,width=.475\textwidth}
\caption{\label{fig:distributionsNMSSM}The invariant mass distributions for the two NMSSM benchmarks described in the text. The integrated luminosity is in this case 3000 fb$^{-1}$, corresponding to the nominal reach of the HL-LHC phase. The \ul{left} panels correspond to a case with two light stops with minimal mixing, while the \ul{right} panels have a light and a heavy stop. Both have a trilinear Higgs coupling that is larger than that of the SM.}
\end{figure}

In the left panels of \Cref{fig:distributionsNMSSM} we show a case with two stops having masses of approximately $600\GeV$, $\tan\beta=1.35$, $\lambda=0.64$, the trilinear Higgs coupling being $50\%$ larger than in the SM, trilinear Higgs-stop couplings being significantly smaller than those of the MSSM and the top Yukawa being close to the SM value, $\kappa_{t}\simeq 0.985$.
We see that there is a deficit of events at low $M_{hh}$: this is due to the enhanced trilinear Higgs coupling. In addition, the excess at the squark threshold has become smaller than the uncertainties\footnote{We remark, however, that the uncertainties in the plots are at LO and therefore bound to significantly reduce once higher-orders corrections are taken into account.} (while still being around twice the SM rate), which emphasises the significance of the bubble and triangle diagrams.

In the right panel of \Cref{fig:distributionsNMSSM} we have a ''MSSM-like'' case in the NMSSM. The stop masses and mixings are similar to those of the MSSM, but we have $\tan\beta=1.38$ and $\lambda=0.69$, which again lead to a significant deviation from the SM in the trilinear Higgs coupling, being about $1.6$ times the SM value.
This results in a deficit of events at low $M_{hh}$ and an excess of events at high $M_{hh}$, the intermediate range being SM-like due to the cancellation of the various BSM amplitudes. As the $bb\gamma\gamma$ channel is sensitive to low invariant masses and the $4b$ channel is sensitive to high invariant masses, we should see a deficit in the former and an excess in the latter.

If such a case were interpreted just through the modification of SM couplings, $\kappa_t$ and $\kappa_{\lambda}$, the results between different channels would show an inconsistency, which implies that the model is not sufficient to describe the physics. Such an effective description can be somewhat misleading, unless a complete EFT basis is used, even in the low $M_{hh}$ region as the interference between the squark diagrams with the SM ones can lead to a $30\%$ deviation in the event rate as our MSSM benchmark (with essentially $\kappa_t=\kappa_{\lambda}=1$) shows. In the case of a large coupling deviation the $\kappa$ framework leads to a better estimate, though even there we would have some uncertainty from the non-decoupling squark effects (difference between green (MB) and red (Signal) curves in the upper right panel of \Cref{fig:distributionsNMSSM}).

In the lowest panels of \Cref{fig:distributionsNMSSM} the ratios of the event rate compared to the SM are shown. The enhanced trilinear Higgs couplings lead to a deficit of about $60$--$70\%$ in the lowest mass bins. Such a precision would be needed to show that a first order EW phase transition is possible without additional light degrees of freedom.

\section{Prospects at the HL-LHC}

\begin{figure}[h]
\epsfig{file=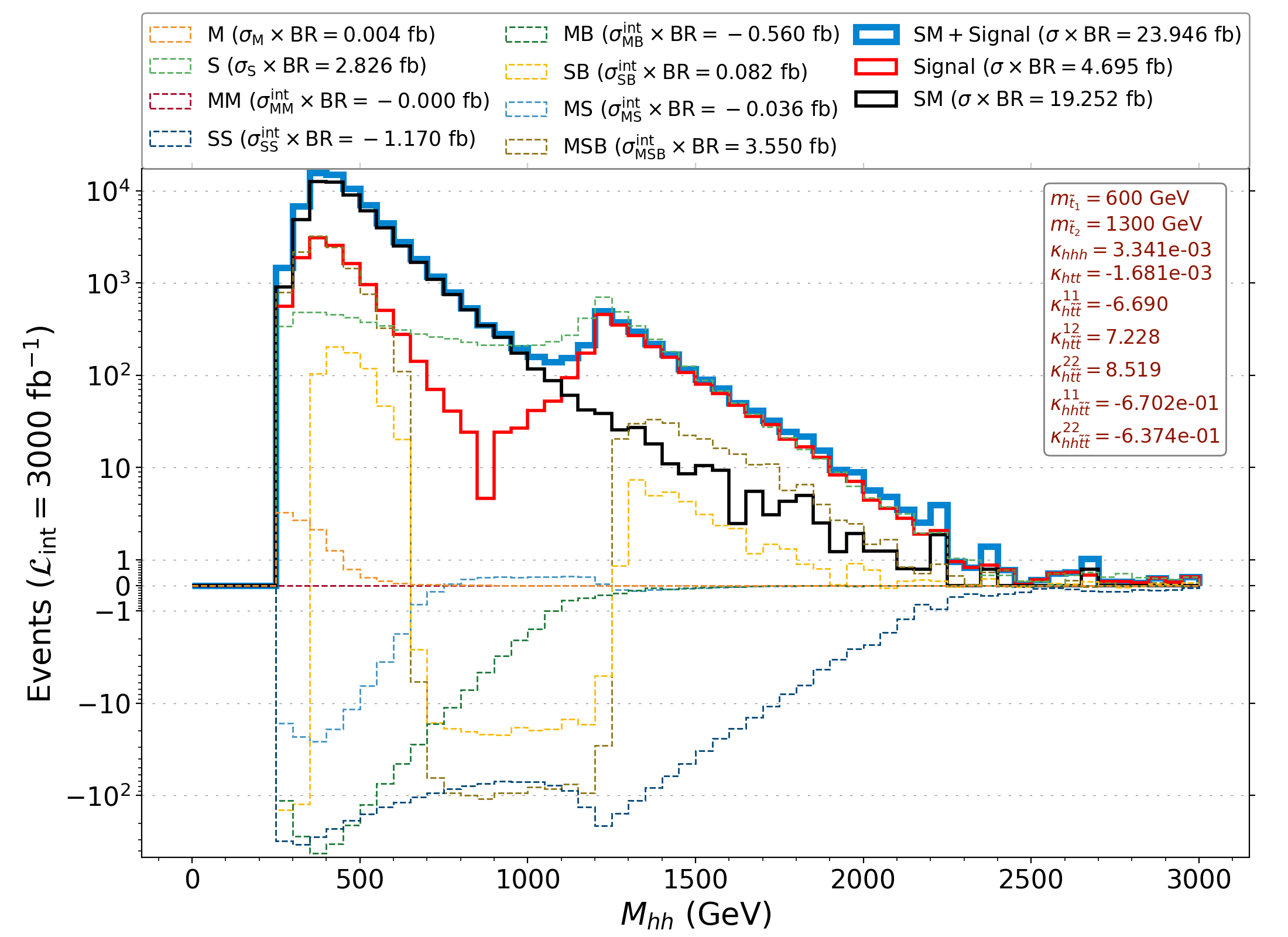,width=.48\textwidth}
\epsfig{file=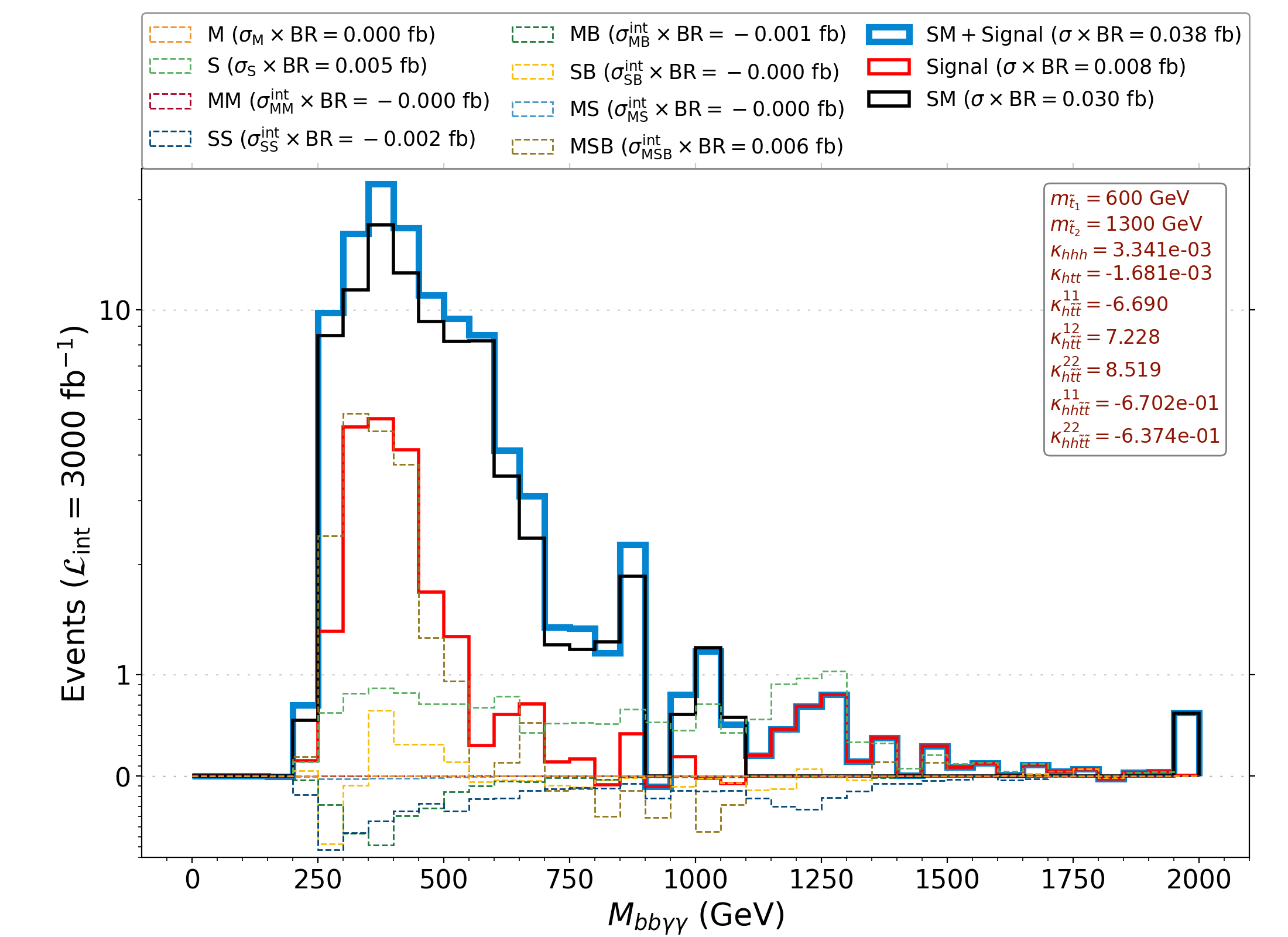,width=.48\textwidth}\\
\epsfig{file=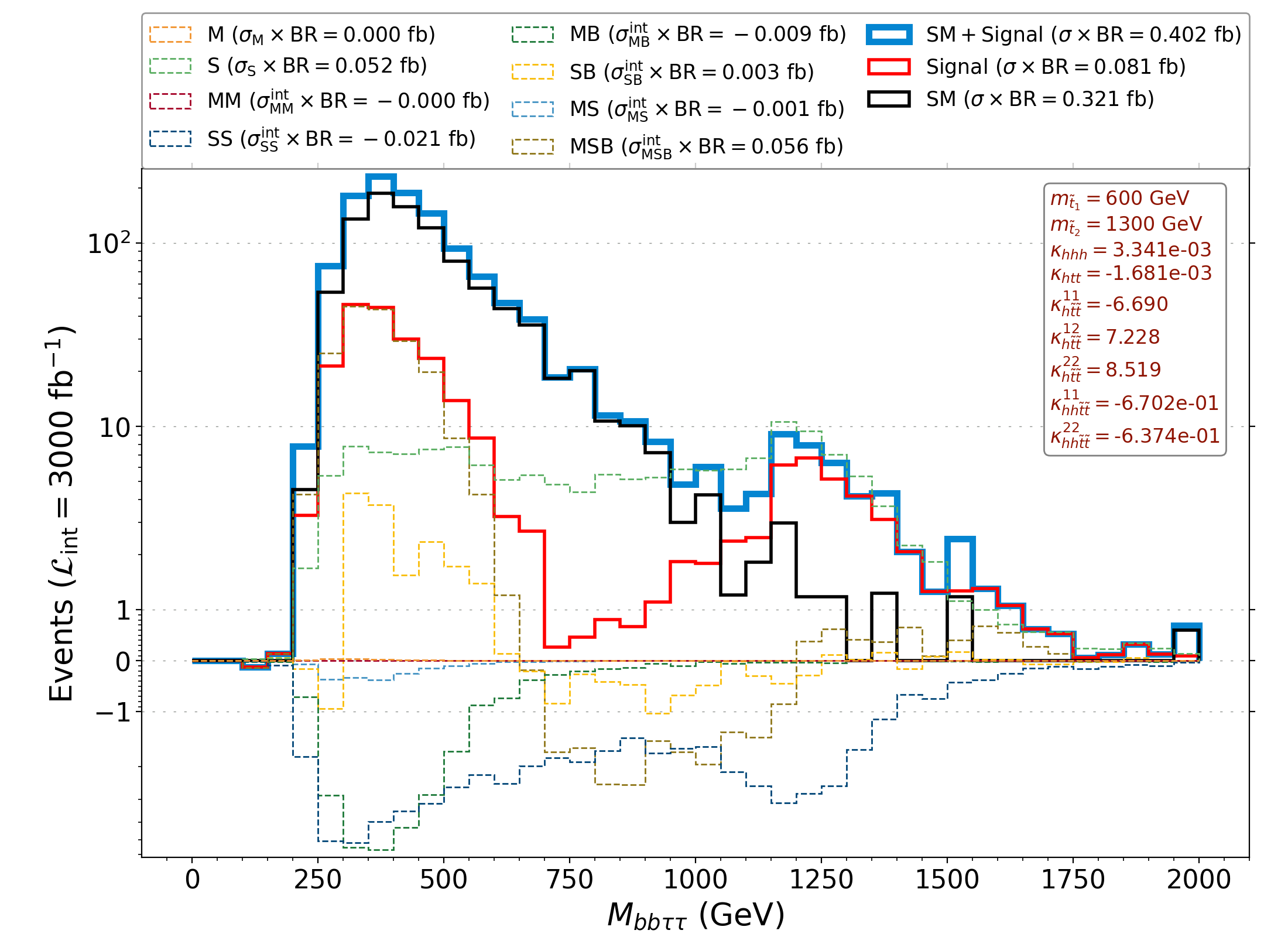,width=.48\textwidth}
\epsfig{file=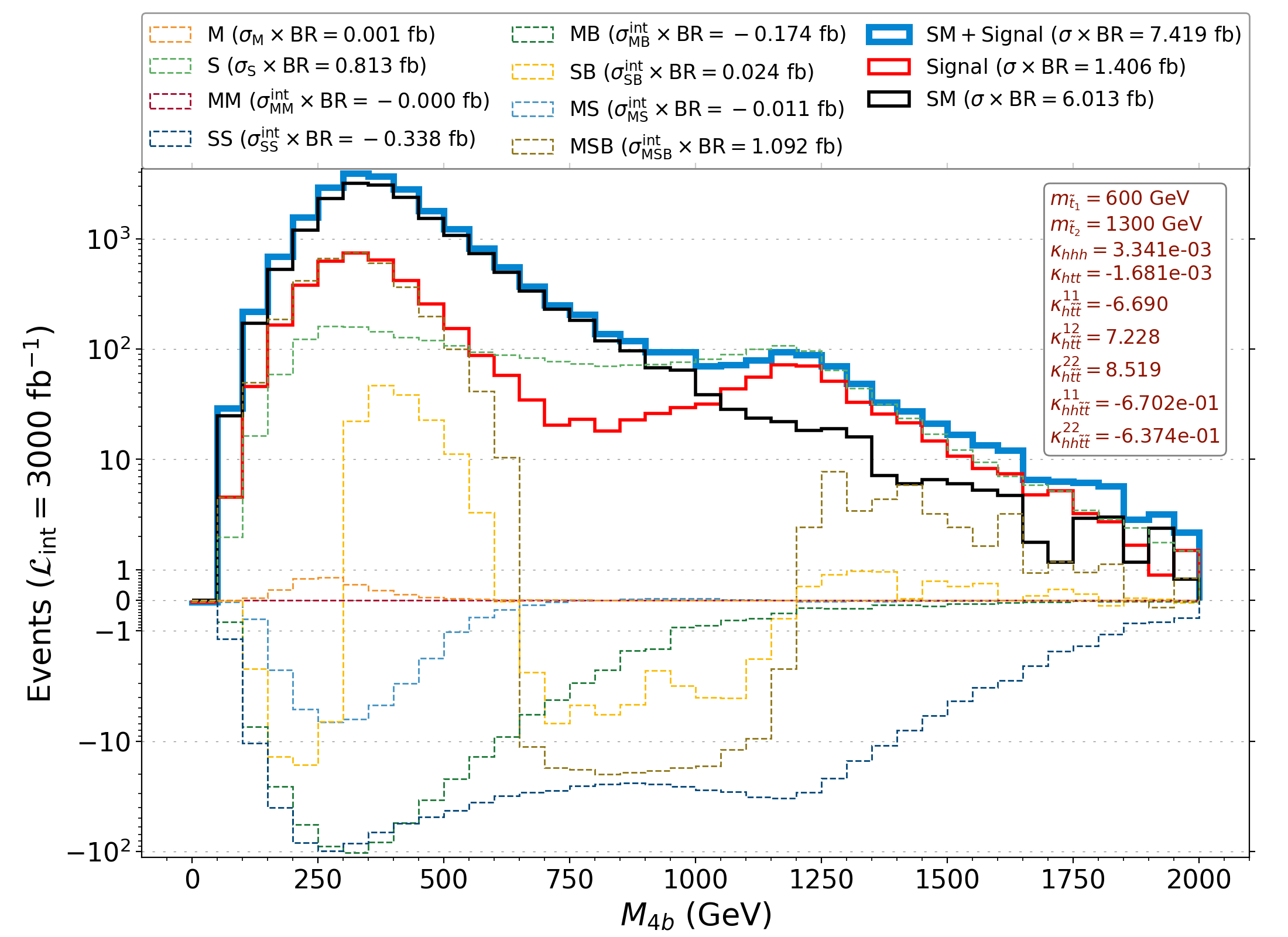,width=.48\textwidth}
\caption{\label{fig:distributionsreco3000}Invariant mass distributions for the MSSM benchmark point and for 3000 fb$^{-1}$. The \ul{top left} panel shows the distribution before Higgs decays, the other panels show the distributions for the 3 final states described in \Cref{sec:resultsreco} and after application of the cuts of \Cref{tab:cuts}. }
\end{figure}

To estimate the prospects for model exclusion/discovery around the 1 TeV region one can, e.g., use information from recent searches for resonant di-Higgs production at the LHC. One such example is an ATLAS search in the $b\bar{b}b\bar{b}$ final state using 139 fb$^{-1}$ of data collected at $\sqrt{s}=13$ TeV \cite{ATLAS:4b}. In the $M_{hh}$ mass range 1--1.2 TeV ATLAS expects $O(10)$ signal events over a background of O(100) events resulting in an excluded cross section around 10 pb. The signal acceptance times efficiency is around $10\%$ given a fiducial preselection. Comparing to the particle level preselected yields in \Cref{fig:distributionsrecobbbb} it seems not possible to make any exclusion of the benchmark point using only current LHC Run 2 data. However, with 20 times the data expected from HL-LHC as shown in \Cref{fig:distributionsreco3000}, compared to the ATLAS search example, the situation for exclusion looks much more promising. 

\begin{figure}[h]
\epsfig{file=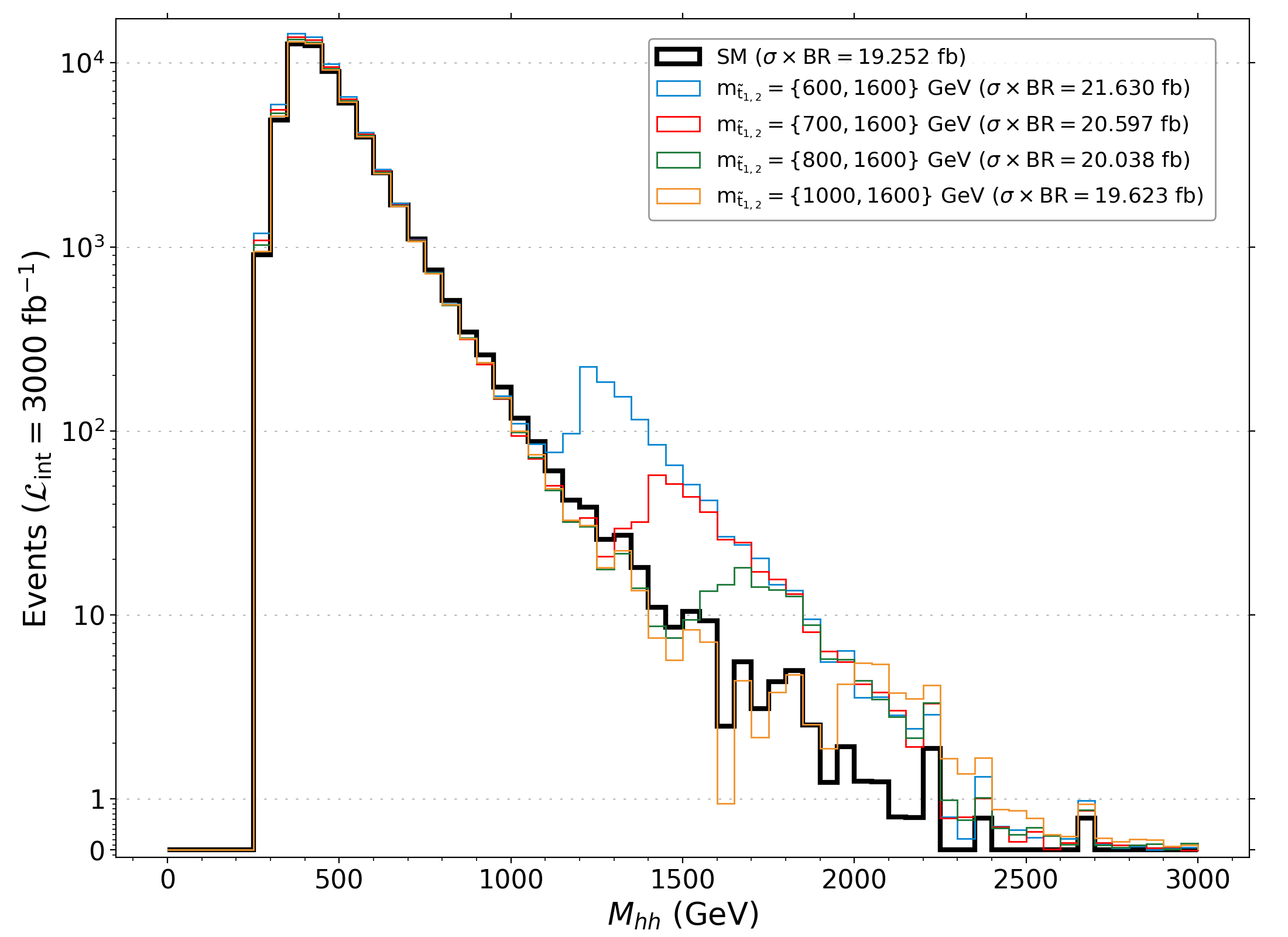,width=.48\textwidth}
\caption{\label{fig:MSSMmassdependence}Invariant mass distributions for 3000 fb$^{-1}$ and for different MSSM benchmark points characterised by different $\tilde t_1$ masses, from 600 GeV to 1 TeV, and $m_{\tilde t_2}=1600$ GeV.}
\end{figure}

This is of course true for the MSSM benchmark point of \cref{tab:MSSMBP}, which has been selected due to its high cross-section. It is important to stress that the cross-section strongly depends on the stop masses, as can be seen in the central panel of \cref{fig:scatterplot}, and that when the mass of the lightest stop approaches 1 TeV, the MSSM cross-section collapses to values similar to the SM case.
The threshold effects corresponding to $2m_{\tilde t_1}$ in the di-Higgs invariant mass distribution, however, can be seen also when the stop mass approaches 1 TeV: 
the excess on the stop threshold is in fact be compensated by a smooth and continuous depletion at smaller invariant mass values, with higher event counts than the peak, leading to a total cross-section which is similar to the SM one. This can be seen in \cref{fig:MSSMmassdependence} for different representative MSSM benchmark points characterised by the same value of $m_{\tilde t_2}$ and different values of $m_{\tilde t_1}$, from 600 GeV to 1000 TeV (couplings are not reported as not relevant in this context). For $m_{\tilde t_1}\gtrsim 800$ GeV the number of expected events is however already small (some units) at parton level, and therefore the actual observability of higher masses at HL-LHC is rather challenging, if just not possible.

The NMSSM benchmarks also introduce large effects at low invariant $M_{hh}$ mass. If one focuses only on the $M(hh)$ region below the stop mass where an EFT descriptions is valid, the NP contributions are dominated by the modified Higgs triple gauge coupling interfering with the rest of the SM background which is visible in \Cref{fig:distributionsNMSSM} (labelled MB). This effect is measured in the experiments using coupling modifiers ($\kappa$-framework) or more consistently with an EFT, and it can, as previously mentioned, have important cosmological implications. The current estimated HL-LHC 95\% exclusion limits for $\kappa_{\lambda}$ using e.g. the ATLAS experiment is $[0,2.5]$ \cite{ATL-PHYS-PUB-2022-053}, while the NMSSM benchmarks have $\kappa_{\lambda}=1.6$. This indicates that the low mass effects might be detectable during HL-LHC with more accurate theory modelling such as an EFT or techniques outlined in this paper, reduced experimental systematic uncertainties and combinations across different experiments.

Note that the sensitivity estimates made in this section do not include any di-Higgs signal corrections to account for missing higher order contributions ($K$-Factors). For NP processes these are estimated to be in the range $K\simeq 1.6-2.4$ \cite{Buchalla_2018}.

\section{Conclusions}
 In summary, we have provided a proof-of-concept that non-resonant di-Higgs production at the LHC (involving the SM-like Higgs boson already discovered) can be used as a proxy to viable models of the EW scale, such as SUSY, via loop effects induced by low-mass (i.e., non-decoupled) stops that can appear through both an increase (or decrease) of the inclusive (integrated) rates and significant shape changes in exclusive (differential) distributions of the $gg\to hh$ process. 
 Chiefly, amongst the latter, the invariant mass of the Higgs boson pair may display a large enhancement or depletion just above the $2m_h$ threshold (where the production rates are maximal) and a local maximum when $m_{hh}\approx 2m_{{\tilde t}_1}$ (in correspondence to a loop threshold involving the lightest stop state), both of which can be used for diagnostic purposes of the underlying EWSB scenario, separately or simultaneously depending on the benchmark point being targeted. While near the $2m_t$ threshold the effects are more spread out, the event rate for $gg\to hh$ can be enhanced by up to an order of magnitude compared to that of the SM at the $2m_{\tilde{t}}$ threshold.

 Quite remarkably, this can happen in the absence of stop cascade signals at the LHC, owing to compressed sparticle mass spectra which can easily be obtained, in compliance with current exclusion limits, in two popular SUSY realisations, minimal and non-minimal, like the MSSM and NMSSM, respectively, which we have adopted here. 
 So that, in the end, the effects investigated in this paper could well be the first evidence of SUSY (at least of the squark sector of it) at the LHC during the HL-LHC run. (Unfortunately, Runs 2 and 3 cannot afford one with significant sensitivity in this respect.)

 In order to prove this, we have resorted to an exact computational framework implementing a simplified model approach that, on the one hand, can easily be translated into any fundamental theory responsible for EWSB dynamics and, on the other hand, dispenses with the necessity of invoking any EFT limit of it, as it can compute (event level) integrated and differential rates for essentially any BSM spectrum of masses and couplings. In fact, the framework relies on a deconstruction method which makes it flexible to study a wide range of scenarios beyond those described here. 

 In this paper, we have shown all this being true while maintaining the $hh$ pair strictly on-shell. However, we have also argued that such stop loop effects might persist after backgrounds (both irreducible and reducible) are accounted for in the most popular channels leading to di-Higgs detection, following a reconstruction-level MC analysis that we have performed. This might be the case even considering both statistic and systematic uncertainties on the signal, which in our analysis are relatively large due to its LO nature. Amongst these, upon an initial assessment based on existing literature, the $4b$ channel is likely to be the one affording an actual experimental study with the best sensitivity (above and beyond $b\bar b\gamma\gamma$ and $b\bar b\tau^+\tau^-$), so long that the QCD-induced background can be brought under control. We postpone to a future paper the detailed detector level MC analysis of these final states, including dedicated selections targeting the described kinematic features. 

\section*{Acknowledgements}
This work would not have been possible without financing from the Knut and Alice Wallenberg
foundation under the grant KAW 2017.0100, which supported LP, JS and (partially) SM within the SHIFT project. SM is also supported in part through the
NExT Institute and the STFC Consolidated Grant No. ST/L000296/1. 
HW is supported by Carl Trygger Foundation under grant No. CTS18:164.
LP acknowledges the use of the Fysikum HPC Cluster at Stockholm University.
All authors thank Arnaud Ferrari, Venugopal Ellajosyula, Jos\'e Eliel Camargo-Molina and Rikard Enberg for discussions.

\appendix

\section{Simulation syntax in {\sc MG5\_aMC} for di-Higgs with the deconstruction method}
\label{app:MCsyntax}

In this appendix we describe how to obtain the various terms of \Cref{tab:deconstructedtopologies} using the {\sc UFO} model in \url{https://hepmdb.soton.ac.uk/hepmdb:0223.0337}. All our simulations have been made using {\sc MG5\_aMC} {\bf version 3.4.1}.\\

The {\sc UFO} model contains 4 coloured scalars, named \cbc{\verb~sq1, sq2, sq3~} and \cbc{\verb~sq4~}. 
Their trilinear couplings with the Higgs boson are proportional to the dimensionless parameters \cbc{\verb~KHSQ1SQ1, KHSQ1SQ2\dots ~}, which multiply a VEV factor, according to \Cref{eq:LagNPstop}. Each coupling is associated to a specific coupling order \cbc{\verb~HSQ1SQ1, HSQ1SQ2\dots ~} which can be invoked when performing simulations in {\sc MG5\_aMC} to restrict the number of possible topologies.
The quadrilinear couplings between the squarks and the Higgs boson are proportional to the dimensionless parameters \cbc{\verb~KHHSQ1SQ1, KHHSQ1SQ2\dots ~}, according to \Cref{eq:LagNPstop}. Also for the quadrilinears, specific coupling orders have been defined, \cbc{\verb~HHSQ1SQ1, HHSQ1SQ2\dots ~}, which allow to select or remove these couplings.
Finally, the modifications to the SM $hhh$, $htt$ and $hbb$ couplings are labelled as \cbc{\verb~KHHH~}, \cbc{\verb~KHTT~} and \cbc{\verb~KHBB~} respectively, with coupling orders \cbc{\verb~HHH~}, \cbc{\verb~HTTMOD~} and \cbc{\verb~HBBMOD~} respectively. The convention for these couplings is again the same of \Cref{eq:LagNPstop}.\\

In order to reproduce the results of this paper, the model has to be imported in {\sc MG5\_aMC} with a restriction which removes \cbc{\verb~sq3~, \verb~sq4~} and their corresponding interactions. 

\begin{tcolorbox}[top=-6pt,bottom=-6pt,left=-8pt,right=-8pt,width=\textwidth]
{\color{coolblack}
\small
\begin{lstlisting}
import model SQ4_diagonalCKM_4FNS_NLO_UFO-sq1sq2
\end{lstlisting}
}
\end{tcolorbox}

While not strictly necessary, the restriction greatly speeds up the calculations by ignoring irrelevant interactions and propagators. Alternatively it would be enough to set all coupling orders associated to \cbc{\verb~sq3~} and \cbc{\verb~sq4~} to zero in the topology-generation phase, or set all the numerical values of their couplings to zero when generating events.
The {\bf baseline syntax} to generate di-Higgs is completely analogous to what one does in the SM:

\begin{tcolorbox}[top=-6pt,bottom=-6pt,left=-8pt,right=-8pt,width=\textwidth]
{\color{coolblack}
\small
\begin{lstlisting}
generate p p > h h [QCD]
\end{lstlisting}
}
\end{tcolorbox}

However, this syntax alone is \ul{not sufficient} (and {\sc MG5\_aMC} would raise an error message), as the other couplings orders of the model have to be specified. In our analysis we want to isolate specific topologies to apply the deconstruction method: a careful tuning of the coupling orders allow us to do it in an efficient and unambiguous way. 

The {\bf SM irreducible di-Higgs background} can be obtained in different ways, either by adding to the baseline syntax all coupling orders associated to NP set to 0:

\begin{tcolorbox}[top=-6pt,bottom=-6pt,left=-8pt,right=-8pt,width=\textwidth]
{\color{coolblack}
\small
\begin{lstlisting}
aEW=2 HHH=0 HTTMOD=0 HBBMOD=0 HSQ1SQ1=0 HSQ2SQ2=0 HSQ1SQ2=0 HHSQ1SQ1=0 HHSQ2SQ2=0 HHSQ1SQ2=0
\end{lstlisting}
}
\end{tcolorbox}
\noindent or by removing all squarks from propagation and setting only the SM modified coupling orders to 0: 
\begin{tcolorbox}[top=-6pt,bottom=-6pt,left=-8pt,right=-8pt,width=\textwidth]
{\color{coolblack}
\small
\begin{lstlisting}
\ sq1 sq2 aEW=2 HHH=0 HTTMOD=0 HBBMOD=0
\end{lstlisting}
}
\end{tcolorbox}

The {\bf signal and interferences} can then be simulated using different combinations of the coupling orders. One example for each component of \Cref{eq:sigmahats} should be sufficient to describe the whole procedure.\\

\noindent {\bf M}: $\kappa_{hhh}^2 \hat\sigma_1$

\begin{tcolorbox}[top=-6pt,bottom=-6pt,left=-8pt,right=-8pt,width=\textwidth]
{\color{coolblack}
\small
\begin{lstlisting}
aEW=1 HHH=1 HTTMOD=0 HBBMOD=0 HSQ1SQ1=0 HSQ2SQ2=0 HSQ1SQ2=0 HHSQ1SQ1=0 HHSQ2SQ2=0 HHSQ1SQ2=0
\end{lstlisting}
}
\end{tcolorbox}

\noindent {\bf S}: $\kappa_{h\tilde t\tilde t}^{11} \kappa_{h\tilde t\tilde t}^{22} \hat\sigma_{5i}^{\rm int}(m_{\tilde t_{1,2}})$

\begin{tcolorbox}[top=-6pt,bottom=-6pt,left=-8pt,right=-8pt,width=\textwidth]
{\color{coolblack}
\small
\begin{lstlisting}
aEW=1 HHH=0 HTTMOD=0 HBBMOD=0 HSQ1SQ1^2==1 HSQ2SQ2^2==1 HSQ1SQ2=0 HHSQ1SQ1=0 HHSQ2SQ2=0 HHSQ1SQ2=0
\end{lstlisting}
}
\end{tcolorbox}

\noindent {\bf MB}: $\kappa_{htt} \hat\sigma_{2B}^{\rm int}$

\begin{tcolorbox}[top=-6pt,bottom=-6pt,left=-8pt,right=-8pt,width=\textwidth]
{\color{coolblack}
\small
\begin{lstlisting}
QED^2==3 HHH=0 HTTMOD^2==1 HBBMOD=0 HSQ1SQ1=0 HSQ2SQ2=0 HSQ1SQ2=0 HHSQ1SQ1=0 HHSQ2SQ2=0 HHSQ1SQ2=0
\end{lstlisting}
}
\end{tcolorbox}

\noindent {\bf SB}: $\kappa_{h\tilde t \tilde t}^{11} \hat\sigma_{5B}^{\rm int}(m_{\tilde t_1})$

\begin{tcolorbox}[top=-6pt,bottom=-6pt,left=-8pt,right=-8pt,width=\textwidth]
{\color{coolblack}
\small
\begin{lstlisting}
QED^2==3 HHH=0 HTTMOD=0 HBBMOD=0 HSQ1SQ1^2==1 HSQ2SQ2=0 HSQ1SQ2=0 HHSQ1SQ1=0 HHSQ2SQ2=0 HHSQ1SQ2=0
\end{lstlisting}
}
\end{tcolorbox}

\noindent {\bf MM}: $\kappa_{hhh}^2 \kappa_{htt}\hat\sigma_{1,3}^{\rm int}$

\begin{tcolorbox}[top=-6pt,bottom=-6pt,left=-8pt,right=-8pt,width=\textwidth]
{\color{coolblack}
\small
\begin{lstlisting}
QED^2==1 HHH^2==2 HTTMOD^2==1 HBBMOD=0 HSQ1SQ1=0 HSQ2SQ2=0 HSQ1SQ2=0 HHSQ1SQ1=0 HHSQ2SQ2=0 HHSQ1SQ2=0
\end{lstlisting}
}
\end{tcolorbox}

\noindent {\bf SS}: $\kappa_{hhh} \kappa_{h\tilde t \tilde t }^{11} (\kappa_{h\tilde t \tilde t }^{12})^2 \hat\sigma_{6,7o}^{\rm int}(m_{\tilde t_{1,2}})$

\begin{tcolorbox}[top=-6pt,bottom=-6pt,left=-8pt,right=-8pt,width=\textwidth]
{\color{coolblack}
\small
\begin{lstlisting}
QED=0 HHH^2==1 HTTMOD=0 HBBMOD=0 HSQ1SQ1^2==1 HSQ2SQ2=0 HSQ1SQ2^2==2 HHSQ1SQ1=0 HHSQ2SQ2=0 HHSQ1SQ2=0
\end{lstlisting}
}
\end{tcolorbox}

\noindent {\bf MS}: $\kappa_{hhh} \kappa_{htt} \kappa_{hh\tilde t\tilde t}^{11} \hat\sigma_{3,8}^{\rm int}(m_{t_1})$

\begin{tcolorbox}[top=-6pt,bottom=-6pt,left=-8pt,right=-8pt,width=\textwidth]
{\color{coolblack}
\small
\begin{lstlisting}
QED=0 HHH^2==1 HTTMOD^2==1 HBBMOD=0 HSQ1SQ1=0 HSQ2SQ2=0 HSQ1SQ2=0 HHSQ1SQ1^2==1 HHSQ2SQ2=0 HHSQ1SQ2=0
\end{lstlisting}
}
\end{tcolorbox}

\noindent {\bf MSB}: $\kappa_{hhh} (\kappa_{h\tilde t \tilde t }^{11})^2 \hat\sigma_{1,7d-5,6}^{\rm int}(m_{\tilde t_1})$

\begin{tcolorbox}[top=-6pt,bottom=-6pt,left=-8pt,right=-8pt,width=\textwidth]
{\color{coolblack}
\small
\begin{lstlisting}
QED^2==1 HHH^2==1 HTTMOD=0 HBBMOD=0 HSQ1SQ1^2==2 HSQ2SQ2=0 HSQ1SQ2=0 HHSQ1SQ1=0 HHSQ2SQ2=0 HHSQ1SQ2=0
\end{lstlisting}
}
\end{tcolorbox}

\section{Trilinear Higgs coupling in the MSSM and the NMSSM}\label{sec:trilinear}

In the SM the Higgs potential is particularly simple. The Higgs self-interactions arise solely from the term $\lambda |H|^{4}$. Since $\lambda$ also determines the Higgs mass, we get a unique prediction
\begin{equation}
    \lambda_{SM}=\frac{g^{2}m_{h}^{2}}{8m_{W}^{2}}.
\end{equation}

In the MSSM it is easier to achieve the $125$~GeV mass in the large $\tan\beta$ regime. In such a case the SM-like Higgs in the alignment limit is practically the neutral component of the $H_{u}$ doublet. Thus the only relevant term of the loop-corrected scalar potential that is relevant for the SM-like Higgs mass generation is of the form $\lambda_{\mathrm{eff}}|H_{u}|^{4}$ and hence necessarily $\lambda_{\mathrm{eff}}\simeq \lambda_{SM}$ in the MSSM. This is why you never get large deviations in the quartic or trilinear Higgs coupling in the MSSM.

In the NMSSM the situation is different. In addition to the standard MSSM D-terms, you get an extra quartic term in the Higgs potential, $|\lambda|^{2}|H_{u}H_{d}|^{2}$, which is relevant to the Higgs mass generation, if $\tan\beta$ is low or moderate. This leads to the well-known mass bound of Eq. (\ref{eq:NMSSMbound}). From the previous argument it is obvious that in the large $\tan \beta$ region it is possible to get the correct Higgs mass, but no deviations from the SM value of the Higgs self-coupling. With $\tan\beta$ close to one the situation is different.

We now assume that the singlet state is decoupled, so that the soft trilinear interaction $A_{\lambda}H_{u}H_{d}S$ is irrelevant for the Higgs self-coupling. At tree-level the quartic self-coupling and the Higgs masses arise from the following terms of the scalar potential
\begin{equation}\label{eq:higgspotential}
V(H_{u}^{0},H_{d}^{0})=\frac{g^{2}+g^{\prime 2}}{8}\Big{|}|H_{u}^{0}|^{2}-|H_{d}^{0}|^{2}\Big{|}^{2}
+\frac{1}{4}|\lambda|^{2}|H_{u}^{0}H_{d}^{0}|^{2}+\left( A_{\lambda}\frac{v_{s}}{\sqrt{2}} +\kappa\lambda \frac{v_{s}^{2}}{2}\right) H_{u}^{0}H_{d}^{0}\ldots
\end{equation}

When the mass matrix is computed, the contribution of the bilinear term comes in a form that leaves no contribution to the SM-like Higgs mass, just like the soft SUSY breaking $B_{\mu}H_{u}H_{d}$ term does not contribute to the Higgs mass in the MSSM. Terms of the form of $|H_{u}^{0}H_{d}^{0}|^{2}$ do not contribute to the diagonal elements of the mass matrix as they get eliminated by the tadpole equations. Hence we get two types of contributions to the mass matrix that affect the SM-like Higgs mass: diagonal elements from $|H_{u,d}^{0}|^{4}$, which come with a factor (from differentiating the potential) of $4\times 3$ and off-diagonal elements from $|H_{u}^{0}H_{d}^{0}|^{2}$, which come with a factor of $2\times 2$, in addition there are factors depending on $\beta$. As we typically have $\tan\beta > 1$, the most important contributions are those from $|H_{u}^{0}|^{4}$ and $|H_{u}^{0}H_{d}^{0}|^{2}$, which have relative weights of $\sin^{2}\beta$ and $\cos^{2}\beta$, respectively. Hence any effect from the $|\lambda|^{2}$ term gets suppressed as $\tan\beta$ increases.

In the trilinear Higgs coupling the relative factors are different, one gets a factor of $4$ for the $|H_{u}^{0}|^{4}$ term and two contributions with factors of $2$ from $|H_{u}^{0}H_{d}^{0}|^{2}$, again the former having a factor of $\sin^{2}\beta$ and the latter having a factor $\cos^{2}\beta$. Therefore the relative contribution of the $|H_{u}^{0}H_{d}^{0}|^{2}$ term to the trilinear Higgs self interaction is larger than to its mass. Hence if $\lambda$ is larger than the gauge couplings, it will enhance the Higgs trilinear from its SM value and if it is smaller, it will reduce it. However, this effect is suppressed by $\cos^{2}\beta$, so you can only have a significant deviation in the region $1\lesssim \tan\beta \lesssim 3$. Obviously, the precise computation of the trilinear Higgs coupling will need the loop corrections to the scalar potential (\ref{eq:higgspotential}), but this tree-level discussion should allow to understand the results.

The enhancement is relatively easy to combine with any stop masses, as the tree-level contribution reduces the need of large loop corrections, so you can always find a suitable level of stop mixing. The parameter space, where the Higgs trilinear is smaller than in the SM is very narrow. The best chances are when the singlet-dominated state is lighter than the SM-like Higgs so that the mixing can push the SM-like Higgs mass up \cite{Badziak:2013bda}. Even in that case large loop corrections are typically needed, so heavy stops and sbottoms are preferred. In addition, as the singlet is not decoupled, you cannot neglect the trilinear soft interaction term $A_{\lambda}H_{u}H_{d}S$, so that scenario is more complex than the case where the trilinear coupling is enhanced.

\bibliography{stopdihiggs}

\begin{thebibliography}{50}%
\makeatletter
\providecommand \@ifxundefined [1]{%
 \@ifx{#1\undefined}
}%
\providecommand \@ifnum [1]{%
 \ifnum #1\expandafter \@firstoftwo
 \else \expandafter \@secondoftwo
 \fi
}%
\providecommand \@ifx [1]{%
 \ifx #1\expandafter \@firstoftwo
 \else \expandafter \@secondoftwo
 \fi
}%
\providecommand \natexlab [1]{#1}%
\providecommand \enquote  [1]{``#1''}%
\providecommand \bibnamefont  [1]{#1}%
\providecommand \bibfnamefont [1]{#1}%
\providecommand \citenamefont [1]{#1}%
\providecommand \href@noop [0]{\@secondoftwo}%
\providecommand \href [0]{\begingroup \@sanitize@url \@href}%
\providecommand \@href[1]{\@@startlink{#1}\@@href}%
\providecommand \@@href[1]{\endgroup#1\@@endlink}%
\providecommand \@sanitize@url [0]{\catcode `\\12\catcode `\$12\catcode
  `\&12\catcode `\#12\catcode `\^12\catcode `\_12\catcode `\%12\relax}%
\providecommand \@@startlink[1]{}%
\providecommand \@@endlink[0]{}%
\providecommand \url  [0]{\begingroup\@sanitize@url \@url }%
\providecommand \@url [1]{\endgroup\@href {#1}{\urlprefix }}%
\providecommand \urlprefix  [0]{URL }%
\providecommand \Eprint [0]{\href }%
\providecommand \doibase [0]{http://dx.doi.org/}%
\providecommand \selectlanguage [0]{\@gobble}%
\providecommand \bibinfo  [0]{\@secondoftwo}%
\providecommand \bibfield  [0]{\@secondoftwo}%
\providecommand \translation [1]{[#1]}%
\providecommand \BibitemOpen [0]{}%
\providecommand \bibitemStop [0]{}%
\providecommand \bibitemNoStop [0]{.\EOS\space}%
\providecommand \EOS [0]{\spacefactor3000\relax}%
\providecommand \BibitemShut  [1]{\csname bibitem#1\endcsname}%
\let\auto@bib@innerbib\@empty
\bibitem [{\citenamefont {Plehn}\ \emph {et~al.}(1996)\citenamefont {Plehn},
  \citenamefont {Spira},\ and\ \citenamefont {Zerwas}}]{Plehn:1996wb}%
  \BibitemOpen
  \bibfield  {author} {\bibinfo {author} {\bibfnamefont {T.}~\bibnamefont
  {Plehn}}, \bibinfo {author} {\bibfnamefont {M.}~\bibnamefont {Spira}}, \ and\
  \bibinfo {author} {\bibfnamefont {P.~M.}\ \bibnamefont {Zerwas}},\ }\href
  {\doibase 10.1016/0550-3213(96)00418-X} {\bibfield  {journal} {\bibinfo
  {journal} {Nucl. Phys. B}\ }\textbf {\bibinfo {volume} {479}},\ \bibinfo
  {pages} {46} (\bibinfo {year} {1996})},\ \bibinfo {note} {[Erratum:
  Nucl.Phys.B 531, 655--655 (1998)]},\ \Eprint
  {http://arxiv.org/abs/hep-ph/9603205} {arXiv:hep-ph/9603205} \BibitemShut
  {NoStop}%
\bibitem [{\citenamefont {Dawson}\ \emph {et~al.}(1998)\citenamefont {Dawson},
  \citenamefont {Dittmaier},\ and\ \citenamefont {Spira}}]{Dawson:1998py}%
  \BibitemOpen
  \bibfield  {author} {\bibinfo {author} {\bibfnamefont {S.}~\bibnamefont
  {Dawson}}, \bibinfo {author} {\bibfnamefont {S.}~\bibnamefont {Dittmaier}}, \
  and\ \bibinfo {author} {\bibfnamefont {M.}~\bibnamefont {Spira}},\ }\href
  {\doibase 10.1103/PhysRevD.58.115012} {\bibfield  {journal} {\bibinfo
  {journal} {Phys. Rev. D}\ }\textbf {\bibinfo {volume} {58}},\ \bibinfo
  {pages} {115012} (\bibinfo {year} {1998})},\ \Eprint
  {http://arxiv.org/abs/hep-ph/9805244} {arXiv:hep-ph/9805244} \BibitemShut
  {NoStop}%
\bibitem [{\citenamefont {Glover}\ and\ \citenamefont {van~der
  Bij}(1988)}]{Glover:1987nx}%
  \BibitemOpen
  \bibfield  {author} {\bibinfo {author} {\bibfnamefont {E.~W.~N.}\
  \bibnamefont {Glover}}\ and\ \bibinfo {author} {\bibfnamefont {J.~J.}\
  \bibnamefont {van~der Bij}},\ }\href {\doibase 10.1016/0550-3213(88)90083-1}
  {\bibfield  {journal} {\bibinfo  {journal} {Nucl. Phys. B}\ }\textbf
  {\bibinfo {volume} {309}},\ \bibinfo {pages} {282} (\bibinfo {year}
  {1988})}\BibitemShut {NoStop}%
\bibitem [{\citenamefont {Sirunyan}\ \emph
  {et~al.}(2019{\natexlab{a}})\citenamefont {Sirunyan} \emph
  {et~al.}}]{CMS:2018sxu}%
  \BibitemOpen
  \bibfield  {author} {\bibinfo {author} {\bibfnamefont {A.~M.}\ \bibnamefont
  {Sirunyan}} \emph {et~al.} (\bibinfo {collaboration} {CMS}),\ }\href
  {\doibase 10.1007/JHEP04(2019)112} {\bibfield  {journal} {\bibinfo  {journal}
  {JHEP}\ }\textbf {\bibinfo {volume} {04}},\ \bibinfo {pages} {112} (\bibinfo
  {year} {2019}{\natexlab{a}})},\ \Eprint {http://arxiv.org/abs/1810.11854}
  {arXiv:1810.11854 [hep-ex]} \BibitemShut {NoStop}%
\bibitem [{\citenamefont {Aaboud}\ \emph {et~al.}(2019)\citenamefont {Aaboud}
  \emph {et~al.}}]{ATLAS:2018ili}%
  \BibitemOpen
  \bibfield  {author} {\bibinfo {author} {\bibfnamefont {M.}~\bibnamefont
  {Aaboud}} \emph {et~al.} (\bibinfo {collaboration} {ATLAS}),\ }\href
  {\doibase 10.1007/JHEP05(2019)124} {\bibfield  {journal} {\bibinfo  {journal}
  {JHEP}\ }\textbf {\bibinfo {volume} {05}},\ \bibinfo {pages} {124} (\bibinfo
  {year} {2019})},\ \Eprint {http://arxiv.org/abs/1811.11028} {arXiv:1811.11028
  [hep-ex]} \BibitemShut {NoStop}%
\bibitem [{\citenamefont {Sirunyan}\ \emph
  {et~al.}(2019{\natexlab{b}})\citenamefont {Sirunyan} \emph
  {et~al.}}]{CMS:2019noi}%
  \BibitemOpen
  \bibfield  {author} {\bibinfo {author} {\bibfnamefont {A.~M.}\ \bibnamefont
  {Sirunyan}} \emph {et~al.} (\bibinfo {collaboration} {CMS}),\ }\href
  {\doibase 10.1007/JHEP10(2019)125} {\bibfield  {journal} {\bibinfo  {journal}
  {JHEP}\ }\textbf {\bibinfo {volume} {10}},\ \bibinfo {pages} {125} (\bibinfo
  {year} {2019}{\natexlab{b}})},\ \Eprint {http://arxiv.org/abs/1904.04193}
  {arXiv:1904.04193 [hep-ex]} \BibitemShut {NoStop}%
\bibitem [{\citenamefont {Aad}\ \emph {et~al.}(2020)\citenamefont {Aad} \emph
  {et~al.}}]{ATLAS:2019vwv}%
  \BibitemOpen
  \bibfield  {author} {\bibinfo {author} {\bibfnamefont {G.}~\bibnamefont
  {Aad}} \emph {et~al.} (\bibinfo {collaboration} {ATLAS}),\ }\href {\doibase
  10.1016/j.physletb.2019.135145} {\bibfield  {journal} {\bibinfo  {journal}
  {Phys. Lett. B}\ }\textbf {\bibinfo {volume} {801}},\ \bibinfo {pages}
  {135145} (\bibinfo {year} {2020})},\ \Eprint
  {http://arxiv.org/abs/1908.06765} {arXiv:1908.06765 [hep-ex]} \BibitemShut
  {NoStop}%
\bibitem [{\citenamefont {Sirunyan}\ \emph
  {et~al.}(2021{\natexlab{a}})\citenamefont {Sirunyan} \emph
  {et~al.}}]{CMS:2020tkr}%
  \BibitemOpen
  \bibfield  {author} {\bibinfo {author} {\bibfnamefont {A.~M.}\ \bibnamefont
  {Sirunyan}} \emph {et~al.} (\bibinfo {collaboration} {CMS}),\ }\href
  {\doibase 10.1007/JHEP03(2021)257} {\bibfield  {journal} {\bibinfo  {journal}
  {JHEP}\ }\textbf {\bibinfo {volume} {03}},\ \bibinfo {pages} {257} (\bibinfo
  {year} {2021}{\natexlab{a}})},\ \Eprint {http://arxiv.org/abs/2011.12373}
  {arXiv:2011.12373 [hep-ex]} \BibitemShut {NoStop}%
\bibitem [{\citenamefont {Aad}\ \emph {et~al.}(2022{\natexlab{a}})\citenamefont
  {Aad} \emph {et~al.}}]{ATLAS:2021ifb}%
  \BibitemOpen
  \bibfield  {author} {\bibinfo {author} {\bibfnamefont {G.}~\bibnamefont
  {Aad}} \emph {et~al.} (\bibinfo {collaboration} {ATLAS}),\ }\href {\doibase
  10.1103/PhysRevD.106.052001} {\bibfield  {journal} {\bibinfo  {journal}
  {Phys. Rev. D}\ }\textbf {\bibinfo {volume} {106}},\ \bibinfo {pages}
  {052001} (\bibinfo {year} {2022}{\natexlab{a}})},\ \Eprint
  {http://arxiv.org/abs/2112.11876} {arXiv:2112.11876 [hep-ex]} \BibitemShut
  {NoStop}%
\bibitem [{\citenamefont {Sirunyan}\ \emph {et~al.}(2022)\citenamefont
  {Sirunyan} \emph {et~al.}}]{CMS:2022hgz}%
  \BibitemOpen
  \bibfield  {author} {\bibinfo {author} {\bibfnamefont {A.~M.}\ \bibnamefont
  {Sirunyan}} \emph {et~al.} (\bibinfo {collaboration} {CMS}),\ }\href@noop {}
  {\  (\bibinfo {year} {2022})},\ \Eprint {http://arxiv.org/abs/2206.09401}
  {arXiv:2206.09401 [hep-ex]} \BibitemShut {NoStop}%
\bibitem [{\citenamefont {Aad}\ \emph {et~al.}(2022{\natexlab{b}})\citenamefont
  {Aad} \emph {et~al.}}]{ATLAS:2022xzm}%
  \BibitemOpen
  \bibfield  {author} {\bibinfo {author} {\bibfnamefont {G.}~\bibnamefont
  {Aad}} \emph {et~al.} (\bibinfo {collaboration} {ATLAS}),\ }\href@noop {} {\
  (\bibinfo {year} {2022}{\natexlab{b}})},\ \Eprint
  {http://arxiv.org/abs/2209.10910} {arXiv:2209.10910 [hep-ex]} \BibitemShut
  {NoStop}%
\bibitem [{\citenamefont {{ATLAS Collaboration}}(2023)}]{ATLAS:2023qzf}%
  \BibitemOpen
  \bibfield  {author} {\bibinfo {author} {\bibnamefont {{ATLAS
  Collaboration}}},\ }\href@noop {} {\  (\bibinfo {year} {2023})},\ \Eprint
  {http://arxiv.org/abs/2301.03212} {arXiv:2301.03212 [hep-ex]} \BibitemShut
  {NoStop}%
\bibitem [{\citenamefont {Moretti}\ and\ \citenamefont
  {Khalil}(2019)}]{Moretti:2019ulc}%
  \BibitemOpen
  \bibfield  {author} {\bibinfo {author} {\bibfnamefont {S.}~\bibnamefont
  {Moretti}}\ and\ \bibinfo {author} {\bibfnamefont {S.}~\bibnamefont
  {Khalil}},\ }\href@noop {} {\emph {\bibinfo {title} {{Supersymmetry Beyond
  Minimality: From Theory to Experiment}}}}\ (\bibinfo  {publisher} {CRC
  Press},\ \bibinfo {year} {2019})\BibitemShut {NoStop}%
\bibitem [{\citenamefont {Gianotti}\ \emph {et~al.}(2005)\citenamefont
  {Gianotti} \emph {et~al.}}]{Gianotti:2002xx}%
  \BibitemOpen
  \bibfield  {author} {\bibinfo {author} {\bibfnamefont {F.}~\bibnamefont
  {Gianotti}} \emph {et~al.},\ }\href {\doibase 10.1140/epjc/s2004-02061-6}
  {\bibfield  {journal} {\bibinfo  {journal} {Eur. Phys. J. C}\ }\textbf
  {\bibinfo {volume} {39}},\ \bibinfo {pages} {293} (\bibinfo {year} {2005})},\
  \Eprint {http://arxiv.org/abs/hep-ph/0204087} {arXiv:hep-ph/0204087}
  \BibitemShut {NoStop}%
\bibitem [{\citenamefont {Donini}(1996)}]{Donini:1995wh}%
  \BibitemOpen
  \bibfield  {author} {\bibinfo {author} {\bibfnamefont {A.}~\bibnamefont
  {Donini}},\ }\href {\doibase 10.1016/0550-3213(96)00081-8} {\bibfield
  {journal} {\bibinfo  {journal} {Nucl. Phys. B}\ }\textbf {\bibinfo {volume}
  {467}},\ \bibinfo {pages} {3} (\bibinfo {year} {1996})},\ \Eprint
  {http://arxiv.org/abs/hep-ph/9511289} {arXiv:hep-ph/9511289} \BibitemShut
  {NoStop}%
\bibitem [{\citenamefont {Sirunyan}\ \emph
  {et~al.}(2021{\natexlab{b}})\citenamefont {Sirunyan} \emph
  {et~al.}}]{CMS:2021beq}%
  \BibitemOpen
  \bibfield  {author} {\bibinfo {author} {\bibfnamefont {A.~M.}\ \bibnamefont
  {Sirunyan}} \emph {et~al.} (\bibinfo {collaboration} {CMS}),\ }\href
  {\doibase 10.1103/PhysRevD.104.052001} {\bibfield  {journal} {\bibinfo
  {journal} {Phys. Rev. D}\ }\textbf {\bibinfo {volume} {104}},\ \bibinfo
  {pages} {052001} (\bibinfo {year} {2021}{\natexlab{b}})},\ \Eprint
  {http://arxiv.org/abs/2103.01290} {arXiv:2103.01290 [hep-ex]} \BibitemShut
  {NoStop}%
\bibitem [{\citenamefont {Tumasyan}\ \emph {et~al.}(2021)\citenamefont
  {Tumasyan} \emph {et~al.}}]{CMS:2021eha}%
  \BibitemOpen
  \bibfield  {author} {\bibinfo {author} {\bibfnamefont {A.}~\bibnamefont
  {Tumasyan}} \emph {et~al.} (\bibinfo {collaboration} {CMS}),\ }\href
  {\doibase 10.1140/epjc/s10052-021-09721-5} {\bibfield  {journal} {\bibinfo
  {journal} {Eur. Phys. J. C}\ }\textbf {\bibinfo {volume} {81}},\ \bibinfo
  {pages} {970} (\bibinfo {year} {2021})},\ \Eprint
  {http://arxiv.org/abs/2107.10892} {arXiv:2107.10892 [hep-ex]} \BibitemShut
  {NoStop}%
\bibitem [{\citenamefont {Aad}\ \emph {et~al.}(2021)\citenamefont {Aad} \emph
  {et~al.}}]{ATLAS:2020xzu}%
  \BibitemOpen
  \bibfield  {author} {\bibinfo {author} {\bibfnamefont {G.}~\bibnamefont
  {Aad}} \emph {et~al.} (\bibinfo {collaboration} {ATLAS}),\ }\href {\doibase
  10.1007/JHEP04(2021)174} {\bibfield  {journal} {\bibinfo  {journal} {JHEP}\
  }\textbf {\bibinfo {volume} {04}},\ \bibinfo {pages} {174} (\bibinfo {year}
  {2021})},\ \Eprint {http://arxiv.org/abs/2012.03799} {arXiv:2012.03799
  [hep-ex]} \BibitemShut {NoStop}%
\bibitem [{\citenamefont {Cao}\ \emph {et~al.}(2013)\citenamefont {Cao},
  \citenamefont {Heng}, \citenamefont {Shang}, \citenamefont {Wan},\ and\
  \citenamefont {Yang}}]{Cao:2013si}%
  \BibitemOpen
  \bibfield  {author} {\bibinfo {author} {\bibfnamefont {J.}~\bibnamefont
  {Cao}}, \bibinfo {author} {\bibfnamefont {Z.}~\bibnamefont {Heng}}, \bibinfo
  {author} {\bibfnamefont {L.}~\bibnamefont {Shang}}, \bibinfo {author}
  {\bibfnamefont {P.}~\bibnamefont {Wan}}, \ and\ \bibinfo {author}
  {\bibfnamefont {J.~M.}\ \bibnamefont {Yang}},\ }\href {\doibase
  10.1007/JHEP04(2013)134} {\bibfield  {journal} {\bibinfo  {journal} {JHEP}\
  }\textbf {\bibinfo {volume} {04}},\ \bibinfo {pages} {134} (\bibinfo {year}
  {2013})},\ \Eprint {http://arxiv.org/abs/1301.6437} {arXiv:1301.6437
  [hep-ph]} \BibitemShut {NoStop}%
\bibitem [{\citenamefont {Batell}\ \emph {et~al.}(2015)\citenamefont {Batell},
  \citenamefont {McCullough}, \citenamefont {Stolarski},\ and\ \citenamefont
  {Verhaaren}}]{Batell:2015koa}%
  \BibitemOpen
  \bibfield  {author} {\bibinfo {author} {\bibfnamefont {B.}~\bibnamefont
  {Batell}}, \bibinfo {author} {\bibfnamefont {M.}~\bibnamefont {McCullough}},
  \bibinfo {author} {\bibfnamefont {D.}~\bibnamefont {Stolarski}}, \ and\
  \bibinfo {author} {\bibfnamefont {C.~B.}\ \bibnamefont {Verhaaren}},\ }\href
  {\doibase 10.1007/JHEP09(2015)216} {\bibfield  {journal} {\bibinfo  {journal}
  {JHEP}\ }\textbf {\bibinfo {volume} {09}},\ \bibinfo {pages} {216} (\bibinfo
  {year} {2015})},\ \Eprint {http://arxiv.org/abs/1508.01208} {arXiv:1508.01208
  [hep-ph]} \BibitemShut {NoStop}%
\bibitem [{\citenamefont {Huang}\ \emph {et~al.}(2018)\citenamefont {Huang},
  \citenamefont {Joglekar}, \citenamefont {Li},\ and\ \citenamefont
  {Wagner}}]{Huang:2017nnw}%
  \BibitemOpen
  \bibfield  {author} {\bibinfo {author} {\bibfnamefont {P.}~\bibnamefont
  {Huang}}, \bibinfo {author} {\bibfnamefont {A.}~\bibnamefont {Joglekar}},
  \bibinfo {author} {\bibfnamefont {M.}~\bibnamefont {Li}}, \ and\ \bibinfo
  {author} {\bibfnamefont {C.~E.~M.}\ \bibnamefont {Wagner}},\ }\href {\doibase
  10.1103/PhysRevD.97.075001} {\bibfield  {journal} {\bibinfo  {journal} {Phys.
  Rev. D}\ }\textbf {\bibinfo {volume} {97}},\ \bibinfo {pages} {075001}
  (\bibinfo {year} {2018})},\ \Eprint {http://arxiv.org/abs/1711.05743}
  {arXiv:1711.05743 [hep-ph]} \BibitemShut {NoStop}%
\bibitem [{\citenamefont {Hollik}\ and\ \citenamefont
  {Penaranda}(2002)}]{Hollik:2001px}%
  \BibitemOpen
  \bibfield  {author} {\bibinfo {author} {\bibfnamefont {W.}~\bibnamefont
  {Hollik}}\ and\ \bibinfo {author} {\bibfnamefont {S.}~\bibnamefont
  {Penaranda}},\ }\href {\doibase 10.1007/s100520100862} {\bibfield  {journal}
  {\bibinfo  {journal} {Eur. Phys. J. C}\ }\textbf {\bibinfo {volume} {23}},\
  \bibinfo {pages} {163} (\bibinfo {year} {2002})},\ \Eprint
  {http://arxiv.org/abs/hep-ph/0108245} {arXiv:hep-ph/0108245} \BibitemShut
  {NoStop}%
\bibitem [{\citenamefont {Dobado}\ \emph {et~al.}(2002)\citenamefont {Dobado},
  \citenamefont {Herrero}, \citenamefont {Hollik},\ and\ \citenamefont
  {Penaranda}}]{Dobado:2002jz}%
  \BibitemOpen
  \bibfield  {author} {\bibinfo {author} {\bibfnamefont {A.}~\bibnamefont
  {Dobado}}, \bibinfo {author} {\bibfnamefont {M.~J.}\ \bibnamefont {Herrero}},
  \bibinfo {author} {\bibfnamefont {W.}~\bibnamefont {Hollik}}, \ and\ \bibinfo
  {author} {\bibfnamefont {S.}~\bibnamefont {Penaranda}},\ }\href {\doibase
  10.1103/PhysRevD.66.095016} {\bibfield  {journal} {\bibinfo  {journal} {Phys.
  Rev. D}\ }\textbf {\bibinfo {volume} {66}},\ \bibinfo {pages} {095016}
  (\bibinfo {year} {2002})},\ \Eprint {http://arxiv.org/abs/hep-ph/0208014}
  {arXiv:hep-ph/0208014} \BibitemShut {NoStop}%
\bibitem [{\citenamefont {Wu}\ \emph {et~al.}(2015)\citenamefont {Wu},
  \citenamefont {Yang}, \citenamefont {Yuan},\ and\ \citenamefont
  {Zhang}}]{Wu:2015nba}%
  \BibitemOpen
  \bibfield  {author} {\bibinfo {author} {\bibfnamefont {L.}~\bibnamefont
  {Wu}}, \bibinfo {author} {\bibfnamefont {J.~M.}\ \bibnamefont {Yang}},
  \bibinfo {author} {\bibfnamefont {C.-P.}\ \bibnamefont {Yuan}}, \ and\
  \bibinfo {author} {\bibfnamefont {M.}~\bibnamefont {Zhang}},\ }\href
  {\doibase 10.1016/j.physletb.2015.06.020} {\bibfield  {journal} {\bibinfo
  {journal} {Phys. Lett. B}\ }\textbf {\bibinfo {volume} {747}},\ \bibinfo
  {pages} {378} (\bibinfo {year} {2015})},\ \Eprint
  {http://arxiv.org/abs/1504.06932} {arXiv:1504.06932 [hep-ph]} \BibitemShut
  {NoStop}%
\bibitem [{\citenamefont {Cepeda}\ \emph {et~al.}(2019)\citenamefont {Cepeda}
  \emph {et~al.}}]{Cepeda:2019klc}%
  \BibitemOpen
  \bibfield  {author} {\bibinfo {author} {\bibfnamefont {M.}~\bibnamefont
  {Cepeda}} \emph {et~al.},\ }\href {\doibase 10.23731/CYRM-2019-007.221}
  {\bibfield  {journal} {\bibinfo  {journal} {CERN Yellow Rep. Monogr.}\
  }\textbf {\bibinfo {volume} {7}},\ \bibinfo {pages} {221} (\bibinfo {year}
  {2019})},\ \Eprint {http://arxiv.org/abs/1902.00134} {arXiv:1902.00134
  [hep-ph]} \BibitemShut {NoStop}%
\bibitem [{\citenamefont {Drees}(1989)}]{Drees:1988fc}%
  \BibitemOpen
  \bibfield  {author} {\bibinfo {author} {\bibfnamefont {M.}~\bibnamefont
  {Drees}},\ }\href {\doibase 10.1142/S0217751X89001448} {\bibfield  {journal}
  {\bibinfo  {journal} {Int. J. Mod. Phys. A}\ }\textbf {\bibinfo {volume}
  {4}},\ \bibinfo {pages} {3635} (\bibinfo {year} {1989})}\BibitemShut
  {NoStop}%
\bibitem [{\citenamefont {Beuria}\ \emph {et~al.}(2015)\citenamefont {Beuria},
  \citenamefont {Chatterjee}, \citenamefont {Datta},\ and\ \citenamefont
  {Rai}}]{Beuria:2015mta}%
  \BibitemOpen
  \bibfield  {author} {\bibinfo {author} {\bibfnamefont {J.}~\bibnamefont
  {Beuria}}, \bibinfo {author} {\bibfnamefont {A.}~\bibnamefont {Chatterjee}},
  \bibinfo {author} {\bibfnamefont {A.}~\bibnamefont {Datta}}, \ and\ \bibinfo
  {author} {\bibfnamefont {S.~K.}\ \bibnamefont {Rai}},\ }\href {\doibase
  10.1007/JHEP09(2015)073} {\bibfield  {journal} {\bibinfo  {journal} {JHEP}\
  }\textbf {\bibinfo {volume} {09}},\ \bibinfo {pages} {073} (\bibinfo {year}
  {2015})},\ \Eprint {http://arxiv.org/abs/1505.00604} {arXiv:1505.00604
  [hep-ph]} \BibitemShut {NoStop}%
\bibitem [{\citenamefont {Alloul}\ \emph {et~al.}(2014)\citenamefont {Alloul},
  \citenamefont {Christensen}, \citenamefont {Degrande}, \citenamefont {Duhr},\
  and\ \citenamefont {Fuks}}]{Alloul:2013bka}%
  \BibitemOpen
  \bibfield  {author} {\bibinfo {author} {\bibfnamefont {A.}~\bibnamefont
  {Alloul}}, \bibinfo {author} {\bibfnamefont {N.~D.}\ \bibnamefont
  {Christensen}}, \bibinfo {author} {\bibfnamefont {C.}~\bibnamefont
  {Degrande}}, \bibinfo {author} {\bibfnamefont {C.}~\bibnamefont {Duhr}}, \
  and\ \bibinfo {author} {\bibfnamefont {B.}~\bibnamefont {Fuks}},\ }\href
  {\doibase 10.1016/j.cpc.2014.04.012} {\bibfield  {journal} {\bibinfo
  {journal} {Comput. Phys. Commun.}\ }\textbf {\bibinfo {volume} {185}},\
  \bibinfo {pages} {2250} (\bibinfo {year} {2014})},\ \Eprint
  {http://arxiv.org/abs/1310.1921} {arXiv:1310.1921 [hep-ph]} \BibitemShut
  {NoStop}%
\bibitem [{\citenamefont {Degrande}\ \emph {et~al.}(2012)\citenamefont
  {Degrande}, \citenamefont {Duhr}, \citenamefont {Fuks}, \citenamefont
  {Grellscheid}, \citenamefont {Mattelaer},\ and\ \citenamefont
  {Reiter}}]{Degrande:2011ua}%
  \BibitemOpen
  \bibfield  {author} {\bibinfo {author} {\bibfnamefont {C.}~\bibnamefont
  {Degrande}}, \bibinfo {author} {\bibfnamefont {C.}~\bibnamefont {Duhr}},
  \bibinfo {author} {\bibfnamefont {B.}~\bibnamefont {Fuks}}, \bibinfo {author}
  {\bibfnamefont {D.}~\bibnamefont {Grellscheid}}, \bibinfo {author}
  {\bibfnamefont {O.}~\bibnamefont {Mattelaer}}, \ and\ \bibinfo {author}
  {\bibfnamefont {T.}~\bibnamefont {Reiter}},\ }\href {\doibase
  10.1016/j.cpc.2012.01.022} {\bibfield  {journal} {\bibinfo  {journal}
  {Comput. Phys. Commun.}\ }\textbf {\bibinfo {volume} {183}},\ \bibinfo
  {pages} {1201} (\bibinfo {year} {2012})},\ \Eprint
  {http://arxiv.org/abs/1108.2040} {arXiv:1108.2040 [hep-ph]} \BibitemShut
  {NoStop}%
\bibitem [{\citenamefont {Alwall}\ \emph {et~al.}(2014)\citenamefont {Alwall},
  \citenamefont {Frederix}, \citenamefont {Frixione}, \citenamefont {Hirschi},
  \citenamefont {Maltoni}, \citenamefont {Mattelaer}, \citenamefont {Shao},
  \citenamefont {Stelzer}, \citenamefont {Torrielli},\ and\ \citenamefont
  {Zaro}}]{Alwall:2014hca}%
  \BibitemOpen
  \bibfield  {author} {\bibinfo {author} {\bibfnamefont {J.}~\bibnamefont
  {Alwall}}, \bibinfo {author} {\bibfnamefont {R.}~\bibnamefont {Frederix}},
  \bibinfo {author} {\bibfnamefont {S.}~\bibnamefont {Frixione}}, \bibinfo
  {author} {\bibfnamefont {V.}~\bibnamefont {Hirschi}}, \bibinfo {author}
  {\bibfnamefont {F.}~\bibnamefont {Maltoni}}, \bibinfo {author} {\bibfnamefont
  {O.}~\bibnamefont {Mattelaer}}, \bibinfo {author} {\bibfnamefont {H.~S.}\
  \bibnamefont {Shao}}, \bibinfo {author} {\bibfnamefont {T.}~\bibnamefont
  {Stelzer}}, \bibinfo {author} {\bibfnamefont {P.}~\bibnamefont {Torrielli}},
  \ and\ \bibinfo {author} {\bibfnamefont {M.}~\bibnamefont {Zaro}},\ }\href
  {\doibase 10.1007/JHEP07(2014)079} {\bibfield  {journal} {\bibinfo  {journal}
  {JHEP}\ }\textbf {\bibinfo {volume} {07}},\ \bibinfo {pages} {079} (\bibinfo
  {year} {2014})},\ \Eprint {http://arxiv.org/abs/1405.0301} {arXiv:1405.0301
  [hep-ph]} \BibitemShut {NoStop}%
\bibitem [{\citenamefont {Bondarenko}\ \emph {et~al.}(2012)\citenamefont
  {Bondarenko}, \citenamefont {Belyaev}, \citenamefont {Blandford},
  \citenamefont {Basso}, \citenamefont {Boos}, \citenamefont {Bunichev} \emph
  {et~al.}}]{hepmdb}%
  \BibitemOpen
  \bibfield  {author} {\bibinfo {author} {\bibfnamefont {M.}~\bibnamefont
  {Bondarenko}}, \bibinfo {author} {\bibfnamefont {A.}~\bibnamefont {Belyaev}},
  \bibinfo {author} {\bibfnamefont {J.}~\bibnamefont {Blandford}}, \bibinfo
  {author} {\bibfnamefont {L.}~\bibnamefont {Basso}}, \bibinfo {author}
  {\bibfnamefont {E.}~\bibnamefont {Boos}}, \bibinfo {author} {\bibfnamefont
  {V.}~\bibnamefont {Bunichev}},  \emph {et~al.},\ }\href
  {https://hepmdb.soton.ac.uk} {\  (\bibinfo {year} {2012})},\ \Eprint
  {http://arxiv.org/abs/1203.1488} {arXiv:1203.1488 [hep-ph]} \BibitemShut
  {NoStop}%
\bibitem [{\citenamefont {Ball}\ \emph {et~al.}(2015)\citenamefont {Ball} \emph
  {et~al.}}]{NNPDF:2014otw}%
  \BibitemOpen
  \bibfield  {author} {\bibinfo {author} {\bibfnamefont {R.~D.}\ \bibnamefont
  {Ball}} \emph {et~al.} (\bibinfo {collaboration} {NNPDF}),\ }\href {\doibase
  10.1007/JHEP04(2015)040} {\bibfield  {journal} {\bibinfo  {journal} {JHEP}\
  }\textbf {\bibinfo {volume} {04}},\ \bibinfo {pages} {040} (\bibinfo {year}
  {2015})},\ \Eprint {http://arxiv.org/abs/1410.8849} {arXiv:1410.8849
  [hep-ph]} \BibitemShut {NoStop}%
\bibitem [{\citenamefont {Porod}(2003)}]{Porod:2003um}%
  \BibitemOpen
  \bibfield  {author} {\bibinfo {author} {\bibfnamefont {W.}~\bibnamefont
  {Porod}},\ }\href {\doibase 10.1016/S0010-4655(03)00222-4} {\bibfield
  {journal} {\bibinfo  {journal} {Comput. Phys. Commun.}\ }\textbf {\bibinfo
  {volume} {153}},\ \bibinfo {pages} {275} (\bibinfo {year} {2003})},\ \Eprint
  {http://arxiv.org/abs/hep-ph/0301101} {arXiv:hep-ph/0301101} \BibitemShut
  {NoStop}%
\bibitem [{\citenamefont {Porod}\ and\ \citenamefont
  {Staub}(2012)}]{Porod:2011nf}%
  \BibitemOpen
  \bibfield  {author} {\bibinfo {author} {\bibfnamefont {W.}~\bibnamefont
  {Porod}}\ and\ \bibinfo {author} {\bibfnamefont {F.}~\bibnamefont {Staub}},\
  }\href {\doibase 10.1016/j.cpc.2012.05.021} {\bibfield  {journal} {\bibinfo
  {journal} {Comput. Phys. Commun.}\ }\textbf {\bibinfo {volume} {183}},\
  \bibinfo {pages} {2458} (\bibinfo {year} {2012})},\ \Eprint
  {http://arxiv.org/abs/1104.1573} {arXiv:1104.1573 [hep-ph]} \BibitemShut
  {NoStop}%
\bibitem [{\citenamefont {Camargo-Molina}\ \emph {et~al.}(2013)\citenamefont
  {Camargo-Molina}, \citenamefont {O'Leary}, \citenamefont {Porod},\ and\
  \citenamefont {Staub}}]{Camargo-Molina:2013sta}%
  \BibitemOpen
  \bibfield  {author} {\bibinfo {author} {\bibfnamefont {J.~E.}\ \bibnamefont
  {Camargo-Molina}}, \bibinfo {author} {\bibfnamefont {B.}~\bibnamefont
  {O'Leary}}, \bibinfo {author} {\bibfnamefont {W.}~\bibnamefont {Porod}}, \
  and\ \bibinfo {author} {\bibfnamefont {F.}~\bibnamefont {Staub}},\ }\href
  {\doibase 10.1007/JHEP12(2013)103} {\bibfield  {journal} {\bibinfo  {journal}
  {JHEP}\ }\textbf {\bibinfo {volume} {12}},\ \bibinfo {pages} {103} (\bibinfo
  {year} {2013})},\ \Eprint {http://arxiv.org/abs/1309.7212} {arXiv:1309.7212
  [hep-ph]} \BibitemShut {NoStop}%
\bibitem [{\citenamefont {Camargo-Molina}\ \emph {et~al.}(2014)\citenamefont
  {Camargo-Molina}, \citenamefont {Garbrecht}, \citenamefont {O'Leary},
  \citenamefont {Porod},\ and\ \citenamefont {Staub}}]{Camargo-Molina:2014pwa}%
  \BibitemOpen
  \bibfield  {author} {\bibinfo {author} {\bibfnamefont {J.~E.}\ \bibnamefont
  {Camargo-Molina}}, \bibinfo {author} {\bibfnamefont {B.}~\bibnamefont
  {Garbrecht}}, \bibinfo {author} {\bibfnamefont {B.}~\bibnamefont {O'Leary}},
  \bibinfo {author} {\bibfnamefont {W.}~\bibnamefont {Porod}}, \ and\ \bibinfo
  {author} {\bibfnamefont {F.}~\bibnamefont {Staub}},\ }\href {\doibase
  10.1016/j.physletb.2014.08.036} {\bibfield  {journal} {\bibinfo  {journal}
  {Phys. Lett. B}\ }\textbf {\bibinfo {volume} {737}},\ \bibinfo {pages} {156}
  (\bibinfo {year} {2014})},\ \Eprint {http://arxiv.org/abs/1405.7376}
  {arXiv:1405.7376 [hep-ph]} \BibitemShut {NoStop}%
\bibitem [{\citenamefont {Deandrea}\ \emph {et~al.}(2021)\citenamefont
  {Deandrea}, \citenamefont {Flacke}, \citenamefont {Fuks}, \citenamefont
  {Panizzi},\ and\ \citenamefont {Shao}}]{Deandrea:2021vje}%
  \BibitemOpen
  \bibfield  {author} {\bibinfo {author} {\bibfnamefont {A.}~\bibnamefont
  {Deandrea}}, \bibinfo {author} {\bibfnamefont {T.}~\bibnamefont {Flacke}},
  \bibinfo {author} {\bibfnamefont {B.}~\bibnamefont {Fuks}}, \bibinfo {author}
  {\bibfnamefont {L.}~\bibnamefont {Panizzi}}, \ and\ \bibinfo {author}
  {\bibfnamefont {H.-S.}\ \bibnamefont {Shao}},\ }\href {\doibase
  10.1007/JHEP08(2021)107} {\bibfield  {journal} {\bibinfo  {journal} {JHEP}\
  }\textbf {\bibinfo {volume} {08}},\ \bibinfo {pages} {107} (\bibinfo {year}
  {2021})},\ \bibinfo {note} {[Erratum: JHEP 11, 028 (2022)]},\ \Eprint
  {http://arxiv.org/abs/2105.08745} {arXiv:2105.08745 [hep-ph]} \BibitemShut
  {NoStop}%
\bibitem [{\citenamefont {Butterworth}\ \emph {et~al.}(2016)\citenamefont
  {Butterworth} \emph {et~al.}}]{Butterworth:2015oua}%
  \BibitemOpen
  \bibfield  {author} {\bibinfo {author} {\bibfnamefont {J.}~\bibnamefont
  {Butterworth}} \emph {et~al.},\ }\href {\doibase
  10.1088/0954-3899/43/2/023001} {\bibfield  {journal} {\bibinfo  {journal} {J.
  Phys. G}\ }\textbf {\bibinfo {volume} {43}},\ \bibinfo {pages} {023001}
  (\bibinfo {year} {2016})},\ \Eprint {http://arxiv.org/abs/1510.03865}
  {arXiv:1510.03865 [hep-ph]} \BibitemShut {NoStop}%
\bibitem [{\citenamefont {Cacciari}\ \emph {et~al.}(2012)\citenamefont
  {Cacciari}, \citenamefont {Salam},\ and\ \citenamefont
  {Soyez}}]{Cacciari:2011ma}%
  \BibitemOpen
  \bibfield  {author} {\bibinfo {author} {\bibfnamefont {M.}~\bibnamefont
  {Cacciari}}, \bibinfo {author} {\bibfnamefont {G.~P.}\ \bibnamefont {Salam}},
  \ and\ \bibinfo {author} {\bibfnamefont {G.}~\bibnamefont {Soyez}},\ }\href
  {\doibase 10.1140/epjc/s10052-012-1896-2} {\bibfield  {journal} {\bibinfo
  {journal} {Eur. Phys. J. C}\ }\textbf {\bibinfo {volume} {72}},\ \bibinfo
  {pages} {1896} (\bibinfo {year} {2012})},\ \Eprint
  {http://arxiv.org/abs/1111.6097} {arXiv:1111.6097 [hep-ph]} \BibitemShut
  {NoStop}%
\bibitem [{\citenamefont {Conte}\ \emph {et~al.}(2013)\citenamefont {Conte},
  \citenamefont {Fuks},\ and\ \citenamefont {Serret}}]{Conte:2012fm}%
  \BibitemOpen
  \bibfield  {author} {\bibinfo {author} {\bibfnamefont {E.}~\bibnamefont
  {Conte}}, \bibinfo {author} {\bibfnamefont {B.}~\bibnamefont {Fuks}}, \ and\
  \bibinfo {author} {\bibfnamefont {G.}~\bibnamefont {Serret}},\ }\href
  {\doibase 10.1016/j.cpc.2012.09.009} {\bibfield  {journal} {\bibinfo
  {journal} {Comput. Phys. Commun.}\ }\textbf {\bibinfo {volume} {184}},\
  \bibinfo {pages} {222} (\bibinfo {year} {2013})},\ \Eprint
  {http://arxiv.org/abs/1206.1599} {arXiv:1206.1599 [hep-ph]} \BibitemShut
  {NoStop}%
\bibitem [{\citenamefont {Conte}\ \emph {et~al.}(2014)\citenamefont {Conte},
  \citenamefont {Dumont}, \citenamefont {Fuks},\ and\ \citenamefont
  {Wymant}}]{Conte:2014zja}%
  \BibitemOpen
  \bibfield  {author} {\bibinfo {author} {\bibfnamefont {E.}~\bibnamefont
  {Conte}}, \bibinfo {author} {\bibfnamefont {B.}~\bibnamefont {Dumont}},
  \bibinfo {author} {\bibfnamefont {B.}~\bibnamefont {Fuks}}, \ and\ \bibinfo
  {author} {\bibfnamefont {C.}~\bibnamefont {Wymant}},\ }\href {\doibase
  10.1140/epjc/s10052-014-3103-0} {\bibfield  {journal} {\bibinfo  {journal}
  {Eur. Phys. J. C}\ }\textbf {\bibinfo {volume} {74}},\ \bibinfo {pages}
  {3103} (\bibinfo {year} {2014})},\ \Eprint {http://arxiv.org/abs/1405.3982}
  {arXiv:1405.3982 [hep-ph]} \BibitemShut {NoStop}%
\bibitem [{\citenamefont {Araz}\ \emph {et~al.}(2021)\citenamefont {Araz},
  \citenamefont {Fuks},\ and\ \citenamefont {Polykratis}}]{Araz:2020lnp}%
  \BibitemOpen
  \bibfield  {author} {\bibinfo {author} {\bibfnamefont {J.~Y.}\ \bibnamefont
  {Araz}}, \bibinfo {author} {\bibfnamefont {B.}~\bibnamefont {Fuks}}, \ and\
  \bibinfo {author} {\bibfnamefont {G.}~\bibnamefont {Polykratis}},\ }\href
  {\doibase 10.1140/epjc/s10052-021-09052-5} {\bibfield  {journal} {\bibinfo
  {journal} {Eur. Phys. J. C}\ }\textbf {\bibinfo {volume} {81}},\ \bibinfo
  {pages} {329} (\bibinfo {year} {2021})},\ \Eprint
  {http://arxiv.org/abs/2006.09387} {arXiv:2006.09387 [hep-ph]} \BibitemShut
  {NoStop}%
\bibitem [{\citenamefont {Cacciari}\ \emph {et~al.}(2008)\citenamefont
  {Cacciari}, \citenamefont {Salam},\ and\ \citenamefont
  {Soyez}}]{Cacciari:2008gp}%
  \BibitemOpen
  \bibfield  {author} {\bibinfo {author} {\bibfnamefont {M.}~\bibnamefont
  {Cacciari}}, \bibinfo {author} {\bibfnamefont {G.~P.}\ \bibnamefont {Salam}},
  \ and\ \bibinfo {author} {\bibfnamefont {G.}~\bibnamefont {Soyez}},\ }\href
  {\doibase 10.1088/1126-6708/2008/04/063} {\bibfield  {journal} {\bibinfo
  {journal} {JHEP}\ }\textbf {\bibinfo {volume} {04}},\ \bibinfo {pages} {063}
  (\bibinfo {year} {2008})},\ \Eprint {http://arxiv.org/abs/0802.1189}
  {arXiv:0802.1189 [hep-ph]} \BibitemShut {NoStop}%
\bibitem [{\citenamefont {{ATLAS
  Collaboration}}(2022)}]{ATL-PHYS-PUB-2022-053}%
  \BibitemOpen
  \bibfield  {author} {\bibinfo {author} {\bibnamefont {{ATLAS
  Collaboration}}},\ }\href {http://cds.cern.ch/record/2841244} {}\bibinfo
  {type} {Tech. Rep.}\ (\bibinfo  {institution} {CERN},\ \bibinfo {address}
  {Geneva},\ \bibinfo {year} {2022})\ \bibinfo {note}
  {{ATL-PHYS-PUB-2022-053}}\BibitemShut {NoStop}%
\bibitem [{\citenamefont {Sirunyan}\ \emph
  {et~al.}(2021{\natexlab{c}})\citenamefont {Sirunyan} \emph
  {et~al.}}]{CMS:2020mpn}%
  \BibitemOpen
  \bibfield  {author} {\bibinfo {author} {\bibfnamefont {A.~M.}\ \bibnamefont
  {Sirunyan}} \emph {et~al.} (\bibinfo {collaboration} {CMS}),\ }\href
  {\doibase 10.1140/epjc/s10052-021-09014-x} {\bibfield  {journal} {\bibinfo
  {journal} {Eur. Phys. J. C}\ }\textbf {\bibinfo {volume} {81}},\ \bibinfo
  {pages} {378} (\bibinfo {year} {2021}{\natexlab{c}})},\ \Eprint
  {http://arxiv.org/abs/2011.03652} {arXiv:2011.03652 [hep-ex]} \BibitemShut
  {NoStop}%
\bibitem [{\citenamefont {Reichert}\ \emph {et~al.}(2018)\citenamefont
  {Reichert}, \citenamefont {Eichhorn}, \citenamefont {Gies}, \citenamefont
  {Pawlowski}, \citenamefont {Plehn},\ and\ \citenamefont
  {Scherer}}]{Reichert:2017puo}%
  \BibitemOpen
  \bibfield  {author} {\bibinfo {author} {\bibfnamefont {M.}~\bibnamefont
  {Reichert}}, \bibinfo {author} {\bibfnamefont {A.}~\bibnamefont {Eichhorn}},
  \bibinfo {author} {\bibfnamefont {H.}~\bibnamefont {Gies}}, \bibinfo {author}
  {\bibfnamefont {J.~M.}\ \bibnamefont {Pawlowski}}, \bibinfo {author}
  {\bibfnamefont {T.}~\bibnamefont {Plehn}}, \ and\ \bibinfo {author}
  {\bibfnamefont {M.~M.}\ \bibnamefont {Scherer}},\ }\href {\doibase
  10.1103/PhysRevD.97.075008} {\bibfield  {journal} {\bibinfo  {journal} {Phys.
  Rev. D}\ }\textbf {\bibinfo {volume} {97}},\ \bibinfo {pages} {075008}
  (\bibinfo {year} {2018})},\ \Eprint {http://arxiv.org/abs/1711.00019}
  {arXiv:1711.00019 [hep-ph]} \BibitemShut {NoStop}%
\bibitem [{\citenamefont {Akula}\ \emph {et~al.}(2017)\citenamefont {Akula},
  \citenamefont {Bal\'azs}, \citenamefont {Dunn},\ and\ \citenamefont
  {White}}]{Akula:2017yfr}%
  \BibitemOpen
  \bibfield  {author} {\bibinfo {author} {\bibfnamefont {S.}~\bibnamefont
  {Akula}}, \bibinfo {author} {\bibfnamefont {C.}~\bibnamefont {Bal\'azs}},
  \bibinfo {author} {\bibfnamefont {L.}~\bibnamefont {Dunn}}, \ and\ \bibinfo
  {author} {\bibfnamefont {G.}~\bibnamefont {White}},\ }\href {\doibase
  10.1007/JHEP11(2017)051} {\bibfield  {journal} {\bibinfo  {journal} {JHEP}\
  }\textbf {\bibinfo {volume} {11}},\ \bibinfo {pages} {051} (\bibinfo {year}
  {2017})},\ \Eprint {http://arxiv.org/abs/1706.09898} {arXiv:1706.09898
  [hep-ph]} \BibitemShut {NoStop}%
\bibitem [{\citenamefont {Aad}\ \emph {et~al.}(2022{\natexlab{c}})\citenamefont
  {Aad} \emph {et~al.}}]{ATLAS:4b}%
  \BibitemOpen
  \bibfield  {author} {\bibinfo {author} {\bibfnamefont {G.}~\bibnamefont
  {Aad}} \emph {et~al.} (\bibinfo {collaboration} {ATLAS}),\ }\href {\doibase
  10.1103/PhysRevD.105.092002} {\bibfield  {journal} {\bibinfo  {journal}
  {Phys. Rev. D}\ }\textbf {\bibinfo {volume} {105}},\ \bibinfo {pages}
  {092002} (\bibinfo {year} {2022}{\natexlab{c}})},\ \Eprint
  {http://arxiv.org/abs/2202.07288} {arXiv:2202.07288 [hep-ex]} \BibitemShut
  {NoStop}%
\bibitem [{\citenamefont {Buchalla}\ \emph {et~al.}(2018)\citenamefont
  {Buchalla}, \citenamefont {Capozi}, \citenamefont {Celis}, \citenamefont
  {Heinrich},\ and\ \citenamefont {Scyboz}}]{Buchalla_2018}%
  \BibitemOpen
  \bibfield  {author} {\bibinfo {author} {\bibfnamefont {G.}~\bibnamefont
  {Buchalla}}, \bibinfo {author} {\bibfnamefont {M.}~\bibnamefont {Capozi}},
  \bibinfo {author} {\bibfnamefont {A.}~\bibnamefont {Celis}}, \bibinfo
  {author} {\bibfnamefont {G.}~\bibnamefont {Heinrich}}, \ and\ \bibinfo
  {author} {\bibfnamefont {L.}~\bibnamefont {Scyboz}},\ }\href {\doibase
  10.1007/jhep09(2018)057} {\bibfield  {journal} {\bibinfo  {journal} {Journal
  of High Energy Physics}\ }\textbf {\bibinfo {volume} {2018}} (\bibinfo {year}
  {2018}),\ 10.1007/jhep09(2018)057}\BibitemShut {NoStop}%
\bibitem [{\citenamefont {Badziak}\ \emph {et~al.}(2013)\citenamefont
  {Badziak}, \citenamefont {Olechowski},\ and\ \citenamefont
  {Pokorski}}]{Badziak:2013bda}%
  \BibitemOpen
  \bibfield  {author} {\bibinfo {author} {\bibfnamefont {M.}~\bibnamefont
  {Badziak}}, \bibinfo {author} {\bibfnamefont {M.}~\bibnamefont {Olechowski}},
  \ and\ \bibinfo {author} {\bibfnamefont {S.}~\bibnamefont {Pokorski}},\
  }\href {\doibase 10.1007/JHEP06(2013)043} {\bibfield  {journal} {\bibinfo
  {journal} {JHEP}\ }\textbf {\bibinfo {volume} {06}},\ \bibinfo {pages} {043}
  (\bibinfo {year} {2013})},\ \Eprint {http://arxiv.org/abs/1304.5437}
  {arXiv:1304.5437 [hep-ph]} \BibitemShut {NoStop}%
\end{thebibliography}%

\end{document}